\documentclass[review]{elsarticle}
 \biboptions{comma,sort&compress}
\usepackage{graphicx}
\usepackage{amsmath}
\usepackage{here}
\usepackage{cuted}
\usepackage{xcolor}

\def\beq{\begin{equation}}
\def\eeq{\end{equation}}
\def\beqn{\begin{equation*}}
\def\eeqn{\end{equation*}}
\def\bea{\begin{eqnarray}}
\def\eea{\end{eqnarray}}
\def\bq{\begin{quote}}
\def\eq{\end{quote}}
\def\ve{\vert}
\def\nnb{\nonumber}
\def\ga{\left(}
\def\dr{\right)}

\def\lrar{\Longrightarrow}
																
\def\nnb{\nonumber}

\def\la{\langle}
\def\ra{\rangle}
\def\nin{\noindent}
\def\ba{\vspace*{-0.2cm}\begin{array}}
\def\ea{\end{array}\vspace*{-0.2cm}}

\def\b{$\bullet~$}

\def\als{\alpha_s}

\def\gg2{ \la\alpha_s G^2 \ra}
\def\gg3{g^3f_{abc}\la G^aG^bG^c \ra}
\def\ggg4{\la\als^2G^4\ra}

\usepackage{amsmath}
\usepackage{slashed}
\usepackage{color}

\begin{document}

\begin{frontmatter}

\title{Improved XTZ masses and mass ratios from  Laplace Sum Rules at NLO}
\author{R. Albuquerque}
\address{Faculty of Technology,Rio de Janeiro State University (FAT,UERJ), Brazil}
\ead{raphael.albuquerque@uerj.br}
\author{S. Narison\corref{cor1}
}
\address{Laboratoire
Univers et Particules de Montpellier (LUPM), CNRS-IN2P3, \\
Case 070, Place Eug\`ene
Bataillon, 34095 - Montpellier, France\\
and\\
Institute of High-Energy Physics of Madagascar (iHEPMAD)\\
University of Ankatso, Antananarivo 101, Madagascar}
\ead{snarison@yahoo.fr}
\author{D. Rabetiarivony}
\ead{rd.bidds@gmail.com}
\address{Institute of High-Energy Physics of Madagascar (iHEPMAD)\\
University of Ankatso, Antananarivo 101, Madagascar}

\begin{abstract}
\noindent
{ We present improved estimates of the couplings, masses and mass ratios of the $Z_Q,X_Q$ and $T_{QQ\bar q\bar q'}$ states ($Q\equiv c,b~;~q,q'\equiv u,d,s$) using (inverse) QCD Laplace sum rules (LSR), their ratios ${\cal R}$ and double ratios DRSR within 
stability criteria, where the NLO factorized PT QCD corrections are included which is important for giving a meaning on the running $\overline {MS}$ heavy quark mass used in the analysis.  We show that combined ${\cal R}$ and DRSR can provide more  precise results. In the 1st part of the paper, we conclude that the observed $X_c(3872)$ and $Z_c(3900)$ are {\it tetramoles} states (superposition of quasi-degenerated molecule and a tetraquark states having (almost) the same coupling to the currents) with the predicted masses\,:
$M_{{\cal T}_{X_c}}=3876(44)$ MeV and $M_{{\cal T}_{Z_c}}=3900(42)$ MeV. In the 2nd part, we focus on the analysis of the four-quark nature of different $T_{QQ\bar q\bar q'}$ $1^+$ and $0^+$ states within the $\bar 3_c3_c$ interpolating currents. 
 The final results from ${\cal R}$ and ${\cal R}\,\oplus$ DRSR are summarized in Table\,\ref{tab:res}.  Combined ${\cal R}$ and DRSR calibrated to the observed   $X_c(3872)$  lead to a precise prediction of e.g. $M_{T_{cc}^{1^+}}$=3886(6) MeV. In a similar way, the DRSR for the $M_{T_{cc}^{0^+}}/M_{T_{cc}^{1^+}}$ calibrated to $M_{T_{cc}^{1^+}}$ gives $M_{T_{cc}^{0^+}}$= 3883(3) MeV. The SU3 breaking ratios $M_{T_{cc\bar s\bar s}^{0^+}}/ M_{T_{cc}^{0^+}}$ lead to the improved mass predictions\,:  $M_{T_{cc\bar s\bar s}^{0^+}}$=3988(12) MeV. In the 3rd part, the analysis is extended  to the beauty mesons, where we find the tetramole masses :  $M_{{\cal T}_{Z_b}}=10579(99)$ MeV and  $M_{X_b}=10545(131)$ MeV. We also observe that the $T^{1^+,0^+}_{bb\bar q\bar q'}$ ($q,q'\equiv u,d,s$) states are (almost) stable (within the errors) against strong interactions. In the 4th part, we (critically) review and correct some recent LSR estimates of the $T^{1^+,0^+}_{QQ\bar q\bar q'}$ masses. 
 Our combined LSR $\oplus$ DRSR  results are confronted with the ones from some other approaches (lattices and quark models) in  Fig.\,\ref{fig:tcc-rev}. }

\end{abstract}
\end{frontmatter}
\pagestyle{plain}
 \section{Introduction}
 Beyond the successful quark model of Gell-Mann\,\cite{GELL}  and Zweig\,\cite{ZWEIG}, Rossi and Veneziano have introduced the four-quark states within the string model\,\cite{ROSSI} in order to describe baryon-antibaryon scattering, while Jaffe\,\cite{JAFFE1} has introduced them  within the bag models for an attempt to explain the complex structure of the $I=1,0$ light scalar mesons (see also\,\cite{ISGUR,ACHASOV,THOOFT,WEINBERG,KNECHT,LATORRE,SNa0})\,\footnote{See however\,\cite{SNS21,VENEZIA,SNG,OCHS,MENES3} for a gluonium interpretation of the light $I=0$ scalar mesons.}

In a series of papers\,\cite{MOLE12,MOLE16,MOLE16X,SU3,4Q,DK,Zc,Zb}, we have used QCD spectral sum rules (QSSR) \`a la SVZ\,\cite{SVZa,ZAKA}\,\footnote{For reviews, see e.g\,\cite{SNB1,SNB2,SNB3,IOFFEb,RRY,DERAF,BERTa,YNDB,PASC,DOSCH,COL}.}  within stability criteria to estimate the masses and couplings of different exotic XYZ states. Compared to the existing papers in the literature, we have emphasized that the inclusion of PT radiative corrections is important for justifying the choice of the input value of the heavy quark mass which plays a capital role in the analysis. In so doing, we have observed that, in the $\overline {MS}$ scheme, this correction is tiny which {\it a posteriori} explains the success of these LO results using the quark mass value in this scheme. 

 More recently, we have applied the LSR\,\cite{SVZa,BELLa,SNR} for interpreting the new states around (6.2-6.9) GeV found by the LHCb-group\,\cite{LHCb1} to be a doubly/fully hidden-charm molecules $(\bar QQ) (Q\bar Q)$ and $( \bar Q \bar Q)(QQ)$ tetraquarks states\,\cite{4Q}, while the new states found by the same group from the $DK$ invariant mass\,\cite{LHCb3} have been interpreted by a $0^+$ and $1^-$ tetramoles (superposition of almost degenerate molecules and tetraquark states having the same quantum numbers and almost the same couplings) slightly mixed with their radial excitations\,\cite{DK}. 
 We have also systematically studied the $Z_c$-like spectra and interpreted the $Z_c(3900)$ and the $Z_{cs}(3983)$ state found by BESIII\,\cite{BES3} as good candidtates for $(1^+)$ tetramole states\,\cite{Zc}. 

Motivated by the recent LHCb discovery of a $1^{+}$ state at 3875 MeV\,\cite{LHCb4}, just below the $D^*D$ threshold, which is a good isoscalar ($I=0$) $T_{cc\bar u\bar d}$ axial vector $(J^P=1^+)$ candidate, we improve in this paper the existing QSSR results by combining the direct mass determinations from the ratios ${\cal R}$ of Inverse Laplace sum rule (LSR)  with the ratio of masses from the double ratio of sum rules (DRSR). In so doing, we start by improving the previous estimate of mass and coupling of the $X_c(3872)$ which will serve as an input in our DRSR approach.   We complete our analysis by  studying the SU3 breakings for $T_{cc\bar s\bar s}$ and $T_{cc\bar s\bar u}$ states. Finally, we extend the whole study to the case of the $T_{bb\bar q\bar q'}$ states. Our results are confronted with the existing LSR results and the ones from some other approaches which are briefly reviewed. 
\section{The QCD Inverse Laplace sum rules (LSR) approach}
We shall be concerned with the two-point correlator :
 \bea
\hspace*{-0.6cm} \Pi^{\mu\nu}_{\cal H}(q^2)&=&i\int \hspace*{-0.15cm}d^4x ~e^{iqx}\la 0\vert {\cal T} {\cal O}^\mu_{\cal H}(x)\ga {\cal O}^\nu_{\cal H}(0)\dr^\dagger \vert 0\ra \nnb\\
&\equiv& -\ga g^{\mu\nu}-\frac{q^\mu q^\nu}{q^2}\dr\Pi^{(1)}_{\cal H}(q^2)+\frac{q^\mu q^\nu}{q^2} \Pi^{(0)}_{\cal H}(q^2)
 \label{eq:2point}
 \eea
built from the local hadronic operators ${\cal O}^\mu_{\cal H}(x)$  (see Table\,\ref{tab:current}).
It obeys the Finite Energy Inverse Laplace Transform Sum Rule (LSR) and their ratios:
\beq
\hspace*{-0.2cm} {\cal L}^c_n\vert_{\cal H}(\tau,\mu)=\int_{(2M_c+m_q+m_{q'})^2}^{t_c}\hspace*{-0.5cm}dt~t^n~e^{-t\tau}\frac{1}{\pi} \mbox{Im}~\Pi^{(1,0)}_{\cal H}(t,\mu)~: ~n=0,1~;~~~~~~~~~~~
 {\cal R}^c_{\cal H}(\tau)=\frac{{\cal L}^c_{1}\vert_{\cal H}} {{\cal L}^c_0\vert_{\cal H}},
\label{eq:lsr}
\eeq
 where $q,q'\equiv u,d,s$,  $M_c$  is the on-shell / pole charm quarm mass and $m_{q,q'}$ (we shall neglect $u,d$ quark masses) the running strange quark mass, $\tau$ is the LSR variable, $t_c$ is  the threshold of the ``QCD continuum" which parametrizes, from the discontinuity of the Feynman diagrams, the spectral function  ${\rm Im}\,\Pi_{\cal H}(t,m_c^2,m_s^2,\mu^2)$.  In the minimal duality ansatz which we shall use in this paper\,\footnote{Parametrization beyond the minimal duality ansatz $\oplus$ uses of high degree moments have been considered in\,\cite{DK,Zc,Zb} to estimate the masses of the 1st radial excitation of four-quark/molecule states
 and in\,\cite{SNS21,SNP21} for studying the gluonia spectra.}:
 \beq
 \frac{1}{\pi}{\rm Im}\,\Pi^{(1,0)}_{\cal H}(t) = f_{\cal H}^2M^8_{\cal H}\,\delta(t-M_{\cal H}^2)+ \frac{1}{\pi}{\rm Im}\,\Pi^{(1,0)}_{\cal H}(t)\vert_{\rm QCD}\,\theta(t-t_c),
\eeq
one can deduce the mass squared from the ratio of LSR at the optimization point $\tau_0$\,:
\beq
 {\cal R}^c_{\cal H}(\tau_0)= M_{\cal H}^2.
\eeq
We shall also work with the double ratio of sum rule (DRSR)\,\cite{DRSR88}\,:
\beq
r_{{\cal H'}/{\cal H}}(\tau_0)\equiv \sqrt{\frac{{\cal R}^c_{\cal H'}}{{\cal R}^c_{\cal H}}}=\frac{M_{\cal H'}}{M_{\cal H}},
\eeq
which can be free from systematics provided that ${\cal R}^c_{\cal H}$ and ${\cal R}^c_{\cal H'}$ optimize at the same values of $\tau$ and of $t_c$:
\beq
 \tau_0\vert_{\cal H}\simeq \tau_0\vert_{\cal H'}~,~~~~~~~~~~~~t_c\vert_{\cal H}\simeq t_c\vert_{\cal H'}~.
\eeq
This DRSR has been used in different channels for predicting successfully the few MeV mass-splittings (SU3-breakings, parity splittings,...) between different hadrons\,\cite{DRSR88,DRSR94,DRSR96,DRSR07,DRSR10,DRSR11,DRSR11a}. In particular, it has been used for four-quark and molecule states in\,\cite{DRSR07,DRSR11,DRSR11a}. In this paper, we extend the previous analysis for improving the existing mass predictions of the $X_Q,Z_Q$ and $T_{QQ\bar q\bar q}$ states and for giving a correlation among them. We also predict the mass-splittings due to SU3 breakings and to spin and parity for the $T_{QQ\bar q\bar q'}$  states.

\section{The stability criteria for extracting the optimal results\label{sec:stability}}
In the LSR analysis, we have three external variables: the LSR variiable $\tau\equiv 1/M^2_B$ where $M_B$ is the original varibale used by SVZ\,\cite{SVZa},   the QCD continuum threshold $t_c$ and the subtraction point $\mu$.  One considers that physical observables like the masses and meson couplings should be independent / minimal sensitive on these parameters. 
\subsection*{\b The $\tau$-stabiity}
\begin{figure}[hbt]
\begin{center}
\centerline {\hspace*{-4.5cm} \bf a)\hspace{8cm} b)}
\includegraphics[width=6cm]{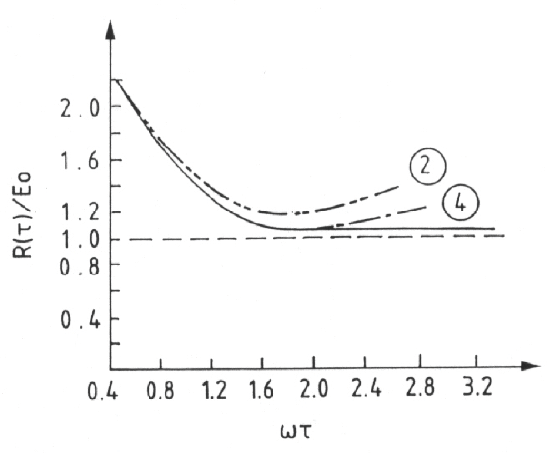} 
\hspace*{2cm}\includegraphics[width=6cm]{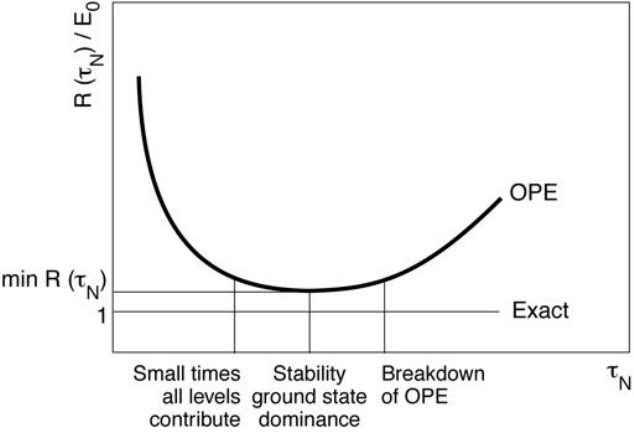}
\centerline {\hspace*{-7.5cm} \bf c) }
\includegraphics[width=10cm]{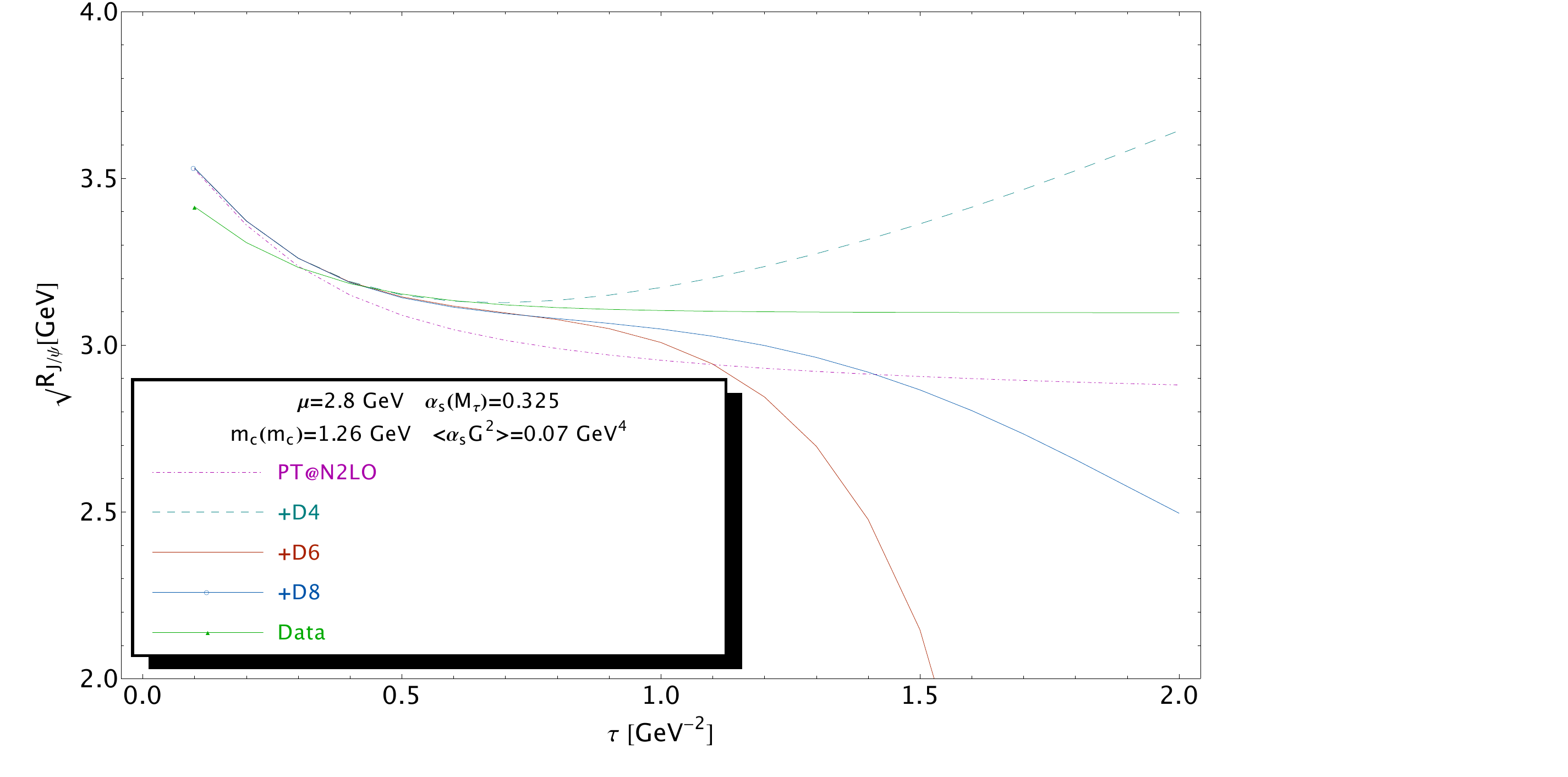}
\vspace*{-0.5cm}
\caption{\footnotesize  a) Harmonic oscillator state for each given truncation of the series compared to the exact solution (horizontal line); b) Schematic presentation of stability of the charmonium ratio of moments; c) Explicit analysis of the $J/\psi$ systems moment for different truncation of the OPE from e.g. \,\cite{SNparam}. } 
\label{fig:oscillo}
\end{center}
\vspace*{-0.5cm}
\end{figure}

 It has been studied  from  the example of the harmonic oscillator in quantum mechanics\,\cite{BERTa,BELLa} and from its analogue  charmonium non-relativistic form of the LSR shown in Fig.\,\ref{fig:oscillo}.  A such quantum mechanic example has been explicitly checked for vector charmonium and bottomium systems where complete data are available (see e.g.\,\cite{SNparam} and  the $J/\psi$ systems in Fig.\,\ref{fig:oscillo}) and in many other examples in\,\cite{SNB1,SNB2} and different original papers by the authors.  

 At this $\tau$ stability point where there is a balance bewteen the low and high-energy region, one can check the lowest ground state dominance of the LSR and the convergence of the OPE. 

\subsection*{\b The $t_c$-stabiity}
 The QCD continuum threshold $t_c$ is (in principle) a free parameter in the analysis though one (intuitively) expects it to be around the mass of the first excitation which cannot be accurate as the QCD continuum is supposed to smear all higher radial exctiations contributions to the spectral function. 

To be  conservative we take $t_c$ from the beginning of $\tau$-stability until the beginning  of $t_c$-stability\,\cite{SNB1,SNB2,SNB3} where the $t_c$-stability region corresponds to a complete dominance of the lowest ground state in the QSSR analysis. This conservative range of $t_c$-values is larger than the usual choice done in the current literature which is often done at the lowest values of $t_c$ where one starts to have the $\tau$-stability. 

\subsection*{\b The $\mu$-stabiity}
This is used to fix in a rigorous optimal way, the arbitrary substraction constant  appearing in the PT calculation of the Wilson coefficients and in the QCD input renormalized parameters. We have observed in our previous analysis for the four-quark and molecule states\,\cite{MOLE16,Zc,Zb,DK} that its value is (almost) universal :
\beq
\mu_c\simeq  4.65(5)~{\rm GeV}~,~~~~~~~~~~~~~~~~~~\mu_b\simeq  5.20(5)~{\rm GeV},
\label{eq:mu}
\eeq
respectively for the charm and beauty states. We shall check this result explicitly in the next sections.

 One can also alternatively eliminate the $\mu$-dependence of the result,  by working with the resummed quantity after applying the homogeneous Renormalization Group equation (RGE) obeyed by the QCD expression of the LSR which is superconvergent\,:
\beq
\Big{\{}-\frac{\partial }{\partial t} +\beta(\alpha_s)\alpha_s \frac{\partial }{\partial \alpha_s}-\sum_i(1+\gamma_m(\alpha_s) x_i\frac{\partial }{\partial x_i}\Big{\}}{\cal L}^c_n(e^t \tau,\alpha_s,x_i,\mu)=0~,
\eeq
where $t\equiv (1/2)L_\tau$, $x_i\equiv m_i/\mu$. The renormalization group improved (RGI) solution is:
\beq
{\cal L}^c_n(e^t \tau,\alpha_s,x_i)= {\cal L}^c_n(t=0,\bar\alpha_s(\tau), \bar x_i(\tau))~,
\eeq
where $\bar\alpha_s(\tau)$ and $\bar x_i(\tau)$ are the running QCD coupling and mass. However, the RGE solution $\mu^2=1/tau$ corresponds to lower values of $\mu\approx 1.6$ GeV where the convergence of the PT series is slower than in the previous case in Eq.\,\ref{eq:mu}. 

An explicit comparison of the results from these two ways can be found in\,\cite{SNFB13}. However, one should remark that the choice $\mu^2=1/tau$ correponds to a value of $\mu$ lower than the optimized one in Eq.\,\ref{eq:mu} where NLO corrections are larger. 

\subsection*{\b Importance of the Figures in the analysis}
We emphasize the importance for showing the different figures for each channels though having similar behaviour as they provide convincing proofs of the choice of the set of external parameters $(\tau,t_c,\mu)$ in the stability region for each channels studied.

\section{The interpolating operators}
 In the first of the paper, we choose to work with the $\bar 3_c3_c$ lowest dimension interpolating currents of the four-quark states  given  in Table\,\ref{tab:current}. 

 Some other choices such as $\bar 6_c6_c$,  $\bar 8_c8_c$ and/or higher dimension operators used in the current literature will be checked and (critically)  reviewed in the second part of the paper. 

 The chiral partner $1^-$ and $0^-$ states and  the molecule assignements of the $T_{QQ\bar q\bar q'}$ states which deserves a particular attention due to the numerous possibilities of such assignements are postponed  in a future publication.
\vspace*{-0.5cm} 
   {\scriptsize
\begin{table}[hbt]
\setlength{\tabcolsep}{0.6pc}
    {\small
  \begin{tabular}{llll}
&\\
\hline
\hline
States&$I(J^P)$& $\bar 3_c3_c$ Four-quark  Currents& Refs.\\
\hline
$Z_c$ &$(1^{+})$
&$ {\cal O}_{A_{cq}}=\epsilon_{ijk}\epsilon_{mnk}\big{
[}(q^T_i\,C\gamma_5\,c_j)(\bar q'_m \gamma_\mu C\, \bar c_n^T)
\,+b,(q^T_i\,C\,c_j)(\bar q'_m\gamma_\mu \gamma_5 C\, 
\bar c_n^T)\big{]}$&\cite{Zc}\\
&&$ {\cal O}_{D^*_qD_q}=(\bar c\gamma_\mu q)(\bar q'\,i\gamma_5c)$  \\
 $X_c $&$(1^{+})$
&$ {\cal O}^3_{X} = \epsilon_{i j k} \:\epsilon_{m n k} \big{[}\left(
  q_i^T\, C \gamma_5 \,c_j \right) \left( \bar{c}_m\, \gamma^\mu
  C \,\bar{q}_n^T\right) + \left(
  q_i^T\, C \gamma^\mu \,c_j \right) \left( \bar{c}_m\, \gamma_5
  C \,\bar{q}_n^T\right)\big{]}$&\cite{DRSR07,DRSR11,DRSR11a,MOLE16} \\
 && $ {\cal O}^6_{X} = \epsilon_{i j k} \:\epsilon_{m n k}\big{[} \left(
  q_i^T\, C \gamma_5 \lambda_{ij}^a\,c_j \right) \left( \bar{c}_m\, \gamma^\mu
  C\lambda_{mn}^a \,\bar{q}_n^T\right) +  \left(
  q_i^T\, C \gamma^\mu \lambda_{ij}^a\,c_j \right) \left( \bar{c}_m\, \gamma_5
  C \lambda_{mn}^a\,\bar{q}_n^T\right)\big{]}$ \\
  && $ {\cal O}_{D^*_qD_q}= \frac{1}{\sqrt{2}}\big{[}(\bar q\gamma_5 c) (\bar c\gamma_\mu q) -  (\bar q\gamma_\mu c) (\bar c\gamma_5 q)\big{]} $ \\
 & & $ {\cal O}_{\psi\pi} = (\bar c\gamma_\mu\lambda^a c)(\bar q\gamma_5\lambda^a q)$\\
 $T_{cc\bar u\bar d}$&$0(1^{+})$
 &  $ {\cal O}_T^{1^+} = \frac{1}{\sqrt{2}}\epsilon_{i j k} \:\epsilon_{m n k} \left(
 c_i^T\, C \gamma^\mu \,c_j \right) \big{[} \left( \bar{u}_m\, \gamma_5
  C \,\bar{d}_n^T\right) -  \left( \bar{d}_m\, \gamma_5
  C \,\bar{u}_n^T\right)\big{]}$&\cite{DRSR11} \\
      $T_{cc\bar u\bar s}$&$\frac{1}{2}(1^{+})$
 &  $ {\cal O}_{T^{1^+}_{us}}   = \epsilon_{i j k} \:\epsilon_{m n k}
    \left( c_i \, C \gamma^{\mu } c_j^T \right) 
    \left( \bar{u}_m \,\gamma_5 C \bar{s}_n^T \right)$\\
  
$T_{cc\bar u\bar d}$&$1(0^{+})$
 &  $ {\cal O}_T^{0^+} = \frac{1}{\sqrt{2}}\epsilon_{i j k} \:\epsilon_{m n k} \left(
 c_i^T\, C \gamma^\mu \,c_j \right) \big{[} \left( \bar{u}_m\, \gamma_\mu
  C \,\bar{d}_n^T\right) + \left( \bar{d}_m\, \gamma_\mu
  C \,\bar{u}_n^T\right)\big{]}$&\cite{DRSR11} \\
  
    $T_{cc\bar u\bar s}$&$\frac{1}{2}(0^{+})$
 &  $ {\cal O}_{T^{0^+}_{us}}   = \epsilon_{i j k} \:\epsilon_{m n k}
    \left( c_i \, C \gamma_{\mu } c_j^T \right) 
    \left( \bar{u}_m \,\gamma^{\mu } C \bar{s}_n^T \right)$\\

  $T_{cc\bar s\bar s}$&$0(0^{+})$
 &  $ {\cal O}_T^{0^+}    = \epsilon_{i j k} \:\epsilon_{m n k}
    \left( c_i \, C \gamma_{\mu } c_j^T \right) 
    \left( \bar{s}_m \,\gamma^{\mu } C \bar{s}_n^T \right)$\\
   \hline\hline
  \vspace*{-0.5cm}
\end{tabular}}
 \caption{Interpolating operators describing the $Z_c,X_c,T_{cc\bar q'\bar q}$ states discussed in this paper where $b=0$ is the optimized mixing parameter\,\cite{Zc}.}  

\label{tab:current}
\end{table}
} 
\section{QCD input parameters}
\nin

The QCD parameters which shall be used here 
are the QCD coupling $\alpha_s$,  the charm quark mass $m_{c}$,
the gluon condensates $ \la\alpha_sG^2\ra$.  Their values  are given in Table\,\ref{tab:param}. 
We shall use  $n_f$=4 and 5 total number of flavours for the numerical value of $a_s\equiv\alpha_s/ \pi$.

{\small
\begin{table}[H]
\setlength{\tabcolsep}{1.9pc}
    {\small
  \begin{tabular}{llll}
&\\
\hline
\hline
Parameters&Values&Sources& Refs.    \\
\hline
$\alpha_s(M_Z)$& $0.1181(16)(3)$&$M_{\chi_{0c,b}-M_{\eta_{c,b}}}$&
\cite{SNparam,SNparam2,SNm20} \\
$\overline{m}_c(m_c)$ [MeV]&$1266(6)$ &$D, B_c \oplus {J/\psi}, \chi_{c1},\eta_{c}$&
\cite{SNm20,SNparam,SNbc20,SNmom18,SNFB13,SNH10,SNH11}\\
$\overline{m}_b(m_b)$ [MeV]&$4196(8)$ &$B_c\oplus{\Upsilon}$&
\cite{SNm20,SNH10,SNH11,SNH12,SNparam,SNbc20,SNmom18,SNFB13}\\
$\hat \mu_q$ [MeV]&$253(6)$ &Light  &\,\cite{SNB1,SNp15} \\
$\hat m_s$ [MeV]&$114(6)$ &Light &\,\cite{SNB1,SNp15} \\
$\kappa\equiv\la \bar ss\ra/\la\bar dd\ra$& $0.74(6)$&Light-Heavy&\cite{SNB1,SNp15,HBARYON1}\\
$M_0^2$ [GeV$^2$]&$0.8(2)$ &Light-Heavy&\,\cite{SNB1,DOSCH,JAMI2a,JAMI2c,HEIDa,HEIDc,SNhl} \\
$\la\alpha_s G^2\ra$ [GeV$^4$]& $6.35(35) 10^{-2}$&Light-Heavy &
 \cite{SNparam,SNm20}\\
${\la g^3  G^3\ra}/{\la\alpha_s G^2\ra}$& $8.2(1.0)$[GeV$^2$]&${J/\psi}$&\cite{SNH10,SNH11}\\
$\rho \alpha_s\la \bar qq\ra^2$ [GeV$^6$]&$5.8(9) 10^{-4}$ &Light,$\tau$-decay&\cite{DOSCH,SNTAU,JAMI2a,JAMI2c,LNT,TARRACH,LAUNERb}\\
\hline\hline
\end{tabular}}
 \caption{QCD input parameters estimated from QSSR (Moments, LSR and ratios of sum rules) used here. 
 }  
\label{tab:param}
\end{table}
} 
 
\section{The $Z_c (1^{+})$ state}
\subsection*{\b Mass and decay constant from LSR}
The extraction of the $Z_c$ mass has been discussed in details in Ref.\,\cite{Zc} using the current in Table\,\ref{tab:current} where the main source of the errors in the mass determination is the localization of the
inflexion point  at which the optimal value is extracted ($\Delta M=40$ MeV) and the trunctation of the OPE ($\Delta M=39$ MeV). The results for a $D^*D$ molecule and for a four-quark state configurations are\,\cite{Zc,MOLE16}:
\beq
M_{D*D}=3912(61)~{\rm MeV},~~~~~~~~~~~~~~M_{A_{cd}}=3889(58)~{\rm MeV},
\eeq
which are almost degenerated (we do not consider the isospin violation). 
\subsection*{\b Ratio $r_{A_{cd}/D*D}$ of masses from DRSR}
\begin{figure}[hbt]
\begin{center}
\includegraphics[width=10cm]{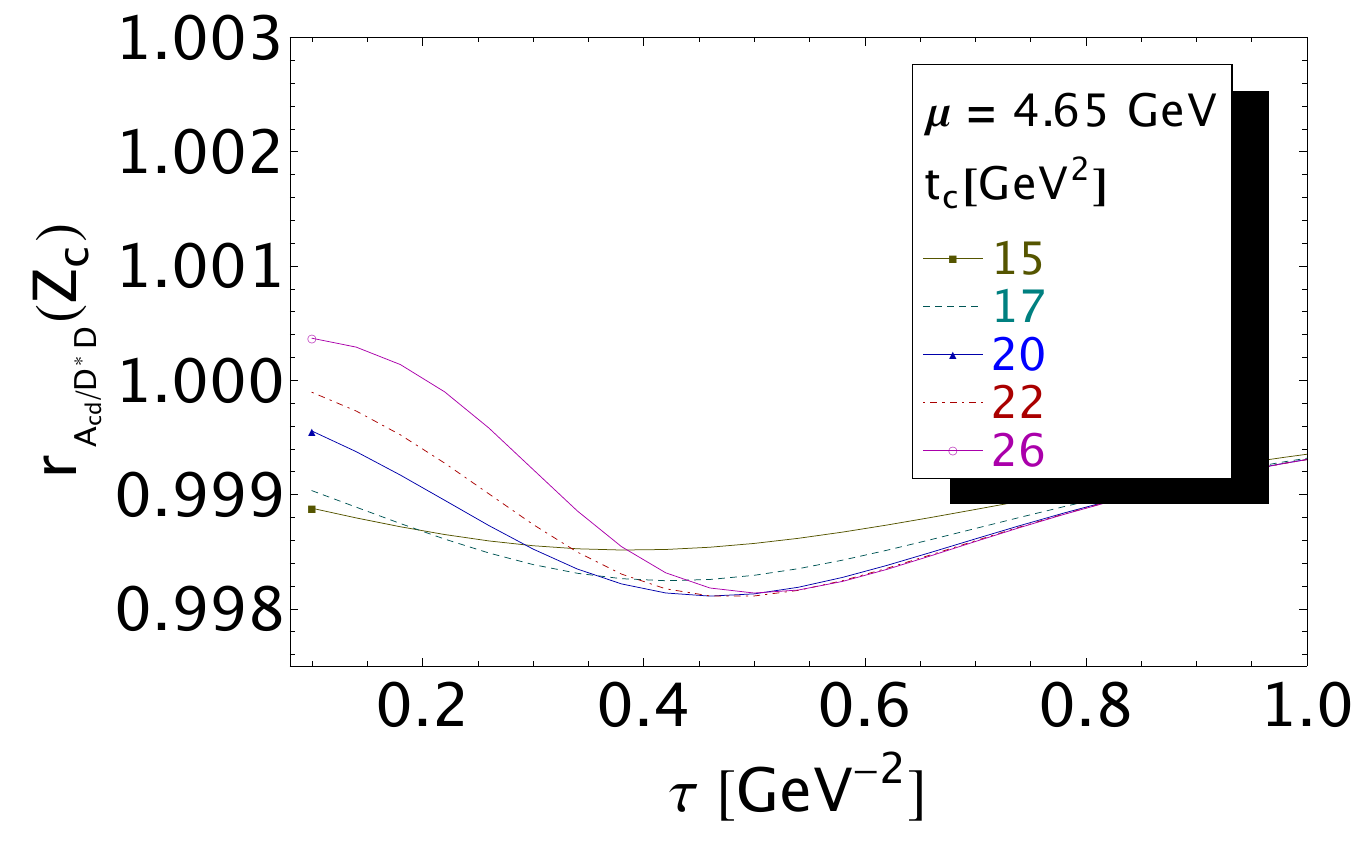}
\vspace*{-0.5cm}
\caption{\footnotesize  $r_{A_{cd}/D*D}$  as function of $\tau$ at NLO for \# values of $t_c$, for $\mu$=4.65 GeV\,\cite{Zc,Zb} and for the QCD inputs in Table\,\ref{tab:param}.} 
\label{fig:racd}
\end{center}
\vspace*{-0.5cm}
\end{figure}

We use the DRSR for studying the ratio of masses. The analysis is shown in Fig.\,\ref{fig:racd}. The optimal result is obtained for the 
sets $(\tau,t_c)=$(0.46,~20) $(\rm GeV^{-2},\rm GeV^2)$ where both present minimum. At these values,  one deduces:
\beq
r_{A_{cd}/D*D}= 0.9981(6)~~~~\lrar~~~~M_{A_{cd}}=3905(61)~{\rm MeV},
\eeq
which consolidates the previous result from a direct determination. 

\subsection*{\b ${\cal T}_{Z_c}$ tetramole}
Noting in\,\cite{Zc,Zb} that the molecule $D^*D$ and the four-quark states are almost degenerated and have almost the same coupling to their respective current, we expect the physically observed state to be their mean which we named {\it tetramole}  (${\cal T}_{Z_c}$). One obtains\,:
\beq
M_{{\cal T}_{Z_c}}=3900(42)~{\rm MeV},~~~~~~~~~~~~~~f_{{\cal T}_{Z_c}}=155(11)~{\rm keV}.
\eeq
which coincides with the experimental $Z_c(3900)$ mass. 

\begin{table*}[hbt]
\setlength{\tabcolsep}{0.25pc}
{\scriptsize{
\begin{tabular*}{\textwidth}{@{}ll ll  ll  ll ll ll ll ll ll ll ll r@{\extracolsep{\fill}}l}
\hline
\hline
        					&\multicolumn{1}{c}{$X_c$}
					&\multicolumn{1}{c}{$T^{1^+}_{cc}$}
					&\multicolumn{1}{c}{$T^{1^+}_{ccqs}$}
					&\multicolumn{1}{c}{$T^{0^+}_{cc}$}
					&\multicolumn{1}{c}{$T^{0^+}_{ccqs}$}
					&\multicolumn{1}{c}{$T^{0^+}_{ccss}$}
					&\multicolumn{1}{c}{$\frac{A_{cd}}{D^*D}$}
					&\multicolumn{1}{c}{$\frac{6}{3}$}
					&\multicolumn{1}{c}{$\frac{\psi\pi}{3}$}
					&\multicolumn{1}{c}{$\frac{T^{1^+}_{cc}}{X_c}$}
					&\multicolumn{1}{c}{$\frac{T^{1^+}_{ccqs}}{T^{1^+}_{ccqq}}$}
					&\multicolumn{1}{c}{$\frac{T^{0^+}_{cc}}{X_c}$}
					&\multicolumn{1}{c}{$\frac{T^{0^+}_{cc}}{T^{1^+}_{cc}}$}
					&\multicolumn{1}{c}{$\frac{T^{0^+}_{ccqs}}{T^{0^+}_{ccqq}}$}
					&\multicolumn{1}{c}{${\frac{T^{0^+}_{ccss}}{T^{0^+}_{ccqq}}}$}

                  \\
\hline
$\bf t_{c}$~ &30 - 46&30 - 46&30 - 46&30 - 46&30 - 46&30 - 46&~~20&20&15 - 20&15 - 20&23 - 32&15 - 20&17 - 22&23 - 32&23 - 32\\
$\bf\tau$~ &$36\ ;37$&$31\ ;34$&$32\ ;35$&$31\ ;34$&$32\ ;35$&$32\ ;35$&$~~46$&$46$&$132;136$&$124;130$&$72\ ;74$&$128 ;132$&$50\ ;74$&$72\ ;74$&$72\ ;74$\\
\hline
\hline
\end{tabular*}
}}
 \caption{{\small Values of the set of LSR parameters $(t_c,\tau)$ in units of (GeV$^2$, GeV$^{-2}\times 10^2$) at the optimization region for the PT series up to NLO and for the OPE truncated at the dimension-six condensates and for $\mu=4.65$  GeV for the charm states.}}
 \vspace*{0.25cm}
\label{tab:tctauc}
\end{table*}

\begin{table}[hbt]
\setlength{\tabcolsep}{0.3pc}
{\scriptsize{
\begin{tabular}{ll ll  ll  ll ll ll ll ll r}
\hline
\hline
                \bf Observables &\multicolumn{1}{c}{$\Delta t_c$}
					&\multicolumn{1}{c}{$\Delta \tau$}
					&\multicolumn{1}{c}{$\Delta \mu$}
					&\multicolumn{1}{c}{$\Delta \alpha_s$}
					&\multicolumn{1}{c}{$\Delta PT$}
					&\multicolumn{1}{c}{$\Delta m_s$}
					&\multicolumn{1}{c}{$\Delta m_c$}
					&\multicolumn{1}{c}{$\Delta \overline{\psi}\psi$}
					&\multicolumn{1}{c}{$\Delta \kappa$}					
					&\multicolumn{1}{c}{$\Delta G^2$}
					&\multicolumn{1}{c}{$\Delta M^{2}_{0}$}
					&\multicolumn{1}{c}{$\Delta \overline{\psi}\psi^2$}
					&\multicolumn{1}{c}{$\Delta G^3$}
					&\multicolumn{1}{c}{$\Delta OPE$}
					&\multicolumn{1}{c}{$\Delta M_{G}$}
					&\multicolumn{1}{r}{Values}\\

\hline
{\bf Coupling} [keV] &&&&&&&&&&&\\
$f_{X_c}$ &1.43&0.17&0.85&4.25&0.40&$\cdots$&2.49&1.67&$\cdots$&0.02&1.89&7.71&0.00&10.9&5.32&183(16) \\
$f_{T^{1^+}_{cc}}$ &7.22&0.55&2.14&10.2&4.02&$\cdots$&6.00&0.00&$\cdots$&0.13&0.00&27.0&0.02&33.6&14.2&491(48) \\
$f_{T^{1^+}_{ccqs}}$ &4.93&0.36&1.42&6.70&3.59&0.13&4.11&0.11&8.22&0.10&0.27&16.0&0.02&20.4&8.65&317(30) \\
$f_{T^{0^+}_{cc}}$ &13.0&0.95&3.75&17.8&4.17&$\cdots$&10.3&0.00&$\cdots$&0.12&0.00&47.2&0.16&58.4&25.3&841(83) \\
$f_{T^{0^+}_{ccqs}}$ &8.73&0.62&2.48&11.7&3.85&0.21&7.06&0.22&14.4&0.009&0.21&28.1&0.13&35.6&14.8&542(53) \\
$f_{T^{0^+}_{ccss}}$ &14.3&0.86&3.29&15.5&4.87&0.93&9.85&0.34&34.3&0.16&0.44&32.6&0.22&41.2&33.7&718(75) \\
\\
{\bf Mass} [MeV] &&&&&&&&&&&&&&\\
$M_{X_c}$ & 17.2&48.6&2.42&13.4&0.02&$\cdots$&5.93&8.48&$\cdots$&0.07&5.58&4.10&0.00&52.9&$\cdots$&3876(76)\\
$M_{T^{1^+}_{cc}}$&8.66&59.4&3.03&12.1&0.07&$\cdots$&5.20&0.00&$\cdots$&0.10&0.00&7.93&0.09&39.4&$\cdots$&3885(74)\\
$M_{T^{1^+}_{ccqs}}$&9.90&56.9&3.13&15.2&0.00&1.63&5.18&0.30&5.49&0.08&0.80&9.64&0.11&65.0&$\cdots$&3940(89)\\
$M_{T^{0^+}_{cc}}$ &6.90&58.2&2.86&12.2&0.00&$\cdots$&4.91&0.00&$\cdots$&0.12&0.00&11.9&0.17&52.9&$\cdots$&3882(81)\\
$M_{T^{0^+}_{ccqs}}$ &8.20&57.8&2.96&14.4&0.02&1.54&4.86&0.10&5.70&0.18&0.39&9.70&0.24&66.1&$\cdots$&3936(90)\\
$M_{T^{0^+}_{ccss}}$&1.00&59.0&3.04&14.8&0.01&3.61&4.65&0.02&7.17&0.26&0.73&9.10&0.36&36.2&$\cdots$&4063(72)\\
\\
{\bf Ratio} &&&&&&&&&&&\\
$r_{A_{cd}/D^*D}$ &0.20&0.20&0.00&0.01&0.00&$\cdots$&0.03&0.06&$\cdots$&0.00&0.48&0.11&0.04&0.28&$\cdots$&0.9983(6) \\
$r_{6/3}$ &0.25&0.25&0.00&0.02&0.00&$\cdots$&0.05&0.08&$\cdots$&0.03&0.84&0.21&0.00&0.35&$\cdots$&0.9969(10) \\
$r_{\psi\pi/3}$ &0.03&0.01&0.01&0.03&0.00&$\cdots$&0.03&0.06&$\cdots$&0.01&0.54&0.15&0.00&0.37&$\cdots$&1.0034(7) \\
$r_{T^{1^+}_{cc}/X_c}$ &0.04&0.01&0.01&0.04&0.00&$\cdots$&0.04&0.09&$\cdots$&0.01&0.58&0.16&0.01&0.76&$\cdots$&1.0035(10) \\
$r_{T^{1^+}_{ccqs}/T^{1^+}_{ccqq}}$ &0.03&0.01&0.02&0.12&0.00&0.76&0.04&0.06&0.39&0.01&0.01&0.29&0.01&0.92&$\cdots$&1.0115(13) \\
$r_{T^{0^+}_{cc}/X_c}$ &0.03&0.00&0.01&0.04&0.00&$\cdots$&0.05&0.07&$\cdots$&0.00&0.56&0.17&0.01&0.76&$\cdots$&1.0033(10) \\
$r_{T^{0^+}_{cc}/T^{1^+}_{cc}}$ &0.10&0.00&0.00&0.02&0.00&$\cdots$&0.01&0.00&$\cdots$&0.05&0.00&0.09&0.02&0.17&$\cdots$&0.9994(2) \\
$r_{T^{0^+}_{ccqs}/T^{0^+}_{ccqq}}$ &0.04&0.02&0.01&0.12&0.00&0.76&0.04&0.06&0.36&0.01&0.02&0.26&0.01&0.85&$\cdots$&1.0113(12) \\
$r_{T^{0^+}_{ccss}/T^{0^+}_{ccqq}}$ &0.14&0.10&0.05&0.27&0.00&1.79&0.10&0.08&0.85&0.04&0.06&0.55&0.04&1.72&$\cdots$&1.0280(27) \\
\\
\hline
\hline
\end{tabular}
}}
 \caption{Sources of errors of $T_{cc}$, $X_c$ and their ratios of masses. We take $\ve \Delta \mu\ve=0.05$ GeV and $\ve \Delta \tau\ve =0.01$ GeV$^{-2}$. For ratios, the errors quoted in the table are multiplied by a factor of $10^3$}

\label{tab:error}
\end{table}

\section{Revisiting the $X_c (1^{+})$ state}
\subsection*{\b Mass and decay constant from the ${\cal O}^3_X$ current using LSR}
\begin{figure}[hbt]
\begin{center}
\centerline {\hspace*{-7.5cm} \bf a)\hspace{8cm} b)}
\includegraphics[width=8cm]{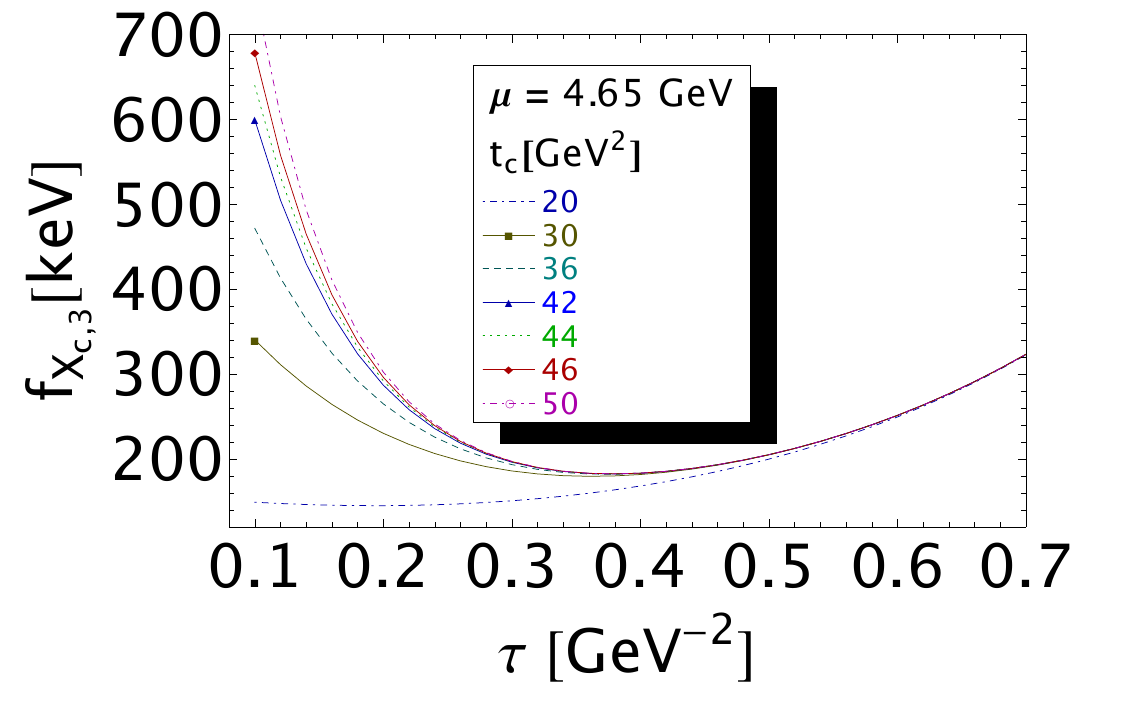}
\includegraphics[width=8cm]{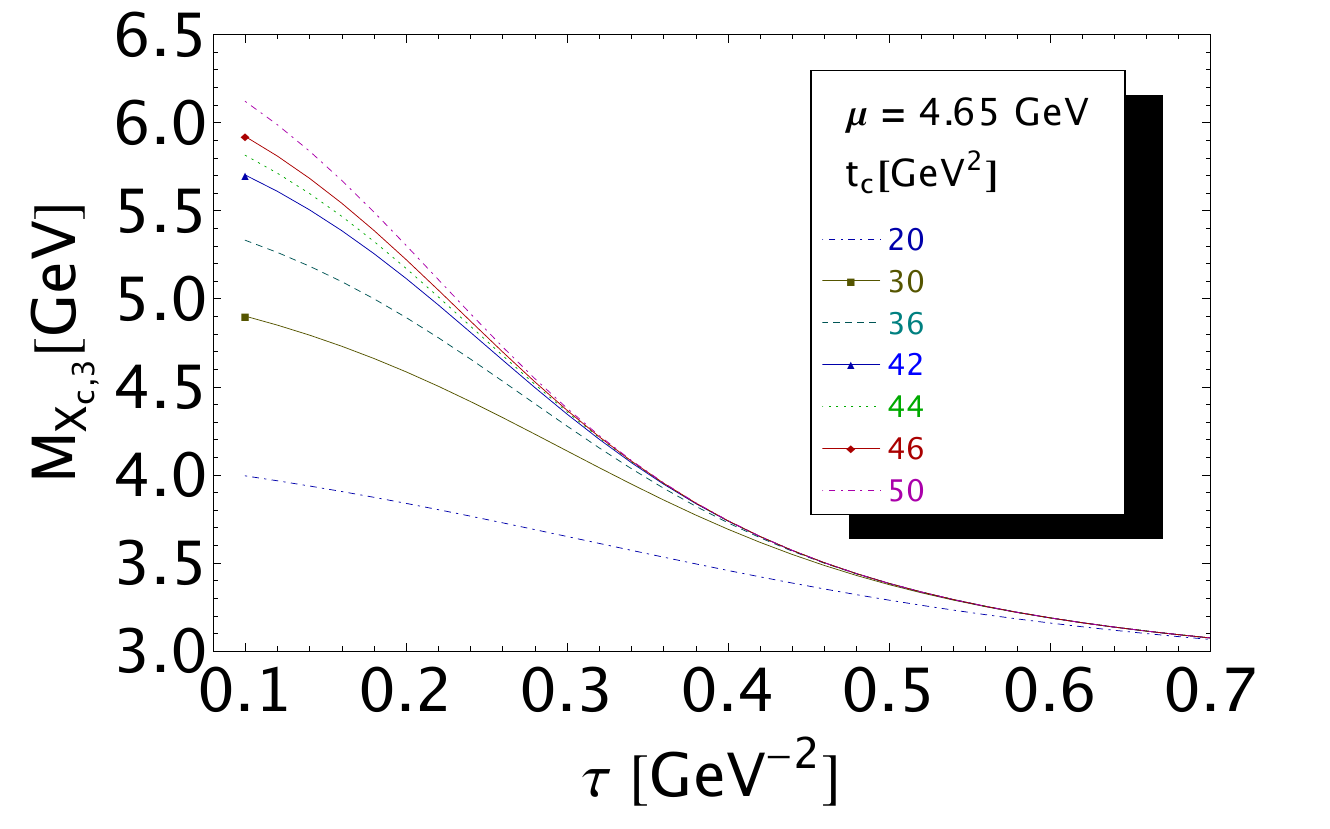}
\vspace*{-0.5cm}
\caption{\footnotesize  $f_{X_{c,3}}$ and $M_{X_{c,3}}$ as function of $\tau$ at NLO for \# values of $t_c$, for $\mu$=4.65 GeV and for the QCD inputs in Table\,\ref{tab:param}.} 
\label{fig:xc3}
\end{center}
\vspace*{-0.5cm}
\end{figure}

The mass and coupling of the $X_c (1^{+})$ have been extracted to lowest order (LO) \,\cite{DRSR07,DRSR11,DRSR11a} using the interpolating four-quark currents given in Table\,\ref{tab:current} and molecule $D^*D$ and $J/\psi\pi$ currents given in the original papers and quoted in Table\,\ref{tab:current}. These early results have been improved in\,\cite{MOLE16} for the ${\cal O}^3_T$ current by including NLO PT corrections in order to justify the use of the running heavy quark mass of the $\overline{MS}$-scheme in the analysis. We have noticed that  the localization of the inflexion point where the mass is extracted is one of the main source of the errors. We repeat the analysis of\,\cite{MOLE16} here by paying attention on this choice of $\tau$. We show the analysis in Fig.\ref{fig:xc3}.
Using the value of the set $(\tau,t_c)=(0.36,30)$ to $(0.37,46)$ $(\rm GeV^{-2},\rm GeV^2)$ corresponding to the $\tau$ minimum of $f_{X_{c,3}}$ which is necessary for a better localization of the inflexion point of $M_{X_{c,3}}$, we obtain :
\beq
f_{X_{c,3}}=183(16)~{\rm keV},~~~~~~~~~~~~~~M_{X_{c,3}}= 3876(76) ~\rm MeV,
\label{eq:x3}
\eeq
 where $f_{X_{c,3}}$ is normalized as $f_\pi=131$ MeV. The set of $(\tau,t_c)$ used in the optimization procedure are given in Table\,\ref{tab:tctauc}. The different sources of errors are given in Table\,\ref{tab:error}.
 One can notice the remarkable agreement of the central value of the mass with the data 3871.69(17) MeV\,\cite{PDG}.

\subsection*{\b $\mu$-dependence of the mass and decay constant from ${\cal O}^3_X$ using LSR}
\begin{figure}[hbt]
\begin{center}
\centerline {\hspace*{-7.5cm} \bf a)\hspace{8cm} b)}
\includegraphics[width=8cm]{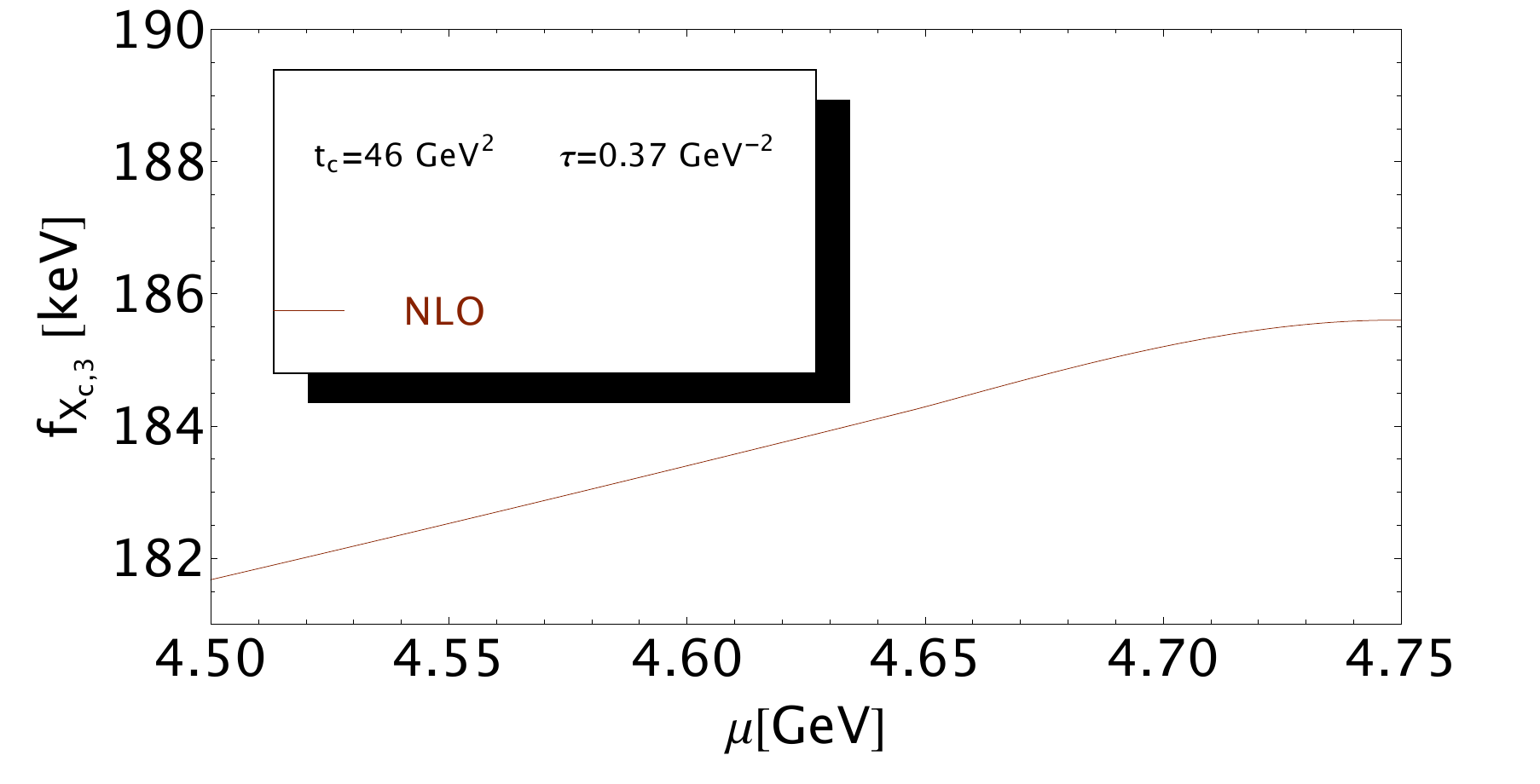}
\includegraphics[width=8cm]{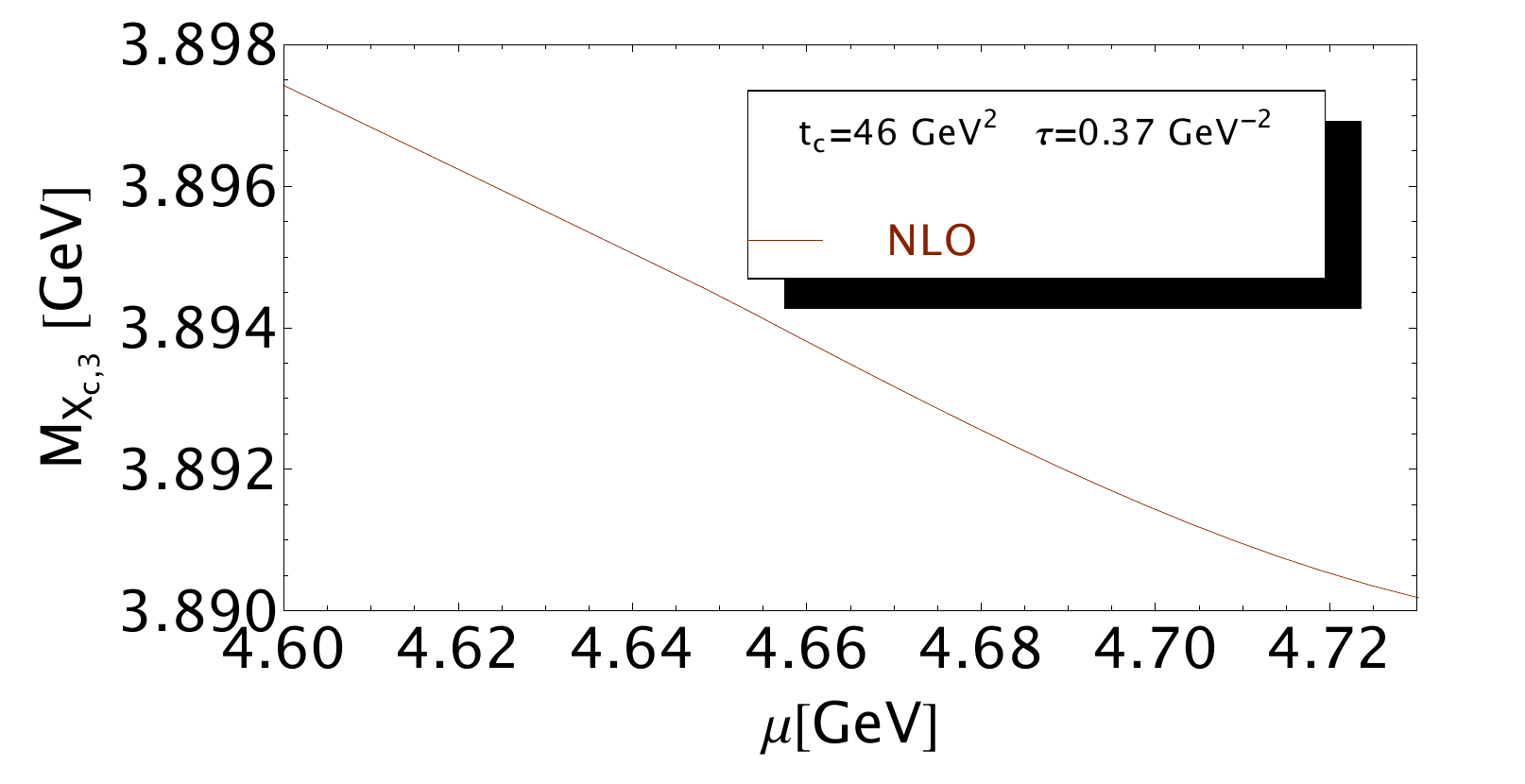}
\vspace*{-0.5cm}
\caption{\footnotesize  $f_{X_{c,3}}$ and $M_{X_{c,3}}$ as function of $\mu$ at NLO for given values of $t_c$and $\tau$ for the QCD inputs in Table\,\ref{tab:param}.} 
\label{fig:xc3mu}
\end{center}
\vspace*{-0.5cm}
\end{figure}

We show in Fig.\,\ref{fig:xc3mu} the $\mu$-dependence of $f_{X_{c,3}}$ and  $M_{X_{c,3}}$ for given values of $t_c=46$ GeV$^2$ and of $\tau=0.37$ GeV$^{-2}$. The optimal result is obtained at:
\beq
\mu_c=(4.65\pm 0.05)~{\rm GeV}~,
\eeq
which appears to be an (almost) universal value for the four-quark and molecule states analysis of the charm quark channels\,\cite{MOLE16,Zc,Zb,DK}. This value of $\mu$ will be used in the analysis of the charm states in the rest of the paper.

\subsection*{\b Mass from the ${\cal O}^6_X$ current using DRSR}
\begin{figure}[hbt]
\begin{center}
\includegraphics[width=10cm]{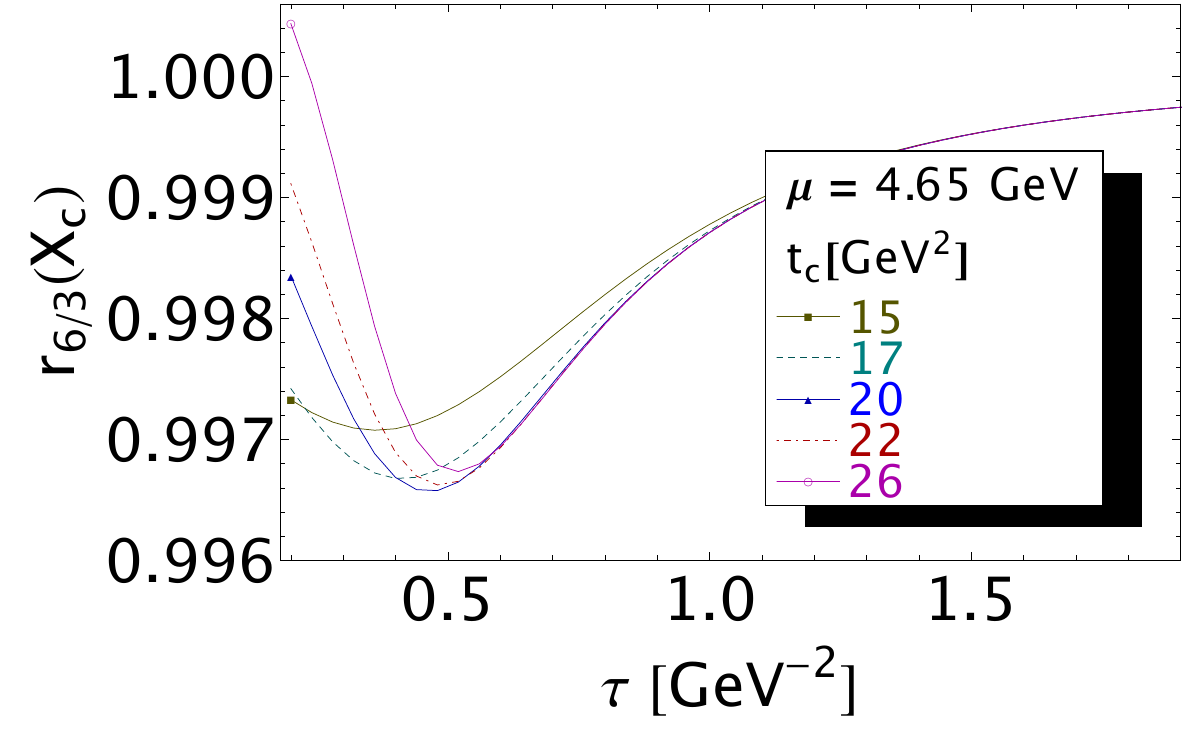}
\vspace*{-0.5cm}
\caption{\footnotesize  $r_{6/3}$  as function of $\tau$ at NLO for \# values of $t_c$, for $\mu$=4.65 GeV and for the QCD inputs in Table\,\ref{tab:param}.} 
\label{fig:r63}
\end{center}
\vspace*{-0.5cm}
\end{figure}

In the following, we improve the analysis in\,\cite{DRSR11} by paying attention on the different sources of errors. 
We consider the double ratio of sum rules (DRSR)  $r_{6/3}$ which we show in Fig.\,\ref{fig:r63}. The optimal result is obtained for the set  $(\tau,t_c)= (0.46,20)$  $(\rm GeV^{-2},\rm GeV^2)$ corresponding to the minimum of $\tau$ and $t_c$ for $r_{6/3}$. One can notice that the stability region is obtained at earlier value of $t_c$ for the DRSR compared to the one for the LSR due to the partial cancellation of the QCD continuum contribution in the DRSR. We obtain:
\beq
r_{6/3}=0.9966(10)~~~~
\lrar~~~~
M_{X_{c,6}}=3863(76)~{\rm MeV},
\eeq
where we have used the previous predicted  mass for $M_{X_{c,3}}$.
\subsection*{\b Mass from the ${\cal O}_{\psi\pi}$ current using DRSR}
\begin{figure}[hbt]
\begin{center}
\includegraphics[width=10cm]{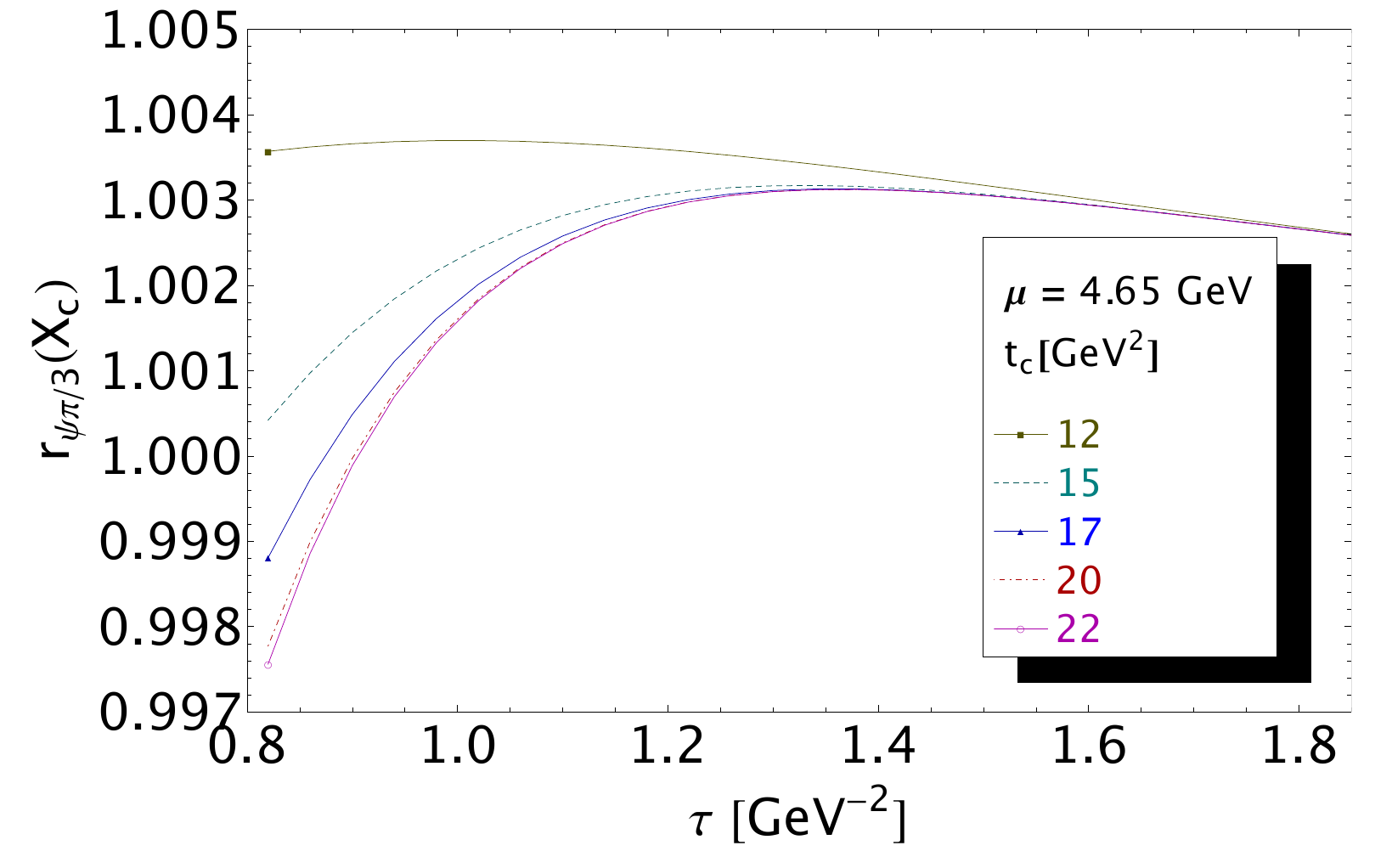}
\vspace*{-0.5cm}
\caption{\footnotesize   $r_{\psi\pi/3}$ as function of $\tau$ at NLO for \# values of $t_c$, for $\mu$=4.65 GeV and for the QCD inputs in Table\,\ref{tab:param}.} 
\label{fig:rpsi3}
\end{center}
\vspace*{-0.5cm}
\end{figure}

The analysis is shown in Fig.\,\ref{fig:rpsi3}. Here, the DRSR presents maximum in $\tau$. The optimal result is obtained for the sets  $(\tau,t_c)=(1.32,15)$ and $(1.36,20)$ in units of $(\rm GeV^{-2},\rm GeV^2)$ corresponding to the region of $\tau$ maximum of $r_{\psi/3}$ and to the stability of $t_c$.
We obtain:
\beq
r_{\psi\pi/3}=1.0034(7)~~~~ \lrar~~~~ M_{X_{c,\psi\pi}}=3889(76)~{\rm MeV},
\eeq
where we have used the previous predicted  mass for $M_{X_{c,3}}$.. Notice that in\,\cite{DRSR11}, the optimal value has been taken in a (misleading) minimum of $\tau$ where the result does not have $t_c$-stability. 
\subsection*{\b Mass from the ${\cal O}_{D^*D}$ current using DRSR}
This current has been studied in Ref.\,\cite{ZANETTI,DRSR11,MOLE16}. Here we use the compact integrated QCD expression of the spectral function from\,\cite{MOLE16} for our analysis of the DRSR $r_{D^*D/3}$.  Inspecting our QCD expression of $X_{c,D^*D}$ from\,\cite{MOLE16} and the one for $Z_{c,D^*D}$ in \cite{Zc}, one can deduce that in our approximation without isospin violation,($m_u=m_d=0$  and $\la\bar uu\ra=\la\bar dd\ra$) the two expressions are identical (isospin symmetry) such that :
\beq
r_{D^*D/3}=1~~~~ \lrar~~~~ M_{X_{c,D^*D}}=M_{Z_{c,D^*D}}=3912(61)~{\rm MeV},
\eeq
We plan to analyze the isospin violation in a future work. 
\subsection*{\b ${\cal T}_{X_c}$ tetramole}
Taking the fact that the different assignements to $X_c$ lead to almost degenerated states and almost the same coupling to the currents, we consider that the observed state is their combination which we call tetramole ${\cal T}_{X_c}$ with the mean mass and coupling\,:
\beq
M_{{\cal T}_{X_c}}=3876(44)~{\rm MeV},~~~~~~~~~~~~~~f_{{\cal T}_{X_c}}=183(16)~{\rm keV}.
\eeq
We have not included the contribution of the $D^*D$ molecule as it does not take into account the isospin violation. 
\section{Conclusion from the $Z_c$ and $X_c$ analysis}

 From the previous discussions, one can notice that the sum rules reproduce quite well the experimental masses of the $X_c(3872)$ and $Z_c(3900)$ within the molecules or/and four-quark state configurations. The DRSR has improved the accuracy of the
predictions compared to the previous ones in the literature\,\footnote{For reviews on previous LO QCD spectral sum rules results  in the literature, see e.g.\,\cite{MOLEREV,ZHUREV,RAPHAEL}. See also Ref.\,\cite{STEELE}.}. 

 However, one can notice as in \cite{DRSR11a} that the alone study of the mass of the $X_c$ and $Z_c$ cannot provide a sharp selection for the four-quark and/or molecule nature of these states without studying in details their decay modes. At the present stage, we can only provide a description of these states as {\it tetramole} ({$\cal T$}) states. 
 
 Another point which deserves future studies is the careful analysis of isospin violation which can differentiate the role of $D^*D, DD,...$ 
in the molecule description of these states. We plan to come back to this point in a future work. 

 In the following part of the paper, we shall definitely use the experimental mass $X_c(3872)$ for a normalization of the DRSR analysis of the $T_{ccqq}$-like states together with the corresponding four-quark current ${\cal O}^3_X$ which provides the best prediction compared to the data (see Eq.\,\ref{eq:x3}). Instead, we could have also choosen to work with the currents $D^*D$ and $A_{cd}$ which also reproduce quite well the experimental $Z_c(3900)$ mass. Unfortunately,  the corresponding  DRSR do not present $\tau$-stability. 
 
 Hereafter, the $X_{c,3}$ state will be also called $X_c$ and will be  identified with the experimental $X_c(3872)$ state. 
\section{The $T_{cc\bar u\bar d}\equiv T_{cc}$  $(1^{+})$ state}
Since, the pioneering work of\,\cite{LEE}, the mass and coupling of $T_{cc\bar q\bar q'}$ and its beauty analogue have been extracted from LSR by different groups\,\cite{DRSR11,WANG-Ta,WANG-Tb,ZHU-T,AGAEV-T,MALT-T}. In this paper, we improve and extend the analysis in\,\cite{DRSR11} using LSR and DRSR by including the factorized NLO PT contributions and by paying more carefully attention on the different sources of the errors.  In the follwing, we shall consider the four-quark currents 
given in Table\,\ref{tab:current}.
\subsection*{\b  Mass and decay constant from LSR at NLO}
\begin{figure}[hbt]
\begin{center}
\centerline {\hspace*{-7.5cm} \bf a)\hspace{8cm} b)}
\includegraphics[width=8cm]{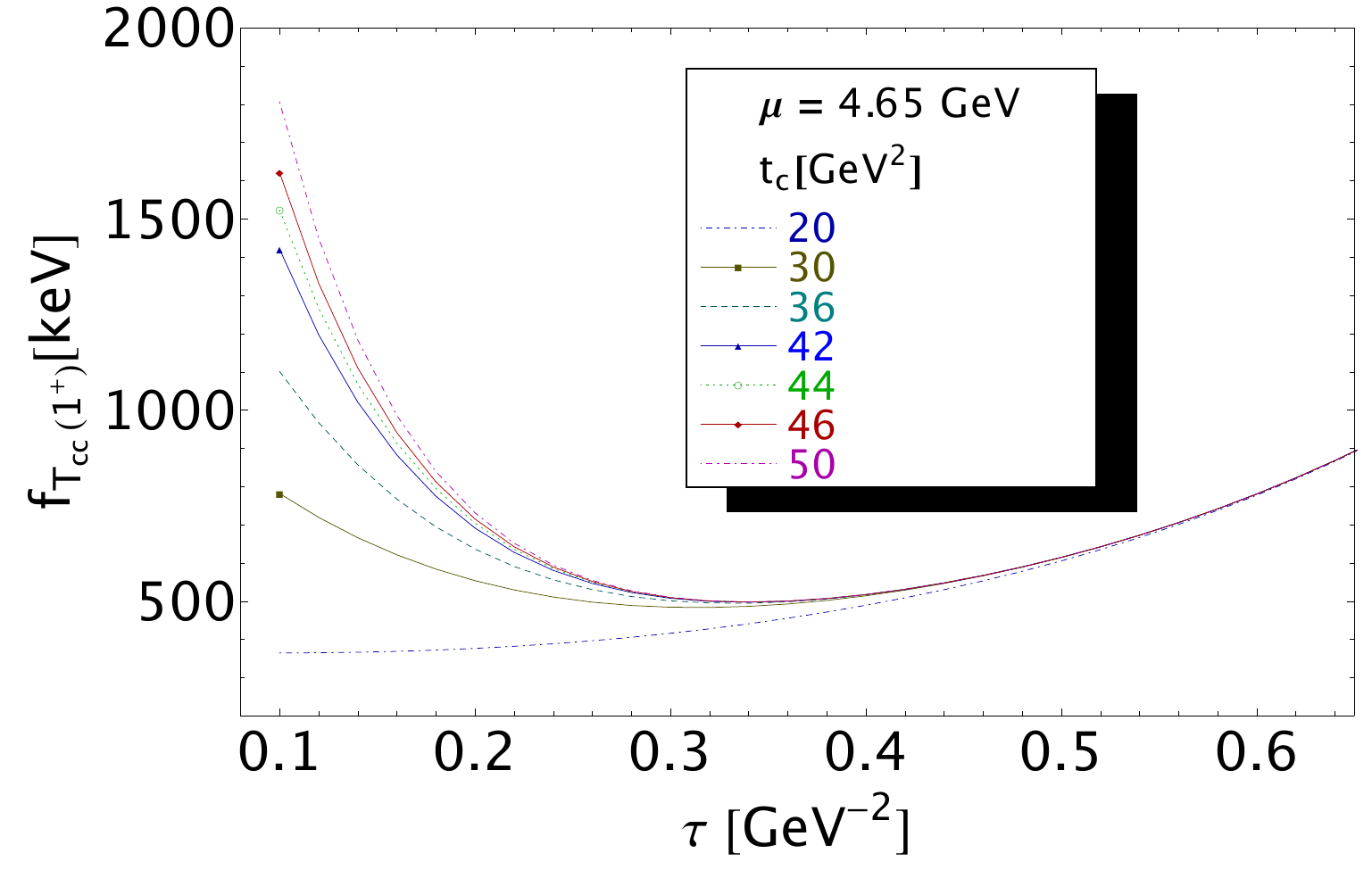}
\includegraphics[width=8cm]{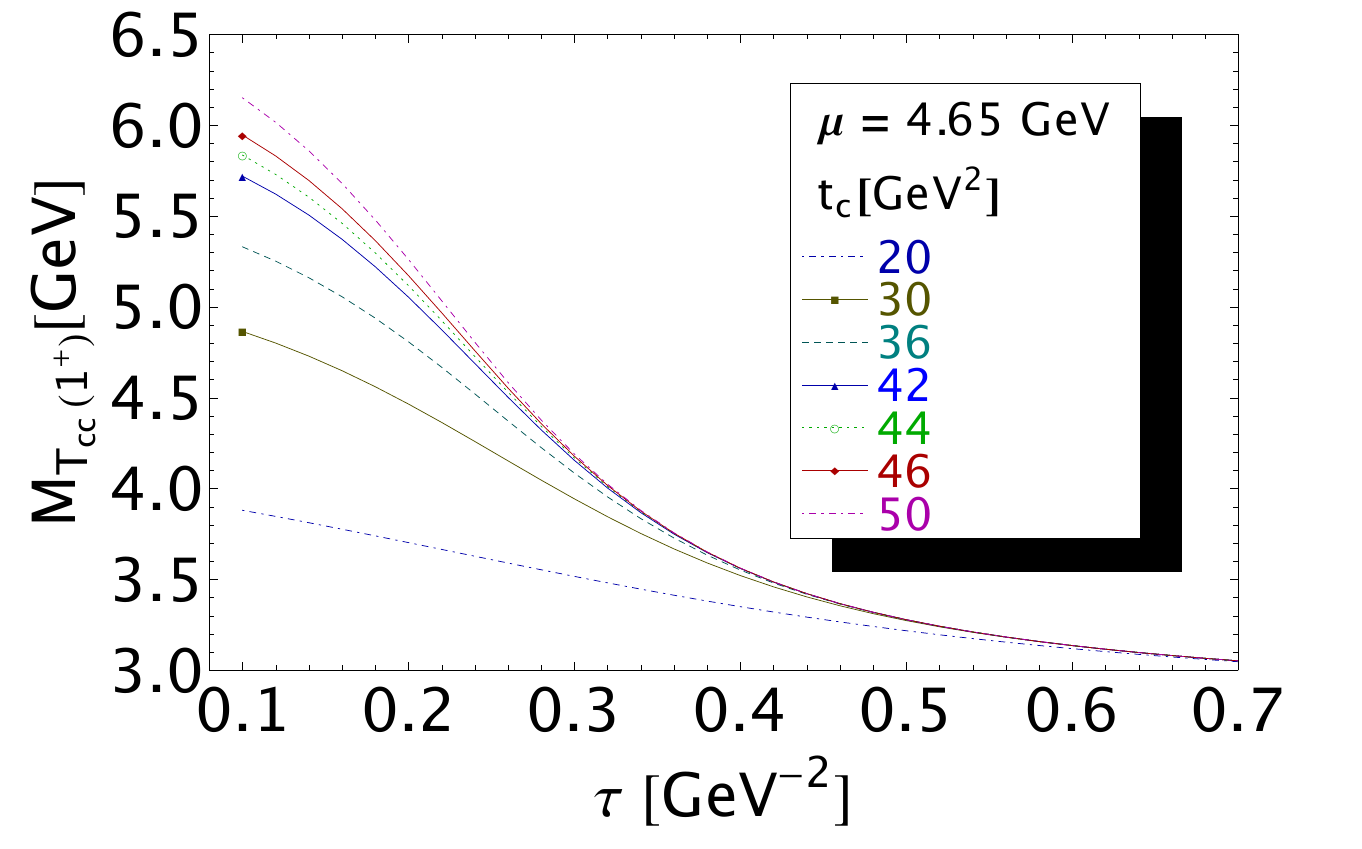}
\vspace*{-0.5cm}
\caption{\footnotesize  $f_{T^{0^+}_{cc}}$ and $M_{T^{0^+}_{cc}}$ as function of $\tau$  for \# values of $t_c$, for $\mu$=4.65 GeV and for the QCD inputs in Table\,\ref{tab:param}.} 
\label{fig:tpcc}
\end{center}
\vspace*{-0.5cm}
\end{figure}

The $\tau$ and $t_c$ behaviours is very similar to the case of $X_c$ and are shown in Fig.\,\ref{fig:tpcc}. The stability region (minimum in $\tau$ for the coupling and inflexion point for the mass) is obtained for the sets $(\tau,t_c)$=(0.31,30) to (0.34,46) in units of (GeV$^{-2}$, GeV$^2$) (see Table\,\ref{tab:tctauc}) from which we deduce:
\beq
f_{T_{cc}}(1^{+})=491(48)~{\rm KeV},~~~~~~~~~~~~M_{T_{cc}}(1^{+})=3885(123)~{\rm MeV}~,
\label{eq:mt+cc}
\eeq
where the mass can be compared with the experimental value $M_{T_{cc}}(1^{+})=3875$ MeV\,\cite{LHCb4}. 

\subsection*{\b  Ratio of masses $r_{T^{1^+}_{cc}/X_c}$ from DRSR}
The result of the analysis is very similar to the one  in Fig.\,\ref{fig:rpsi3}. The optimal result is obtained  for the sets $(\tau,t_c)$=(1.24,15) to (1.30,20)  in units of (GeV$^{-2}$, GeV$^2$) (see Table\,\ref{tab:tctauc}) \,:
\beq
r_{T^{1^+}_{cc}/X_c}=1.0035(10)~~~~ \lrar~~~~ M_{T_{cc}}(1^{+})=3886(4)~{\rm MeV}
\eeq
where we have taken the experimental mass of the $X_c(3872)$\,\cite{PDG}. The result is in perfect agreement with the direct mass determination in Eq.\,\ref{eq:mt+cc} but very accurate as the DRSR is less affected by systematics which tend to cancel out. 
\subsection*{\b  Final prediction for $M_{T_{cc}}(1^{+})$}
As a final prediction, we take the mean of the two previous determinations and take the most precise error:
\beq
M_{T_{cc}}(1^{+})= 3886(4)~{\rm MeV}.
\label{eq:1+cc}
\eeq
This value is comparable with the recent LHCb data $T_{cc}(1^{+})=3875$ MeV which is $(9\pm 4)$ MeV above the $D^*D$ threshold of 3877 MeV\,\cite{PDG}. 
\section{The $T_{cc\bar s\bar u}(1^{+})$   mass}
\subsection*{\b  $r_{T^{1^+}_{cc\bar s\bar u}/T^{1^+}_{cc}}$ ratio of masses}
\begin{figure}[hbt]
\begin{center}
\includegraphics[width=10cm]{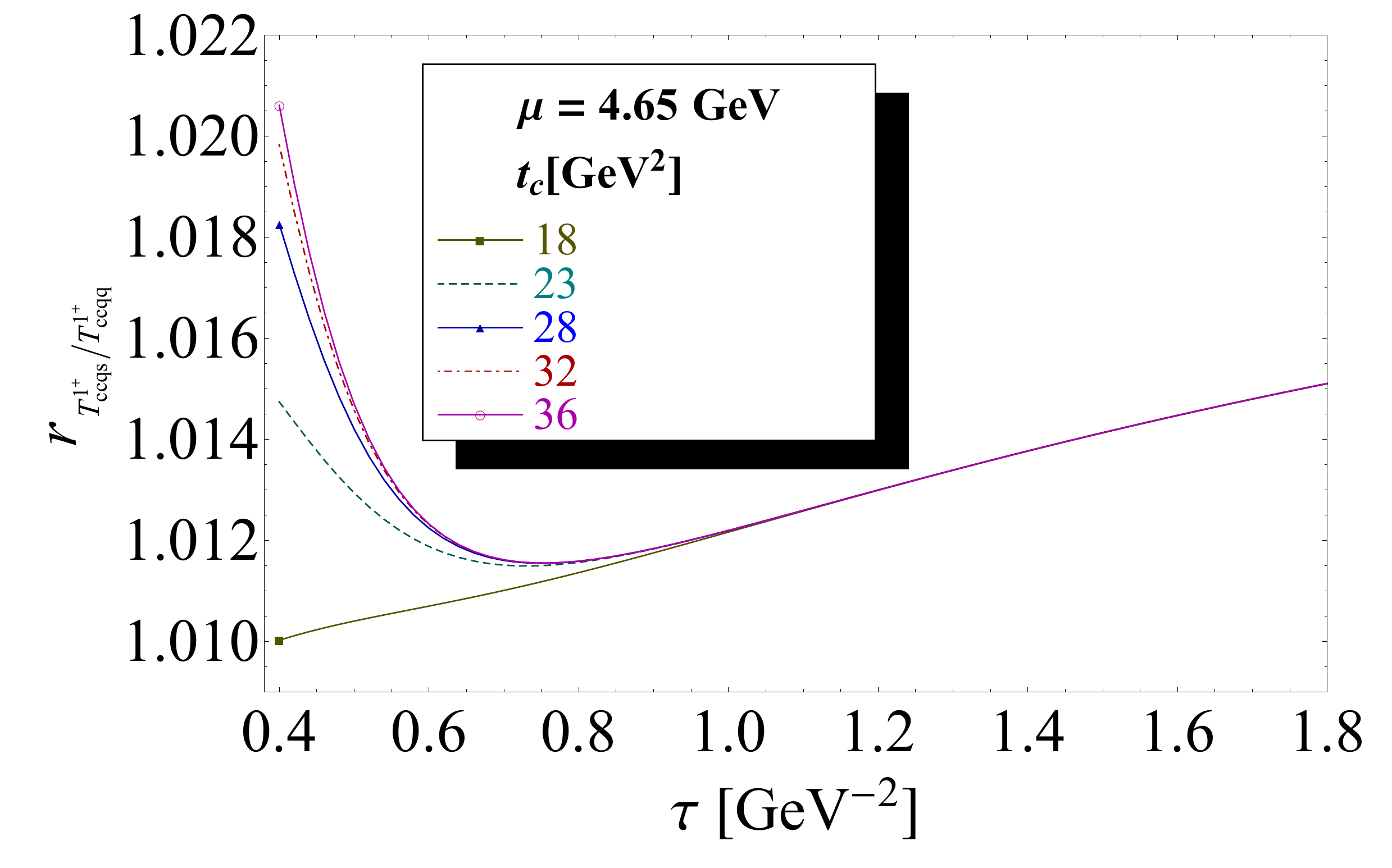}
\vspace*{-0.5cm}
\caption{\footnotesize   $r_{T_{cc\bar s\bar u}/T_{cc}}(1^+)$ ratio of masses as function of $\tau$ at NLO for \# values of $t_c$, for $\mu$=4.65 GeV and for the QCD inputs in Table\,\ref{tab:param}.} 
\label{fig:rcc1us}
\end{center}
\vspace*{-0.5cm}
\end{figure}

We study the SU3 ratio of masses $r_{T^{1^+}_{cc\bar s\bar u}/T^{1^+}_{cc}}$ in Fig.\,\ref{fig:rcc1us}. The optimal result is obtained for the sets $(\tau,t_c)$=(0.72,23) to (0.74,32)  (GeV$^{-2}$, GeV$^2$) at which we deduce:
\beq
r_{T_{cc\bar s\bar u}/T_{cc}(1^+)}=1.0115(13) ~~~~ \lrar~~~~ M_{T_{cc\bar s\bar u}}(1^{+})=3931(7)~{\rm MeV},
\eeq

\subsection*{\b  Decay constant and mass from LSR at NLO}
Here, we extract directly the  $T_{cc\bar s\bar u}$ coupling and mass from the LSR moments and ratio of moments.
The $\tau$ and $t_c$-behaviours are very similar to the one in Fig.\,\ref{fig:xc3}. The optimal result is  obtained for the sets $(\tau,t_c)$=(0.32,30) to (0.35,46) in units of (GeV$^{-2}$, GeV$^2$) (see Table\,\ref{tab:tctauc}) at which the coupling presents minimum and the mass an inflexion point :
\beq
f_{T_{cc\bar s\bar u}}(1^{+})=317(30)~{\rm keV},~~~~~~~~~~~~M_{T_{cc\bar s\bar u}}(1^{+})=3940(89)~{\rm MeV}~,
\eeq
\subsection*{\b  Final result}
As a final result for the mass, we take the mean from the DRSR and LSR ratios:
\beq
 M_{T_{cc\bar s\bar u}}(1^{+})=3931(7)~{\rm MeV}~.
 \eeq

\section{The $T_{cc\bar u\bar d}$ or $T_{cc}$  $(0^{+})$ state}
\subsection*{\b  Mass and decay constant from LSR at NLO}
We pursue the analysis for the case of $0^{+}$ state. The $\tau$ and $t_c$-behaviours are very similar to the ones in Fig.\,\ref{fig:xc3}. The optimal results are obtained with the sets\,:$(\tau,t_c)$=(0.31, 30) to (0.34, 46) (GeV$^{-2}$, GeV$^2$):
\beq
f_{T_{cc}}(0^{+})=841(83)~{\rm KeV},~~~~~~~~~~~~M_{T_{cc}}(0^{+})=3882(129)~{\rm MeV}~,
\label{eq:mt0cc}
\eeq
\subsection*{\b  Ratio of masses $r_{T^{0^+}_{cc}/X_c}$ from DRSR}
The result of the analysis is very similar to the one in Fig.\,\ref{fig:rpsi3} from which we deduce the optimal reults for the sets $(\tau,t_c)$=(1.28, 15) to (1.32, 20)\,(GeV$^{-2}$, GeV$^2$):
\beq
r_{T^{0^+}_{cc}/X_c}=1.0033(10)~~~~ \lrar~~~~ M_{T_{cc}}(0^{+})=3885(4)~{\rm MeV},
\eeq
where $X_c(3872)$ from the data has been used. The result from DRSR agrees completely with the direct determination but more accurate where the sources of the errors can be found in Table\,\ref{tab:error}.  

\subsection*{\b  Ratio of masses $r_{T^{0^+}_{cc}/T^{1^+}_{cc}}$ from DRSR}
\begin{figure}[hbt]
\begin{center}
\includegraphics[width=10cm]{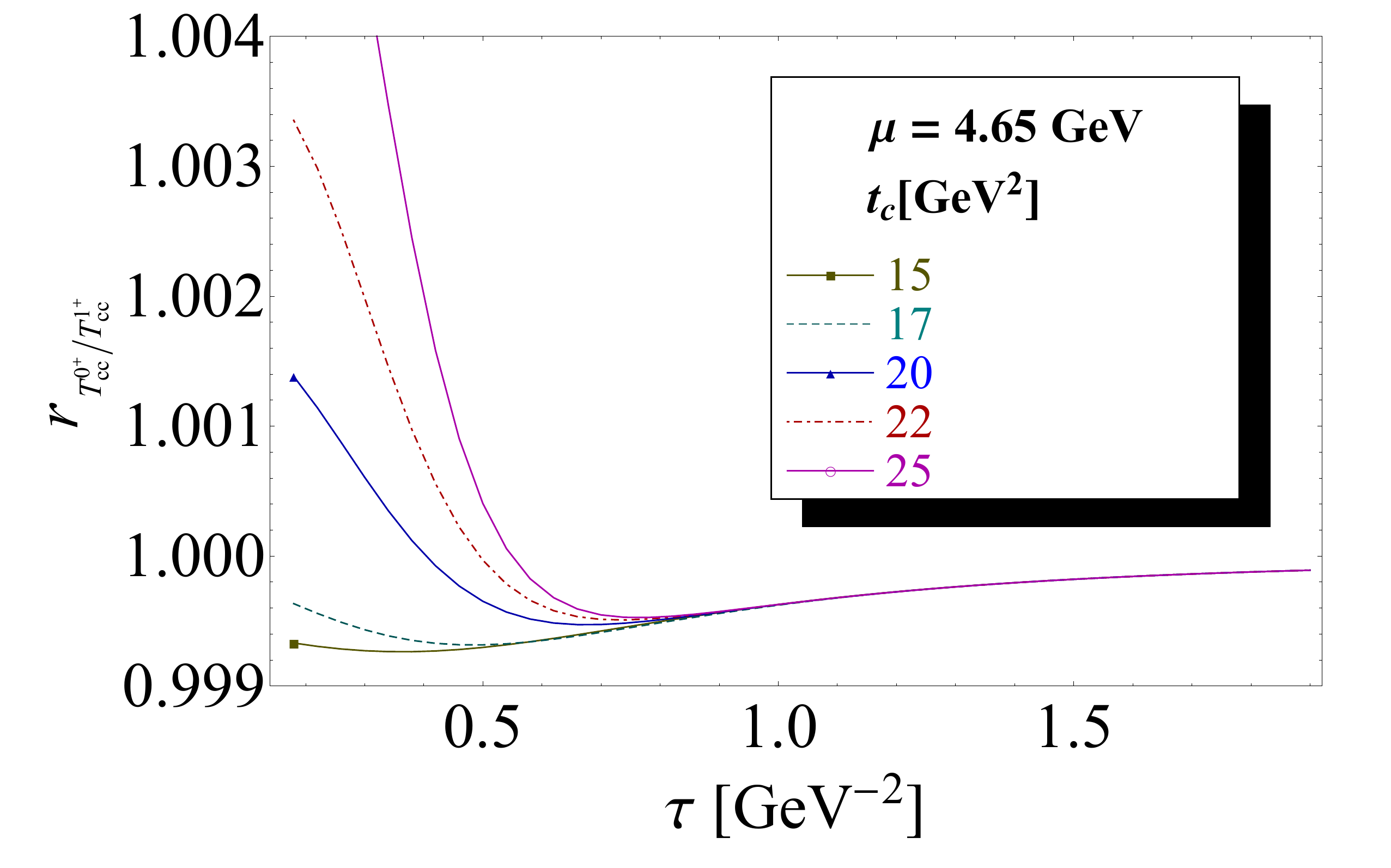}
\vspace*{-0.5cm}
\caption{\footnotesize   $r_{T^{0^+}_{cc}/T^{1^+}_{cc}}$ as a function of $\tau$ at NLO for \# values of $t_c$, for $\mu$=4.65 GeV and for the QCD inputs in Table\,\ref{tab:param}.} 
\label{fig:rtcc10}
\end{center}
\vspace*{-0.5cm}
\end{figure}

The result of the analysis is shown in Fig.\,\ref{fig:rtcc10} from which we deduce for the sets $(\tau,t_c)$=(0.36, 15) to (0.72, 20)\,(GeV$^{-2}$, GeV$^2$):
\beq
r_{T^{0^+}_{cc}/T^{1^+}_{cc}}=0.9994(2)~~~~ \lrar~~~~ M_{T_{cc}}(0^{+})=3878(5)~{\rm MeV},
\eeq
where we have used the mean from the $T_{cc}(1^+)$ mass predicted in Eq.\,\ref{eq:1+cc} and the data 3875 MeV\,\cite{LHCb4}.  
\subsection*{\b Final value of $M_{T^{0^+}_{cc}}$ from LSR $\oplus$ DRSR}
As a final value of $M_{T^{0^+}_{cc}}$, we take the mean of the previous three determinations:
\beq
 M_{T_{cc}}(0^{+})=3883(3)~{\rm MeV}.
 \eeq
\section{The $T_{cc\bar s\bar u}(0^{+})$   mass}
\subsection*{\b  $r_{T^{0^+}_{cc\bar s\bar u}/T^{0^+}_{cc}}$ ratio of masses}
We study the SU3 ratio of masses $r_{T^{0^+}_{cc\bar s\bar u}/T^{0^+}_{cc}}$. The $\tau$ and $t_c$-behaviours are very similar to the $1^+$ case  in Fig.\,\ref{fig:rcc1us}. The optimal result is obtained for the sets $(\tau,t_c)$=(0.72,23) to (0.74,32)  (GeV$^{-2}$, GeV$^2$) at which we deduce:
\beq
r_{T_{cc\bar s\bar u}/T_{cc}(0^+)}=1.0113(12) ~~~~ \lrar~~~~ M_{T_{cc\bar s\bar u}}(0^{+})=3927(6)~{\rm MeV},
\eeq
\subsection*{\b  Decay constant and mass from LSR at NLO}
Here, we extract directly the  $T_{cc\bar s\bar u}$ coupling and mass from the LSR moments and ratio of moments.
The $\tau$ and $t_c$-behaviours are very similar to the ones in Fig.\,\ref{fig:xc3}. We deduce for the the sets $(\tau,t_c)$=(0.32,30) to (0.35,46) in units of (GeV$^{-2}$, GeV$^2$) at which the coupling presents a minimum and the mass an inflexion point\,:
\beq
f_{T_{cc\bar s\bar u}}(0^{+})=542(53)~{\rm keV},~~~~~~~~~~~~M_{T_{cc\bar s\bar u}}(0^{+})=3936(90)~{\rm MeV}~,
\eeq
\subsection*{\b  Final result}
As a final result for the mass, we take the mean from the DRSR and LSR ratios:
\beq
 M_{T_{cc\bar s\bar u}}(0^{+})=3983(7)~{\rm MeV}~.
 \eeq
\section{The $T_{cc\bar s\bar s}(0^{+})$   state}
\subsection*{\b  $r_{T^{0^+}_{cc\bar s\bar s}/T^{0^+}_{cc}}$ ratio of masses}
We study the SU3 ratio of masses $r_{T^{0^+}_{cc\bar s\bar s}/T^{0^+}_{cc}}$. The $\tau$ and $t_c$ behaviours are similar to the ones in Fig.\,\ref{fig:rcc1us}. The optimal result is obtained for the sets $(\tau,t_c)$=(0.72, 23) to (0.74, 32)  (GeV$^{-2}$,  GeV$^2$) at which we deduce:
\beq
r_{T_{cc\bar s\bar s}/T_{cc}(0^+)}=1.0280(27) ~~~~ \lrar~~~~ M_{T_{cc\bar s\bar s}}(0^{+})=3992(11)~{\rm MeV},
\eeq
\subsection*{\b  Decay constant and mass from LSR at NLO}
Here, we extract directly the  $T_{cc\bar s\bar s}$ coupling and mass from the LSR moments and ratio of moments.
The $\tau$ and $t_c$-behaviours are very similar to the ones in Fig.\,\ref{fig:xc3}. We deduce for the sets $(\tau,t_c)$=(0.32, 30) to (0.35, 40) in units of (GeV$^{-2}$,GeV$^2$) at which the coupling presents a minimum and the mass an inflexion point :
\beq
f_{T_{cc\bar s\bar s}}(0^{+})=718(75)~{\rm keV},~~~~~~~~~~~~M_{T_{cc\bar s\bar s}}(0^{+})=4063(125)~{\rm MeV}~,
\eeq
\subsection*{\b  Final result}
As a final result for the mass, we take the mean from the DRSR and LSR ratios:
\beq
 M_{T_{cc\bar s\bar s}}(0^{+})=3993(11)~{\rm MeV}~.
 \eeq
 

We extend the previous analysis for the $b$-quark states
\section {$Z_b$ state}
The direct determination for the molecule and four-quark assignements of the $Z_b$ from LSR at NLO  gives\,\cite{Zb}\,\footnote{For recent reviews on some other works based on QCD spectral sum rules at LO, see e.g.\,\cite{MOLEREV,ZHUREV}.}:
\bea
f_{B^*B}&=&9(2)~{\rm keV},~~~~~~~~~~~~~~~~~~~~~~f_{A_{bd}}=11(2)~{\rm MeV},\\
M_{B^*B}&=&10582(169)~{\rm MeV}~~~~~~~~~~~~~M_{A_{bd}}=10578(123)~{\rm MeV}.
\eea
The corresponding tetramole state ${\cal T}_{Z_b}$ has the mass and coupling:
\beq
M_{{\cal T}_{Z_b}}=10579(99)~{\rm MeV},~~~~~~~~~~~~~~~~f_{{\cal T}_{Z_b}}=10(2)~{\rm keV}.
\eeq
The mass prediction is in the range of the Belle data for  $Z_b(10610)$  and $Z_b(10650)$\,\cite{BELLEZb}. However, due to the large error in the mass prediction, we cannot give a sharp conclusion about the nature of these two states. 
\section {$X_b$ state}
\begin{figure}[hbt]
\begin{center}
\centerline {\hspace*{-7.5cm} \bf a)\hspace{8cm} b)}
\includegraphics[width=8cm]{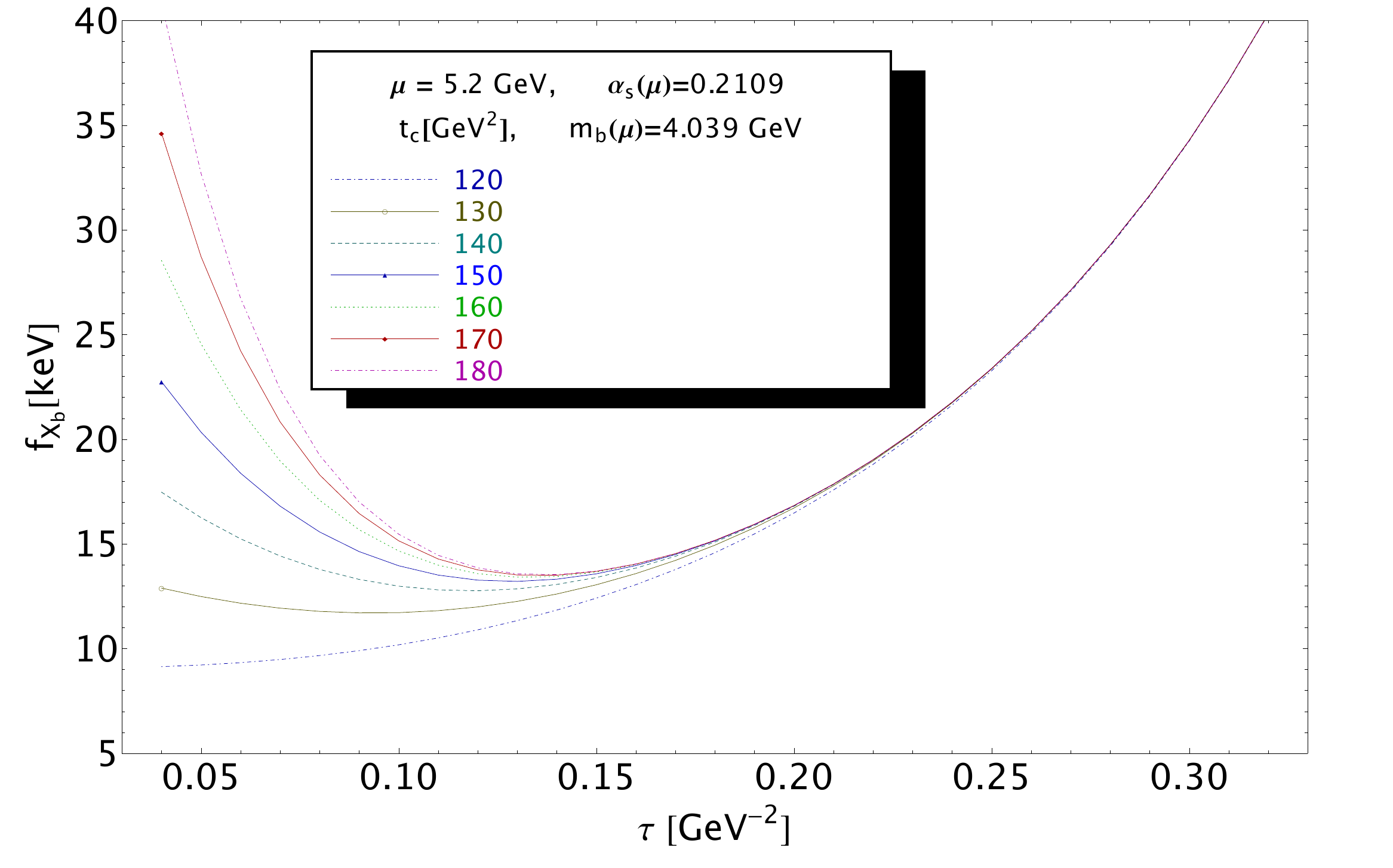}
\includegraphics[width=8cm]{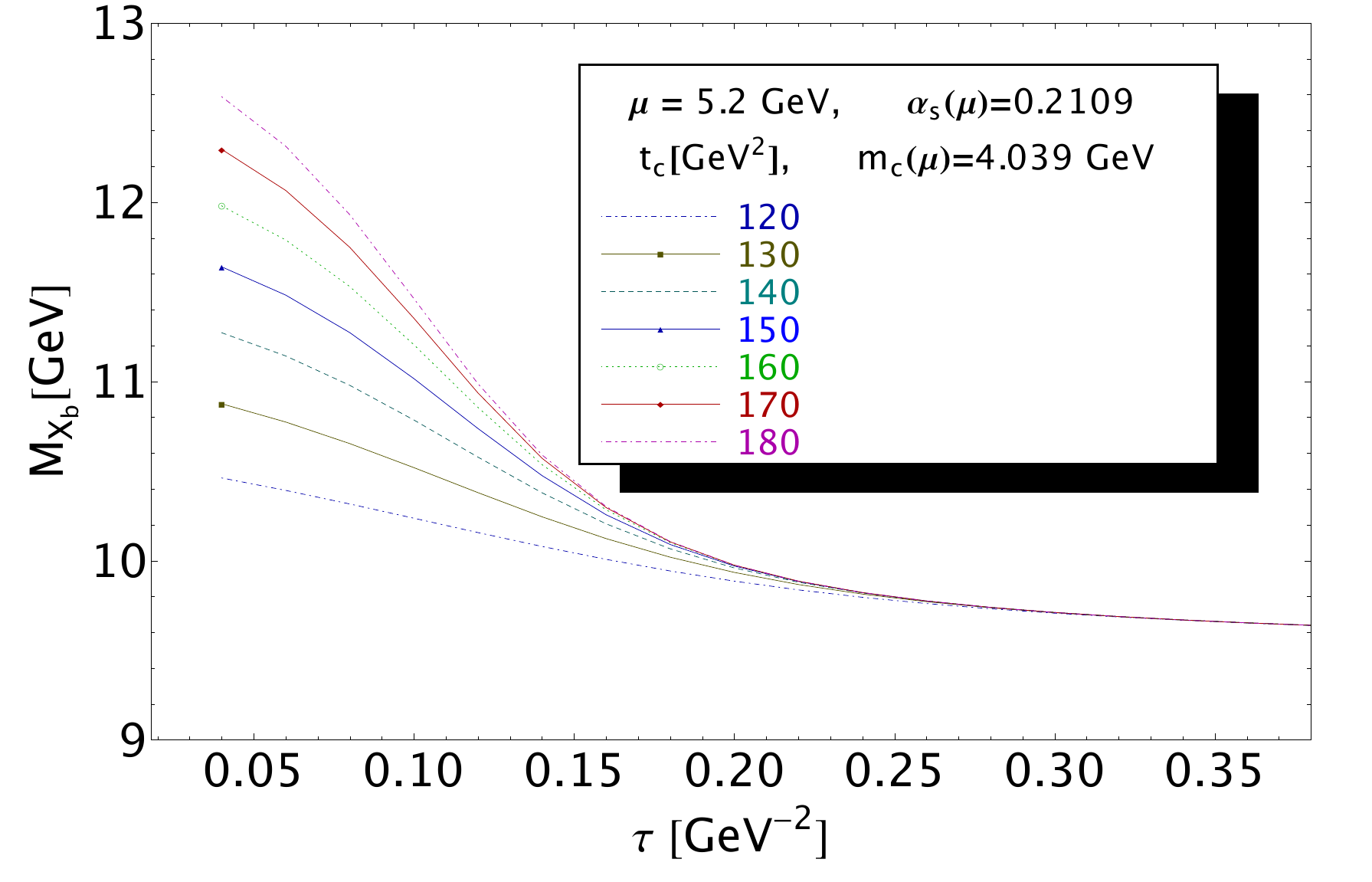}
\vspace*{-0.5cm}
\caption{\footnotesize  $f_{X_{b,3}}$ and $M_{X_{b,3}}$ as function of $\tau$ at NLO for different values of $t_c$ and for $\mu=5.2$ GeV using the QCD inputs in Table\,\ref{tab:param}.} 
\label{fig:xb3}
\end{center}
\vspace*{-0.5cm}
\end{figure}

We have studied this state using LSR at NLO in\,\cite{DRSR07,MOLE16}\,(for other works see e.g. the recent reviews\,\cite{MOLEREV,ZHUREV}).  Here, we update the analysis which is shown in Fig.\ref{fig:xb3} for the four-quark current ${\cal O}_3$. The $(\tau,t_c)$ stabilities are obtained for $(\tau,t_c)$=(0.10, 130) to (0.14, 170) (GeV$^{-2}$, GeV$^2$) where in this region, we deduce the optimal estimate:
\beq
f_{X_{b,3}}=14(3)~{\rm keV},~~~~~~~~~~~~~M_{X_{b,3}}=10545(131)~{\rm MeV},
\label{eq:xb}
\eeq
for a given value of $\mu=5.2$ GeV.

\begin{figure}[hbt]
\begin{center}
\centerline {\hspace*{-7.5cm} \bf a)\hspace{8cm} b)}
\includegraphics[width=8cm]{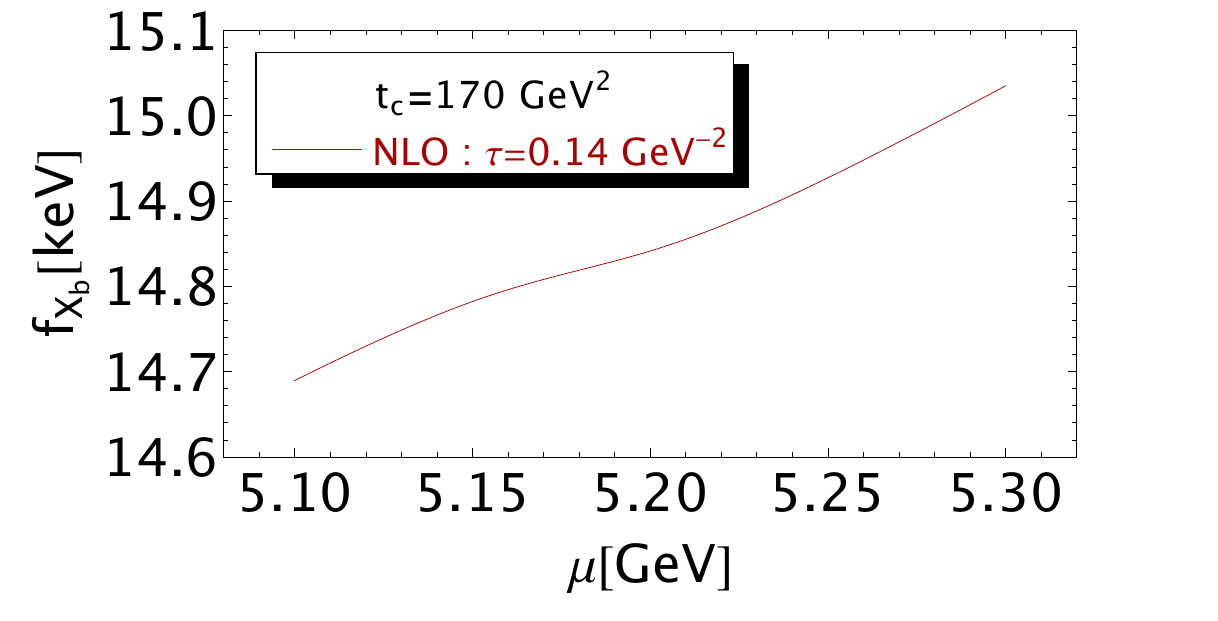}
\includegraphics[width=8cm]{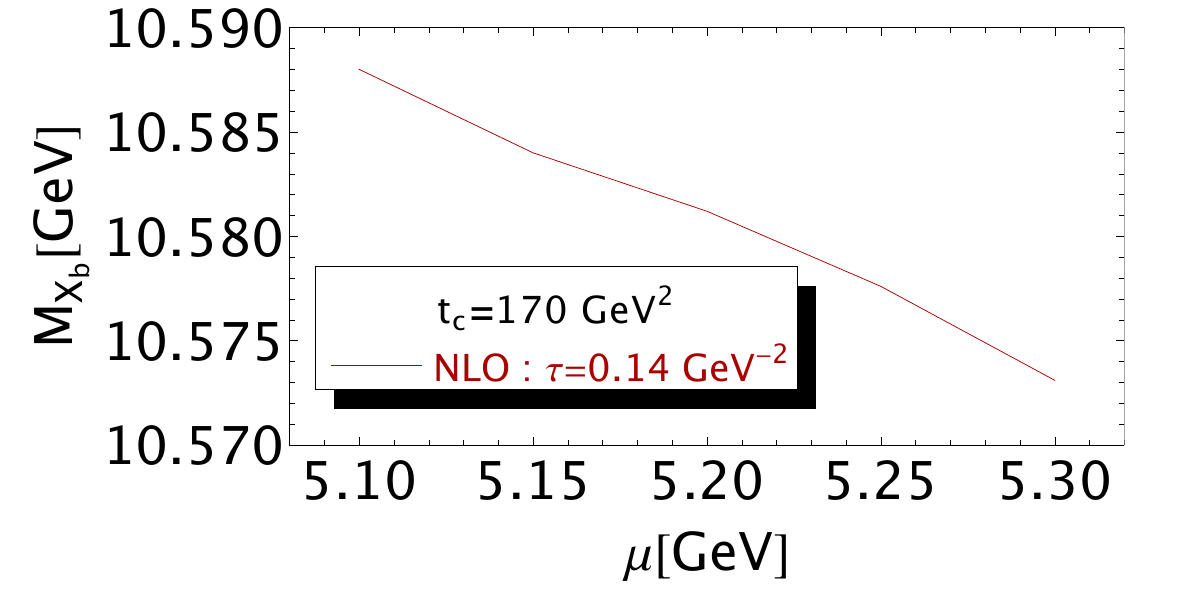}
\vspace*{-0.5cm}
\caption{\footnotesize  $f_{X_{b,3}}$ and $M_{X_{b,3}}$ as function of $\mu$ at NLO for given values of $t_c$and $\tau$ for the QCD inputs in Table\,\ref{tab:param}.} 
\label{fig:xbmu}
\end{center}
\vspace*{-0.5cm}
\end{figure}

 We study the $\mu$ dependence of the result in Fig.\ref{fig:xbmu} at NLO from which we extract  an optimal value at :
\beq
\mu_b=(5.2\pm 0.05)~{\rm GeV}~.
\eeq

This value of $\mu$ is (almost) universal in the $b$-quark channel as it is the same in all our previous works\,\cite{MOLE16,Zc,Zb,DK}. 

This result in Eq.\,\ref{eq:xb}can be compared with the one in\,\cite{DRSR07,MOLE16}, where one can notice that the result obtained in\,\cite{DRSR07} corresponds to a low range of $t_c$-values (104-117) GeV$^2$ outside the optimal region leading to a low value of  $M_{X_b}=10144(106)$ MeV. The one in Ref.\,\cite{MOLE16} is 10701(172) MeV where the relatively high-central value is due to the unprecise choice of $\tau$ at the inflexion point. 

One can notice that the central value of $M_{X_{b,3}}$ in Eq.\,\ref{eq:xb} is below the physical $B^*B$ threshold of 10604 MeV which goes in line with the expectations from some other approaches\,\footnote{For reviews, see e.g.\,\cite{MOLEREV,ZHUREV,MAIANI,RICHARD,SWANSON,DOSCH2,QIANG,BRAMBILLA}.}
\section{$T^{1^+}_{bb}$ state}
\subsection*{\b $T^{1^+}_{bb}/X_b$ mass ratio from DRSR}
\begin{figure}[hbt]
\begin{center}
\includegraphics[width=10cm]{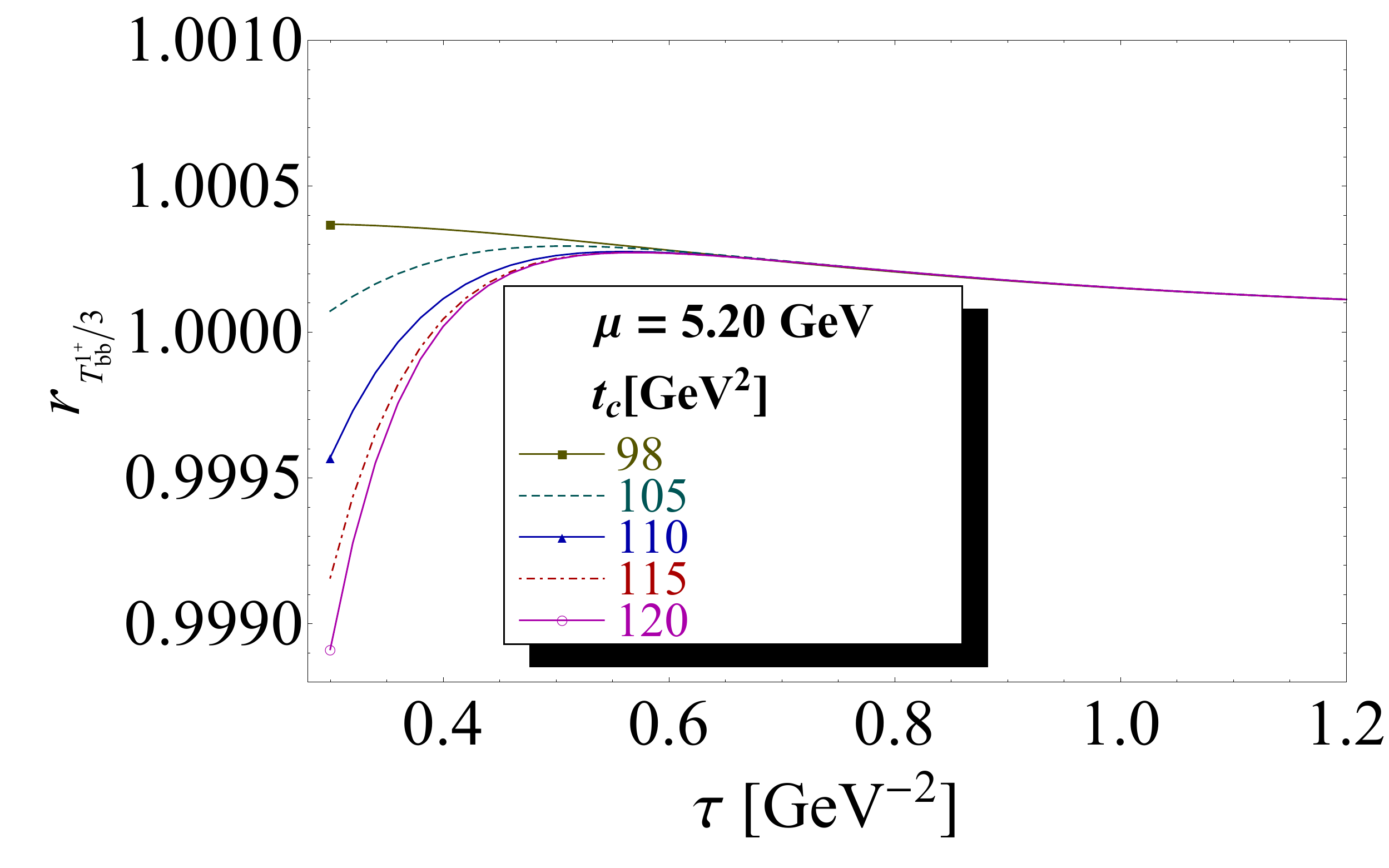}
\vspace*{-0.5cm}
\caption{\footnotesize   $r_{T^{1^+}_{bb}/3}$ as a function of $\tau$ at NLO for \# values of $t_c$, for $\mu$=5.2 GeV and for the QCD inputs in Table\,\ref{tab:param}.} 
\label{fig:rtbb1p}
\end{center}
\vspace*{-0.5cm}
\end{figure}

We show the analysis of the $T^{1^+}_{bb}$ over the $X_{b,3}$ mass  in Fig.\,\ref{fig:rtbb1p}. The optimal result is obtained for the sets : $(\tau,t_c)$=(0.56, 105) to (0.56, 115)  (GeV$^{-2}$, GeV$^2$) from which we deduce:
\beq
r_{T^{1^+}_{bb}/3}=1.0003(1),~~~~~~\lrar~~~~~~M_{T^{1^+}_{bb}}=10548(131)~{\rm MeV}~
\eeq
The sources of the errors are given in Table\,\ref{tab:tbb}. 
%
\subsection*{\b Direct estimate of the $T^{1^+}_{bb}$ coupling and mass from LSR}
\begin{figure}[hbt]
\begin{center}
\centerline {\hspace*{-7.5cm} \bf a)\hspace{8cm} b)}
\includegraphics[width=8cm]{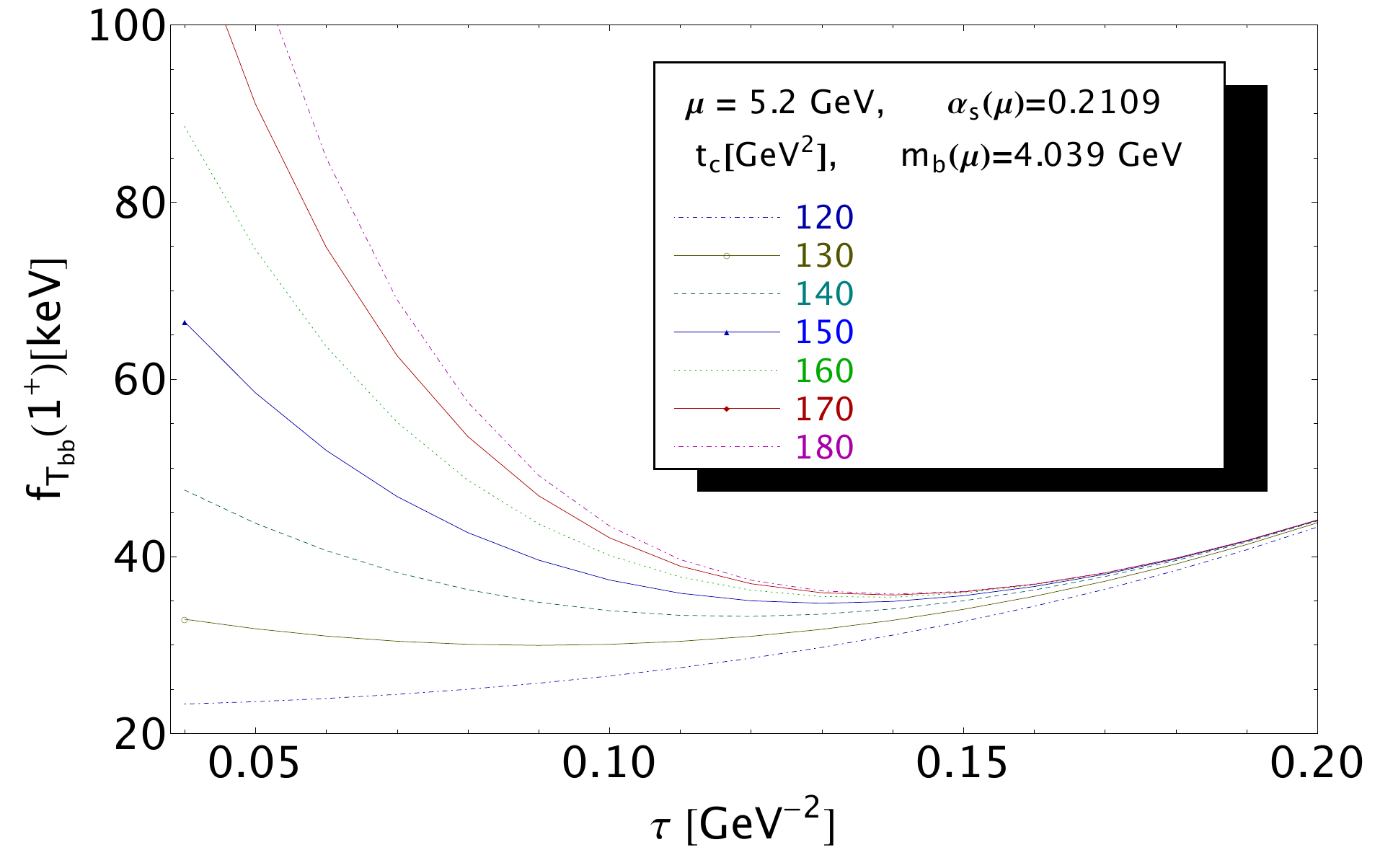}
\includegraphics[width=8cm]{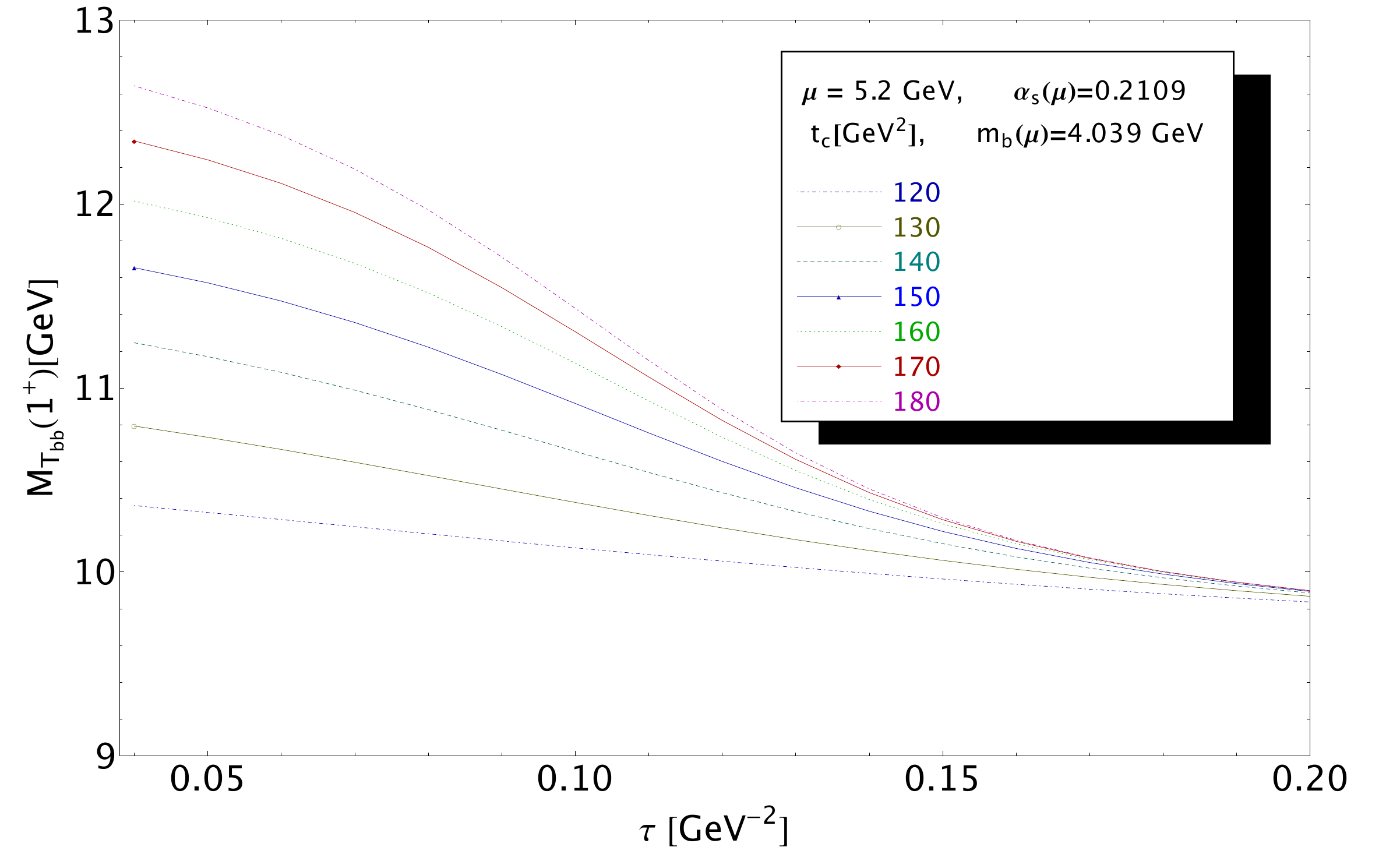}
\vspace*{-0.5cm}
\caption{\footnotesize  $f_{T^{1^+}_{bb}}$ and $M_{T^{1^+}_{bb}}$ as function of $\tau$  for \# values of $t_c$, for $\mu$=5.2 GeV and for the QCD inputs in Table\,\ref{tab:param}.} 
\label{fig:tbb1p}
\end{center}
\vspace*{-0.5cm}
\end{figure}

The analysis of the $T^{1^+}_{bb}$  mass and coupling is shown in  Fig.\,\ref{fig:tbb1p}. The optimal result is obtained for the sets\,: $(\tau,t_c)$=(0.09, 130) to (0.14, 170) (GeV$^{-2}$, GeV$^2$) from which we deduce:
\beq
f_{T^{1^+}_{bb}}=33(7)~{\rm keV},~~~~~~~~~~~~~~~M_{T^{1^+}_{bb}}=10441(147)~{\rm MeV}~,
\eeq
where the different sources of the errors are given in Table\,\ref{tab:tbb}.

\subsection*{\b Final  result for the $T^{1^+}_{bb}$ mass}
As a final result, we take the mean from the LSR and DRSR results from which we obtain:
\beq
M_{T^{1^+}_{bb}}=10501(98)~{\rm MeV}~.
\label{eq:tbb1}
\eeq

\section{The $T^{1^+}_{bb\bar u\bar s}$ state}
\subsection*{\b $T^{1^+}_{bb\bar u\bar s}/T^{1^+}_{bb}$ mass ratio from DRSR}
\begin{figure}[hbt]
\begin{center}
\includegraphics[width=10cm]{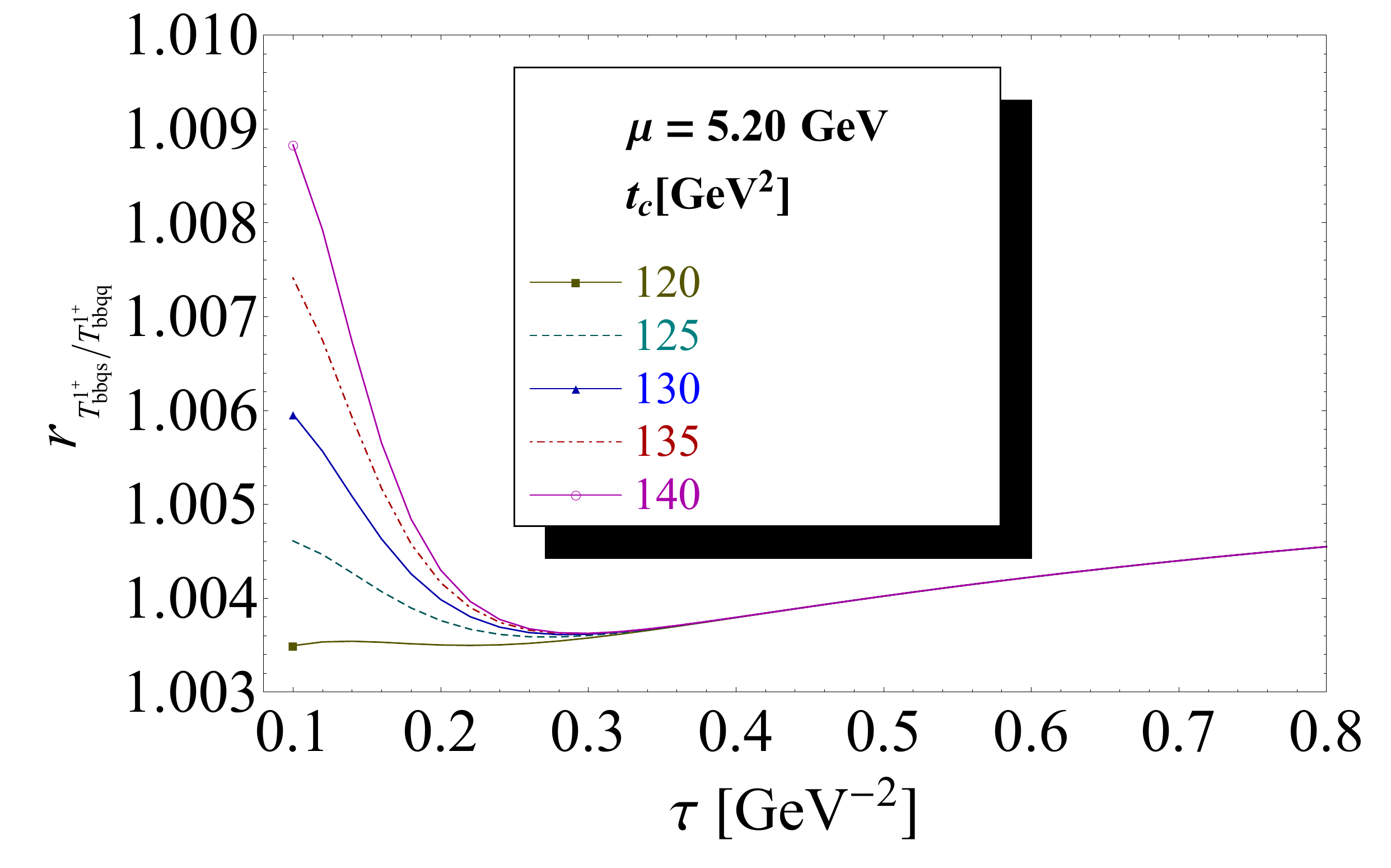}
\vspace*{-0.5cm}
\caption{\footnotesize   $r_{T^{1^+}_{bbus}/T^{1^+}_{bb}}$ as a function of $\tau$ at NLO for \# values of $t_c$, for $\mu$=5.2 GeV and for the QCD inputs in Table\,\ref{tab:param}.} 
\label{fig:rtbbsu1}
\end{center}
\vspace*{-0.5cm}
\end{figure}

We study in Fig.\,\ref{fig:rtbbsu1} the SU3 breakings on the above mass ratio.  The set of $(\tau,t_c)$ values (0.26, 125) to (0.28, 135) (GeV$^{-2}$, GeV$^2$) used to get the optimal result are given in Table\,\ref{tab:tctaub} at which we deduce:
\beq
r_{T^{1^+}_{bb\bar u\bar s}/T^{1^+}_{bb}}=1.0036(3)~~~\lrar~~~M_{T^{1^+}_{bb\bar u\bar s}}=10539(98)~{\rm MeV},
\eeq
where the value of $M_{T^{1^+}_{bb}}$ in Eq.\,\ref{eq:tbb1} has been used. 
\subsection*{\b Direct estimate of the $T^{1^+}_{bb\bar u\bar s}$ coupling and mass from LSR}
Here, we extract directly the mass and coupling of $T^{1^+}_{bb\bar u\bar s}$ from LSR. The analysis is similar to Fig.\,\ref{fig:xb3}. The optimal result is obtained for the set of $(\tau,t_c)$ values (0.10, 130) to (0.15, 170) (GeV$^{-2}$, GeV$^2$) 
We obtain :
\beq
f_{T^{1^+}_{bb\bar u\bar s}}=21(4)~~{\rm keV},~~~~~~~~~~~~~M_{T^{1^+}_{bb\bar u\bar s}}=10476(153)~{\rm MeV},
\eeq
where the different sources of the errors are given in Table\,\ref{tab:tbb}.
\subsection*{\b Final estimate of the $T^{1^+}_{bb\bar u\bar s}$ mass}
Combining the LSR and DRSR results, we deduce:
\beq
~M_{T^{1^+}_{bb\bar u\bar s}}=10521(83)~{\rm MeV}.
\eeq
\begin{table*}[hbt]
\setlength{\tabcolsep}{0.27pc}
{\scriptsize{
\begin{tabular*}{\textwidth}{@{}ll ll  ll  ll ll ll ll ll ll ll r@{\extracolsep{\fill}}l}
\hline
\hline
                   			 &\multicolumn{1}{c}{$X_b$}
					&\multicolumn{1}{c}{$T^{1^+}_{bb}$}
					&\multicolumn{1}{c}{$T^{1^+}_{bbqs}$}
					&\multicolumn{1}{c}{$T^{0^+}_{bb}$}
					&\multicolumn{1}{c}{$T^{0^+}_{bbqs}$}
					&\multicolumn{1}{c}{$T^{0^+}_{bbss}$}		
					&\multicolumn{1}{c}{$\frac{T^{1^+}_{bb}}{X_b}$}
					&\multicolumn{1}{c}{$\frac{T^{1^+}_{bbqs}}{T^{1^+}_{bbqq}}$}
					&\multicolumn{1}{c}{$\frac{T^{0^+}_{bb}}{X_b}$}
					&\multicolumn{1}{c}{$\frac{T^{0^+}_{bb}}{T^{1^+}_{bb}}$}
					&\multicolumn{1}{c}{$\frac{T^{0^+}_{bbqs}}{T^{0^+}_{bbqq}}$}
					&\multicolumn{1}{c}{${\frac{T^{0^+}_{bbss}}{T^{0^+}_{bbqq}}}$}

                  \\
\hline
$\bf t_{c}$ &130 - 170&130 - 170&130 - 170&130 - 170&130 - 170&130 - 170&105 - 115&125 - 135&105 - 115&122&125 - 135&125 - 135\\
$\bf\tau$&$10~~ ;~14$&$~~9~~ ;~14$&$10~~ ;~15$&$~~9~~ ;~14$&$10~~ ;~15$&$10~~ ;~15$&$56~~ ;~56$&$26~~ ;~28$&$58~~ ;~58$&$~~9$&$26~~ ;~28$&$26~~ ;~28$\\
\hline
\hline
\end{tabular*}
 }}
 \caption{{\small Values of the set of LSR parameters $(t_c,\tau)$ at the optimization region for the PT series up to NLO and for the OPE truncated at the dimension-six condensates and for $\mu=5.20$ GeV.}}
 \vspace*{0.25cm}
\label{tab:tctaub}
\end{table*}

\begin{table*}[hbt]
\setlength{\tabcolsep}{0.3pc}
{\scriptsize{
\begin{tabular*}{\textwidth}{@{}ll ll  ll  ll ll ll ll ll l@{\extracolsep{\fill}}l}

\hline
\hline
                \bf Observables &\multicolumn{1}{c}{$\Delta t_c$}
					&\multicolumn{1}{c}{$\Delta \tau$}
					&\multicolumn{1}{c}{$\Delta \mu$}
					&\multicolumn{1}{c}{$\Delta \alpha_s$}
					&\multicolumn{1}{c}{$\Delta PT$}
					&\multicolumn{1}{c}{$\Delta m_s$}
					&\multicolumn{1}{c}{$\Delta m_c$}
					&\multicolumn{1}{c}{$\Delta \overline{\psi}\psi$}
					&\multicolumn{1}{c}{$\Delta \kappa$}					
					&\multicolumn{1}{c}{$\Delta G^2$}
					&\multicolumn{1}{c}{$\Delta M^{2}_{0}$}
					&\multicolumn{1}{c}{$\Delta \overline{\psi}\psi^2$}
					&\multicolumn{1}{c}{$\Delta G^3$}
					&\multicolumn{1}{c}{$\Delta OPE$}
					&\multicolumn{1}{c}{$\Delta M_{G}$}
					&\multicolumn{1}{r}{Values}\\

\hline
{\bf Coupling} [keV] &&&&&&&&&&&\\
$f_{X_b}$ &1.15&0.08&0.18&0.44&0.17&$\cdots$&0.40&0.14&$\cdots$&0.00&0.15&0.58&0.00&3.07&1.36&14(3) \\
$f_{T^{1^+}_{bb}}$ &2.84&0.21&0.43&0.96&0.31&$\cdots$&0.61&0.00&$\cdots$&0.01&0.00&1.80&0.00&3.58&4.00&33(7) \\
$f_{T^{1^+}_{bbqs}}$ &1.76&0.14&0.10&0.67&0.28&0.02&0.42&0.01&0.57&0.01&0.02&1.08&0.00&2.30&2.93&21(4) \\
$f_{T^{0^+}_{bb}}$ &4.42&0.34&0.72&1.61&0.15&$\cdots$&1.00&0.00&$\cdots$&0.01&0.00&3.09&0.00&6.21&6.59&54(11) \\
$f_{T^{0^+}_{bbqs}}$ &2.78&0.25&0.17&1.11&0.33&0.03&0.69&0.02&1.00&0.01&0.02&1.88&0.00&3.97&4.84&35(7) \\
$f_{T^{0^+}_{bbss}}$ &4.30&0.29&0.23&1.52&0.50&0.10&0.97&0.02&2.42&0.01&0.03&2.29&0.00&9.93&5.80&47(13) \\
\\
{\bf Mass} [MeV] &&&&&&&&&&&&&&\\
$M_{X_b}$ &26.0&115&2.92&18.7&0.55&$\cdots$&14.8&8.77&$\cdots$&0.20&1.48&38.9&0.00&32.0&$\cdots$&10545(131)\\
$M_{T^{1^+}_{bb}}$&9.65&119&7.40&15.4&0.26&$\cdots$&7.00&0.00&$\cdots$&0.13&0.00&40.0&0.00&73.8&$\cdots$&10441(147)\\
$M_{T^{1^+}_{bbqs}}$&58.9&109&2.63&16.2&0.10&2.08&7.38&0.33&20.5&0.13&0.58&60.4&0.00&59.3&$\cdots$&10476(153)\\
$M_{T^{0^+}_{bb}}$ &12.6&117&7.40&15.3&0.63&$\cdots$&7.10&0.00&$\cdots$&0.08&0.00&39.0&0.03&74.1&$\cdots$&10419(146)\\
$M_{T^{0^+}_{bbqs}}$ &61.6&108&2.58&16.1&0.10&2.05&7.48&0.30&20.3&0.05&0.30&59.9&0.05&59.7&$\cdots$&10454(153)\\
$M_{T^{0^+}_{bbss}}$&2.50&113&2.60&16.1&0.27&4.78&6.78&0.28&26.5&0.00&0.63&42.2&0.05&30.8&$\cdots$&10538(129)\\
\\
{\bf Ratio} &&&&&&&&&&&\\
$r_{T^{1^+}_{bb}/X_b}$ &0.01&0.00&0.00&0.01&0.00&$\cdots$&0.00&0.00&$\cdots$&0.00&0.10&0.02&0.00&0.01&$\cdots$&1.0003(1) \\
$r_{T^{1^+}_{bbqs}/T^{1^+}_{bbqq}}$ &0.02&0.02&0.01&0.03&0.00&0.24&0.01&0.01&0.08&0.00&0.00&0.06&0.00&0.19&$\cdots$&1.0036(3) \\
$r_{T^{0^+}_{bb}/X_b}$ &0.01&0.00&0.00&0.01&0.00&$\cdots$&0.00&0.01&$\cdots$&0.00&0.09&0.01&0.00&0.02&$\cdots$&1.0003(1) \\
$r_{T^{0^+}_{bb}/T^{1^+}_{bb}}$ &0.02&0.03&0.00&0.00&0.00&$\cdots$&0.00&0.00&$\cdots$&0.01&0.00&0.03&0.00&0.02&$\cdots$&0.9994(1) \\
$r_{T^{0^+}_{bbqs}/T^{0^+}_{bbqq}}$ &0.02&0.02&0.01&0.03&0.00&0.24&0.01&0.02&0.07&0.00&0.01&0.06&0.00&0.18&$\cdots$&1.0035(3) \\
$r_{T^{0^+}_{bbss}/T^{0^+}_{bbqq}}$ &0.06&0.01&0.02&0.07&0.00&0.56&0.02&0.02&0.14&0.01&0.01&0.11&0.01&0.75&$\cdots$&1.0086(10) \\
\\
\hline
\hline
\end{tabular*}
}}
 \caption{Sources of errors of $T_{bb}$, $X_b$ and their ratios of masses. We take $\ve \Delta \mu\ve=0.05$ GeV and $\ve \Delta \tau\ve =0.01$ GeV$^{-2}$. For the ratios, the errors quoted in the table are multiplied by a factor of $10^3$}
 \vspace*{0.25cm} 
\label{tab:tbb}
\end{table*}
\section{The $T^{0^+}_{bb}$ state}
\subsection*{\b $T^{0^+}_{bb}/X_b$ mass ratio from DRSR}
The analysis of the $T^{0^+}_{bb}$ over the $X_{b,3}$ mass  is similar to the $1^+$ case in Fig.\,\ref{fig:rtbb1p}. The optimal result is obtained for the sets : $(\tau,t_c)$=(0.58, 105) to (0.58, 115)  (GeV$^{-2}$, GeV$^2$) from which we deduce:
\beq
r_{T^{0^+}_{bb}/3}=1.0003(1),~~~~~~\lrar~~~~~~M_{T^{0^+}_{bb}}=10501(98)~{\rm MeV}~.
\eeq
The sources of the errors are given in Table\,\ref{tab:tbb}. 
\subsection*{\b $T^{0^+}_{bb}/T^{1^+}_{bb}$ mass ratio from DRSR}
\begin{figure}[hbt]
\begin{center}
\includegraphics[width=10cm]{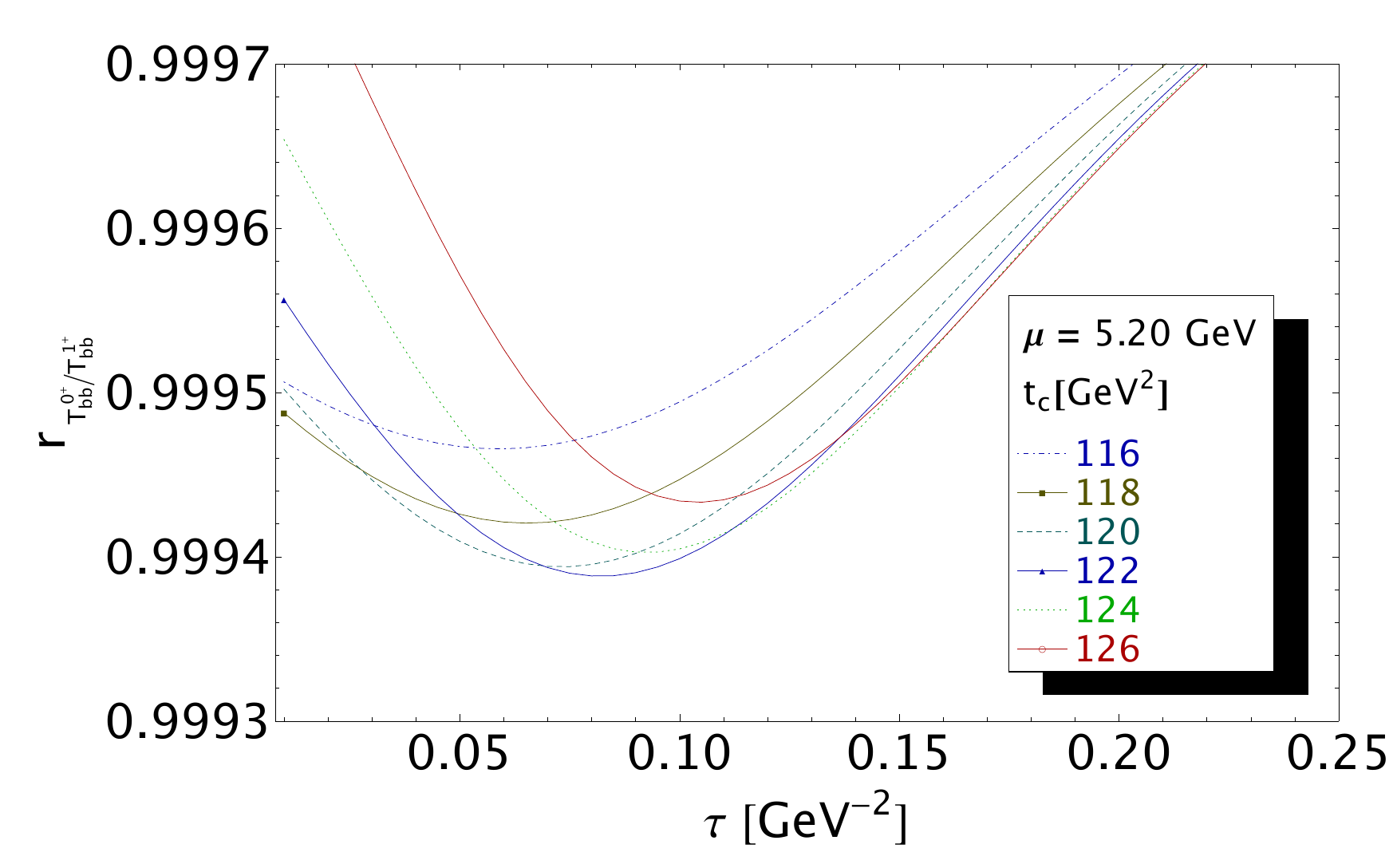}
\vspace*{-0.5cm}
\caption{\footnotesize   $r_{T^{0^+}_{bb}/T^{1^+}_{bb}}$ as a function of $\tau$ at NLO for \# values of $t_c$, for $\mu$=5.2 GeV and for the QCD inputs in Table\,\ref{tab:param}.} 
\label{fig:rtbb01}
\end{center}
\vspace*{-0.5cm}
\end{figure}

The analysis of this mass ratio is given in Fig.\,\ref{fig:rtbb01}. We obtain a minimum at $\tau=0.09$ GeV$^{-2}$ and $t_c=122$ GeV$^2$, which are in the raange of the direct determinations of $M_{T^{1^+}_{bb}}$ and $M_{T^{0^+}_{bb}}$ (see Table\,\ref{tab:tctaub}). At this minimum, we deduce the optimal value:
\beq
r_{T^{0^+}_{bb}/T^{1^+}_{bb}}=0.9994(1)~~~~~~~~~~~\lrar ~~~~~~~~~~~ M_{T^{0^+}_{bb}}=10495(98)~{\rm MeV},
\eeq
after using $M_{T^{1^+}_{bb}}$given in Eq.\,\ref{eq:tbb1}. 
\subsection*{\b Direct estimate of the $T^{0^+}_{bb}$ coupling and mass from LSR}
The analysis of the $T^{0^+}_{bb}$  mass and coupling is similar to the one in Fig.\,\ref{fig:xb3}. The optimal result is obtained for the sets : $(\tau,t_c)$=(0.09, 130) to (0.14, 170) (GeV$^{-2}$, GeV$^2$) from which we deduce:
\beq
f_{T^{0^+}_{bb}}=54(11)~{\rm keV},~~~~~~~~~~~~~~~M_{T^{0^+}_{bb}}=10419(146)~{\rm MeV},
\eeq
where the different sources of the errors are given in Table\,\ref{tab:tbb}.
\subsection*{\b Final  result for the $T^{0+}_{bb}$ mass}
As a final result, we take the mean from the three results from LSR and DRSR from which we obtain:
\beq
M_{T^{0^+}_{bb}}=10484(63)~{\rm MeV}~,
\label{eq:mtbb0}
\eeq
where one can notice an almost degeneracy between the $T_{bb}$ $1^+$ and $0^+$ masses. 
\section{The $T^{0^+}_{bb\bar u\bar s}$ state}
\subsection*{\b $T^{0^+}_{bb\bar u\bar s}/T^{0^+}_{bb}$ mass ratio from DRSR}
We study the SU3 breakings on the above mass ratio.  The analysis is similar to Fig\,\ref{fig:rtbbsu1}. The sets of $(\tau,t_c)$ values used to get the optimal result are $(\tau,t_c)$=(0.26, 125) to (0.28, 135) (GeV$^{-2}$, GeV$^2$) (see Table\,\ref{tab:tctaub}) at which we deduce:
\beq
r_{T^{0^+}_{bb\bar u\bar s}/T^{0^+}_{bb}}=1.0035(3)~~~\lrar~~~M_{T^{0^+}_{bb\bar u\bar s}}=10521(63)~{\rm MeV},
\eeq
where the value of $M_{T^{0^+}_{bb}}$ in Eq.\,\ref{eq:mtbb0} has been used. 
\subsection*{\b Direct estimate of the $T^{0^+}_{bb\bar u\bar s}$ coupling and mass from LSR}
Here, we extract directly the mass and coupling of $T^{0^+}_{bb\bar u\bar s}$from LSR. The analysis is similar to the one in Fig.\,\ref{fig:xb3}. We obtain at $(\tau,t_c)$=(0.10, 130) to (0.15, 170) (GeV$^{-2}$, GeV$^2$)\,:
\beq
f_{T^{0^+}_{bb\bar u\bar s}}=35(7)~~{\rm keV},~~~~~~~~~~~~~M_{T^{0^+}_{bb\bar u\bar s}}=10454(153)~{\rm MeV},
\eeq
where the different sources of the errors are given in Table\,\ref{tab:tbb}.
\subsection*{\b Final estimate of the $T^{0^+}_{bb\bar u\bar s}$ mass}
Combining the LSR and DRSR results, we deduce:
\beq
~M_{T^{0^+}_{bb\bar u\bar s}}=10511(58)~{\rm MeV}.
\eeq
\section{The $T^{0^+}_{bb\bar s\bar s}$ state}
\subsection*{\b $T^{0^+}_{bb\bar s\bar s}/T^{0^+}_{bb}$ mass ratio due to  SU3 breakings from DRSR}
The analysis of the $T^{0^+}_{bb\bar s\bar s}$ over $T^{0^+}_{bb}$ mass  is similar to  Fig.\,\ref{fig:rtbbsu1}. The SU3 breaking parameters used in the analysis are in Table\,\ref{tab:param}. The optimal result is obtained for the sets : $(\tau,t_c)$=(0.26, 125) to (0.28, 135)  (GeV$^{-2}$, GeV$^2$) from which we deduce:
\beq
r_{T^{0^+}_{bbss}/T^{0^+}_{bbqq}}=1.0086(10)~~~~~~\lrar~~~~~~M_{T^{0^+}_{bbss}}=10574(64)~{\rm MeV}~,
\eeq
where the different sources of the errors are given in Table\,\ref{tab:tbb}.
\subsection*{\b Direct estimate of the $T^{0^+}_{bb\bar s\bar s}$ coupling and mass from LSR}
The analysis of the $T^{0^+}_{bb\bar s\bar s}$ coupling and  mass  is similar to Fig.\,\ref{fig:xb3}. The optimal result is obtained for the sets : $(\tau,t_c)$=(0.10, 130) to (0.15, 170)  (GeV$^{-2}$, GeV$^2$) from which we deduce:
\beq
f_{T^{0^+}_{bb\bar s\bar s}}=47(13)~~{\rm keV},~~~~~~~~~~~~~M_{T^{0^+}_{bb\bar s\bar s}}=10538(129)~{\rm MeV},
\eeq
where the different sources of the errors are given in Table\,\ref{tab:tbb}.
\subsection*{\b Final result for the $T^{0^+}_{bb\bar s\bar s}$ mass}
Combining the LSR and DRSR results, we deduce:
\beq
~M_{T^{0^+}_{bb\bar s\bar s}}=10567(57)~{\rm MeV}.
\eeq
\begin{table}[hbt]
\begin{center}
\setlength{\tabcolsep}{0.7pc}
\newlength{\digitwidth} \settowidth{\digitwidth}{\rm 0}
\catcode`?=\active \def?{\kern\digitwidth}
\footnotesize{
\begin{tabular}{llllll  ll   l }
\hline\hline

&&&&\multicolumn{4}{c}{Mass}& \multicolumn{1}{l}{$\Delta E_B$}\\
\cline{5-9} 
States&$J^{P}$& 
Decay&Thresholds&\multicolumn{1}{l}{Data}&\multicolumn{1}{l}{Config.}&\multicolumn{1}{l}{LSR}
&\multicolumn{2}{c}{LSR $\oplus$ DRSR}\\
 \hline
     $Z_c$&$1^{+}$&$\bar{D^0}D^{*+}$& 3876&3900&$D^*D$&3912(61)\, \cite{Zc,MOLE16}&&\\
     &&&&& $\bar  3_c3_c$&3889(58)\cite{Zc,MOLE16}\\
         &&&&& ${\cal T}_{Z_c}$&&{\it 3900(42)}\,\cite{Zc,MOLE16}&$+24(42)$\\
  $Z_b$    &$1^{+}$&$\bar B^0B^{*+}$&10605&& ${\cal T}_{Z_b}$&&{\it 10579(99)}\,\cite{Zc,MOLE16}&$-26(99)$\\    
  \\
  $X_c$&$1^{+}$&$\bar{D^0}D^{*+}$& 3876&3872&$\bar  3_c3_c$&3876(76)& \\
    &&&&& $\bar  6_c6_c$&&3864(76)&\\
    &&&&& $ {\psi\pi}$&&3889(76)&\\
     &&&&& $ {D^*D}$&&3912(61)&\\
    &&&&& ${\cal T}_{X_c}$&&\it 3876(44)&$+0(44)$\\
     $X_b$&$1^{+}$&$\bar B^0B^{*+}$&10605&&$\bar  3_c3_c$&10545(131)&&$-60(131)$ \\
\\
  $T_{cc\bar u\bar d}$&$1^{+}$&$\bar{D^0}D^{*+}$& 3876&3875&$\bar  3_c3_c$&3885(74)&3886(4)&$+14(4)$\\
  $T_{cc\bar u\bar s}$&$1^{+}$&$\bar {D^0_s}D^*$& 3975&&--
&3940(89)&3931(7)& $-44(7)$\\

$T_{cc\bar u\bar d}$&$0^{+}$&$\bar {D^0}D^0$& 3730&&--
&3882(81)&3883(3)& +153(3)\\
  $T_{cc\bar u\bar s}$&$0^{+}$&$\bar {D^0_s}D$& 3833&&--
&3936(90)&3927(6)&$+94(6)$\\

$T_{cc\bar s\bar s}$&$0^{+}$& $ {D^+_s}D^-_s$& 3937&&--&4063(72)&3993(11)&$+56(11)$\\
\\
$T_{bb\bar u\bar d}$&$1^{+}$& $\bar{B^0}B^{*+}$& 10605&&--&10441(147)&10501(98)&$ -104(98)$\\
$T_{bb\bar u\bar s}$&$1^{+}$& $\bar{B^0_s}B^{*+}$& 10692&&--&10476(154)&10521(83)&$ -171(83)$\\

$T_{bb\bar u\bar d}$&$0^{+}$& $\bar {B^0}B^0$ &10559&&--&10419(146)&10484(63)&$ -75(63)$\\
$T_{bb\bar u\bar s}$&$0^{+}$& $\bar {B^0_s}B^0$ &10646&&--&10454(153)&10511(58)&$ -135(56)$\\
$T_{bb\bar s\bar s}$&$0^{+}$& $ {\bar B^0_s}B^0_s$& 10734&&--&10538(129)&10567(57)&$-167(57)$\\

\hline\hline
\end{tabular}
}
 \caption{Summary of the results of the $XZT$ states masses in units of MeV obtained  in this paper from LSR (Tables\,\ref{tab:error} and \ref{tab:tbb}) and DRSR using the currents in Table\,\ref{tab:current}. Our final values are in the column ``LSR $\oplus$ DRSR" which are the mean from LSR with the ones deduced from DRSR.}  
\label{tab:res}
\end{center}
\end{table}
\section{General comments on the LSR results}
 Before comparing the different LSR results, let us address some  general comments\,:
\subsection*{\b Ambiguous quark mass definition at LO}
 As we have continuously  stressed in our previous papers\,\cite{MOLE12,MOLE16,MOLE16X,SU3,4Q,DK,Zc}, the use of the running $\overline{MS}$-scheme mass in the LO expression of the spectral function is not justified as the heavy quark mass which plays a key role in the analysis is ill-defined at LO while the spectral function has been computed within the on-shell scheme where the on-shell heavy quark mass  enters naturally.  To that order, one can equally use the pole / on-shell quark mass. The (lucky) success of the LO results is only due to the (estimated) small NLO corrections in the $\overline{MS}$-scheme where the NLO corrections tend to compensate in the ratio of moments used to extract the ground state mass.  We have demonstrated this fact in our previous papers where we have  used factorization (valid to leading order in $1/N_c$) to estimate the NLO contributions\,\cite{MOLE12,MOLE16,MOLE16X,SU3,4Q,DK,Zc}. One should note that, at this level of ($1/N_c$) approximation for NLO, we cannot differentiate between a meson and a diquark state. 

\subsection*{\b The choice of the interpolating currents}
This choice is not also trivial which may lead to inconsistencies. In the precise case of the $T_{QQ}$ compact four-quark currents $\bar 3_c3_c$used in this paper, we realize that some choices like e.g.:
\bea
 {\cal O}_T^{1^+} &=& \epsilon_{i j k} \:\epsilon_{m n k} \left(
 c_i^T\, C \gamma^\mu \,c_j \right) \big{[} \left( \bar{q}_m\, \gamma_5
  C \,\bar{q}_n^T\right) \nnb\\
  &=&  \frac{1}{\sqrt{2}}\epsilon_{i j k} \:\epsilon_{m n k} \left(
 c_i^T\, C \gamma^\mu \,c_j \right) \big{[} \left( \bar{q}_m\, \gamma_5
  C \,\bar{q'}_n^T\right)+  \left( \bar{q'}_m\, \gamma_5
  C \,\bar{q}_n^T\right)\big{]}\nnb\\
&=& \cdots
\eea
  lead to null contributions due to SU3 symmetry.
\subsection*{\b The QCD expressions of the two-point correlator} 
These expressions are non-trivial such that it is difficult to check carefully the  expressions given by each authors. However, in some papers, we have realized that, besides the error in the calculations, the contributions of some diagrams are missing\,:

 -- From an examination of the QCD expressions of the propagators used as inputs in the calculation, we notice that in Refs.\,\cite{WANG-Ta,WANG-Tb,AGAEV-T,ZHU-T}, the propagators do not induce properly the contributions from the mixed quark-gluon $\la \bar qGq\ra$ and gluon $\la G^2\ra$ condensates in Fig.\,\ref{fig:mixed}. 

-- The missed diagrams  also happen when  the authors include  high dimension operators contributions where (often) the alone contributions of some classes of diagrams are included.  More drastic is the fact that some authors include $D=8,10,...$ condensate contributions but (for consistency) the contribution of the $D=6$ triple gluon condensate $\la G^3\ra$ is not included.  

-- In this and in our previous papers, we do the OPE up to $D=6$ where  ALL POSSIBLE contributions up to $D=6$ dimension are given
in integrated and compact expressions of these horrible unintegrated QCD expressions given in the literature. Such integrated expressions are more easier to use. 
\subsection*{\b Values of the QCD condensates}
-- It is clear from the observation of the violation of the vacuum saturation for the four-quark condensates\,\cite{JAMI2a,JAMI2c,LNT,LAUNERb,SNTAU} and the large  value of the $\la G^3\ra$ triple gluon condensate from charmonium sum rules\,\cite{SNH10,SNH11} which largely deviates from the dilute gas instanton liquid model\,\cite{SVZa}  that the structure and the strength of higher dimension condensates are not trivial (violation of vacuum saturation (see Table\,\ref{tab:param}), mixing under renormalization\,\cite{TARRACH},...) such that the inclusion of only some classes of these high-dimension condensates in the OPE can be misleading. Instead, it may eventually serve as a check of the convergence of the OPE and/or an alternative estimate of the systematic errors. 

-- Some authors continue to use obsolete and inaccurate values of the $\la G^2\ra$ and $\la G^3\ra$ gluon condensates while the vacuum saturation  to estimate the four-quark operators of dimension 6 and higher dimensions ones are used.  The previous condensates have been re-estimated, as mentioned above,  since the former SVZ\,\cite{SVZa} pionner's work. The uses of different  inputs  are a source of discrepancy among the existing results.

\subsection*{\b Sources of the errors}

-- Often, the details of the different sources of  errors in the estimate are not given by the authors such that one has only to believe the errors quoted.There is not also a clear estimate of the systematic errors due to the truncation of the OPE. In our analysis and previous papers, we estimate these unknown remaining terms e.g. as :
\beq
\Delta{\rm OPE}\approx \frac{M_Q^2\tau}{3}C_6\la O_6\ra,
\eeq
where $C_6\la O_6\ra$ is the known contribution due to the dimension-six four-quark and $\la G^3\ra$ gluon condensates, while the factor 1/3 is the suppression factor in the LSR due to the exponential weight in the OPE. It is obvious that the OPE converges faster when the vacuum saturation is used to estimate the high-dimension vacuum  condensates but the validity of a such approximation has been questioned from different phenomeno;ogical analysis from $e^+e^-\to$ hadrons, $\tau$-decay data
and baryon sum rules (see Table\,\ref{tab:param}). . 

-- In our earlier works\,\cite{MOLE16,MOLE16X,SU3}, we have estimated the higher order terms of the PT series by  an estimate of the N2LO terms. We found that these contributions are negligible. 

\subsection*{\b Stability criteria}
The criteria used in many papers are often ad hoc / handwavings where the per cent of the ground state and continuum contributions to the sum rules and the per cent constraint for the convergence of the OPE are fixed by hand inside the choosen {\it sum rule window}. On the contrary, in all our LSR works (for a reviews see e.g.\,\cite{SNB1,SNB2,SNB3})  we use the optimization procedure based on  the minimal sensitivity on the changes of the set $(\tau,t_c,\mu)$ external variables as discussed in Section\,\ref{sec:stability} which is more rigorous. 

\subsection*{\b Concluding remarks}

One may say that reading some recent papers, one has the impression that the field of QCD (spectral) sum rules (QSSR) has not made any progress since its introduction by SVZ in 1979 despite the different active works done in the 80-90 for improving this nice SVZ discovery. Unfortunately, these different efforts seem to be ignored by the new generations of QCD (spectral) sum rules pratictioners  !

\section{Checking the  $T^{1^+,0^+}_{QQ\bar q'\bar q}$  
results of  Wang et al. in Refs.\,\cite{WANG-Ta,WANG-Tb}\label{sec:wang}}
Wang et al. use the same currents as in Table\,\ref{tab:current}.  
\subsection*{\b QCD expressions} 
\begin{figure}[hbt]
\begin{center}
\centerline {\hspace*{-3.8cm} \bf a)\hspace{4.8cm} b)}
\includegraphics[width=4cm]{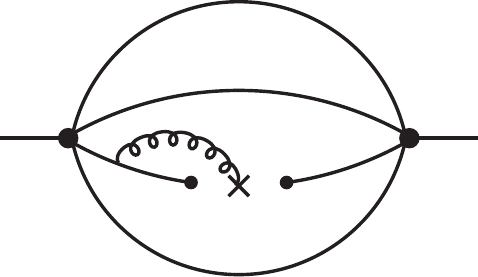}\hspace*{1cm}
\includegraphics[width=4cm]{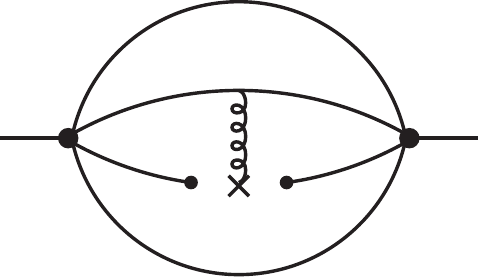}
\hspace*{2cm}
\vspace*{-0.5cm}
\caption{\footnotesize  Mixed quark-gluon condensate: a) self-energy\,; b) gluon exchange.} 
\label{fig:mixed}
\end{center}
\vspace*{-0.5cm}
\end{figure}

\begin{figure}[hbt]
\begin{center}
\includegraphics[width=4cm]{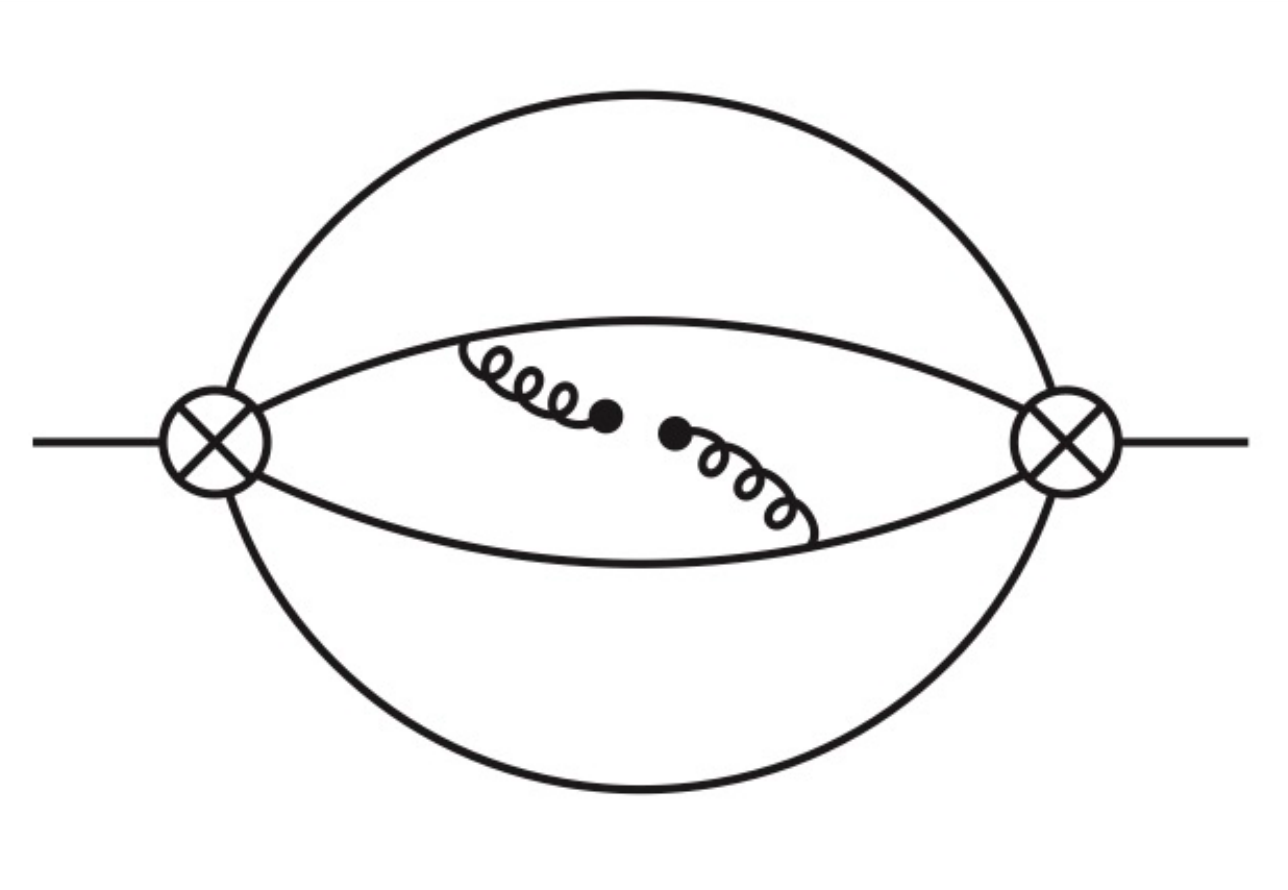}
\hspace*{2cm}
\vspace*{-0.5cm}
\caption{\footnotesize  Gluon condensate from one gluon exchange. } 
\label{fig:gluon}
\end{center}
\vspace*{-0.5cm}
\end{figure}

 Comparing the QCD expressions in the $1^+$ and $0^+$ channels, we find that we disagree for the $\la G^2\ra$ gluon and mixed $\la \bar qG q\ra$ condensates contributions while the $D=6$ $\la G^3\ra$ gluon condensate is missing.  Inspecting the expression of the propagator, we see that the propagator used in\,\cite{WANG-Ta,WANG-Tb} does not induce the contribution of the mixed condensate shown in Fig.\,\ref{fig:mixed}. 
These missed contributions read for the $T_{cc\bar u\bar s}(1^+)$ state:
\beq
 \rho^{\langle \bar{q}Gq \rangle}_{\rm missed}(t) = -\frac{m_Q^2 m_s\,\langle \bar{q}Gq \rangle}{3^2\times 2^5\times \pi^4}\,v \left(2+\frac{1}{x}\right)\ga 1-\frac{3}{2}\dr,
\eeq
with\,: $x=m^2/s$ and  $v=\sqrt{1-4x}$. 
We also suspect that the contribution due to one gluon exchange for the $\la G^2\ra$ is not generated by the propagator which can explain the origin of the discrepancy. This contribution 
is shown in Fig.\,\ref{fig:gluon}

and reads:
\begin{eqnarray}
    \rho^{\langle G^2 \rangle}_{\mathrm{missed}}(t) &=&- 
    \frac{m_Q^4 \langle G^2 \rangle}{3^3 \cdot 2^{11} \pi^6}
    \bigg[ v \Big( 42x + 43 - 88/x + 3/x^2 \Big) + \nonumber\\
    &&
    + 6 {\cal L}_v \Big( 14x^2 +12x -15 + 9\log(x) + 
    8/x \Big) + 108{\cal L_+} \bigg],
\end{eqnarray}

with: 
\begin{equation}
x=m_{Q}^{2}/t ~~ {\rm and} ~~v=\sqrt{1-4x}.
\label{eq:def}
\end{equation}
Hopefully, these missed contributions do not affect in a significant way the numerical results. However, a more precise comparison  cannot be done without an explicit expression of the contribution from each diagrams from the authors.
\subsection*{\b Comparison of the mass results} 
-- Comparing the mass results, we see a good agreement with\,\cite{WANG-Ta} (within the errors) for the $T^{1^+,0^+}_{cc\bar q\bar q'}$ states (Fig.\,\ref{fig:tccf}). However, one should note that the results quoted in the former paper \,\cite{WANG-Tb} give  masses higher (about 480 MeV) than the ones from\,\cite{WANG-Ta}. 

\begin{figure}[hbt]
\begin{center}
\centerline {\hspace*{-7.5cm} \bf a)\hspace{8cm} b)}
\includegraphics[width=8cm]{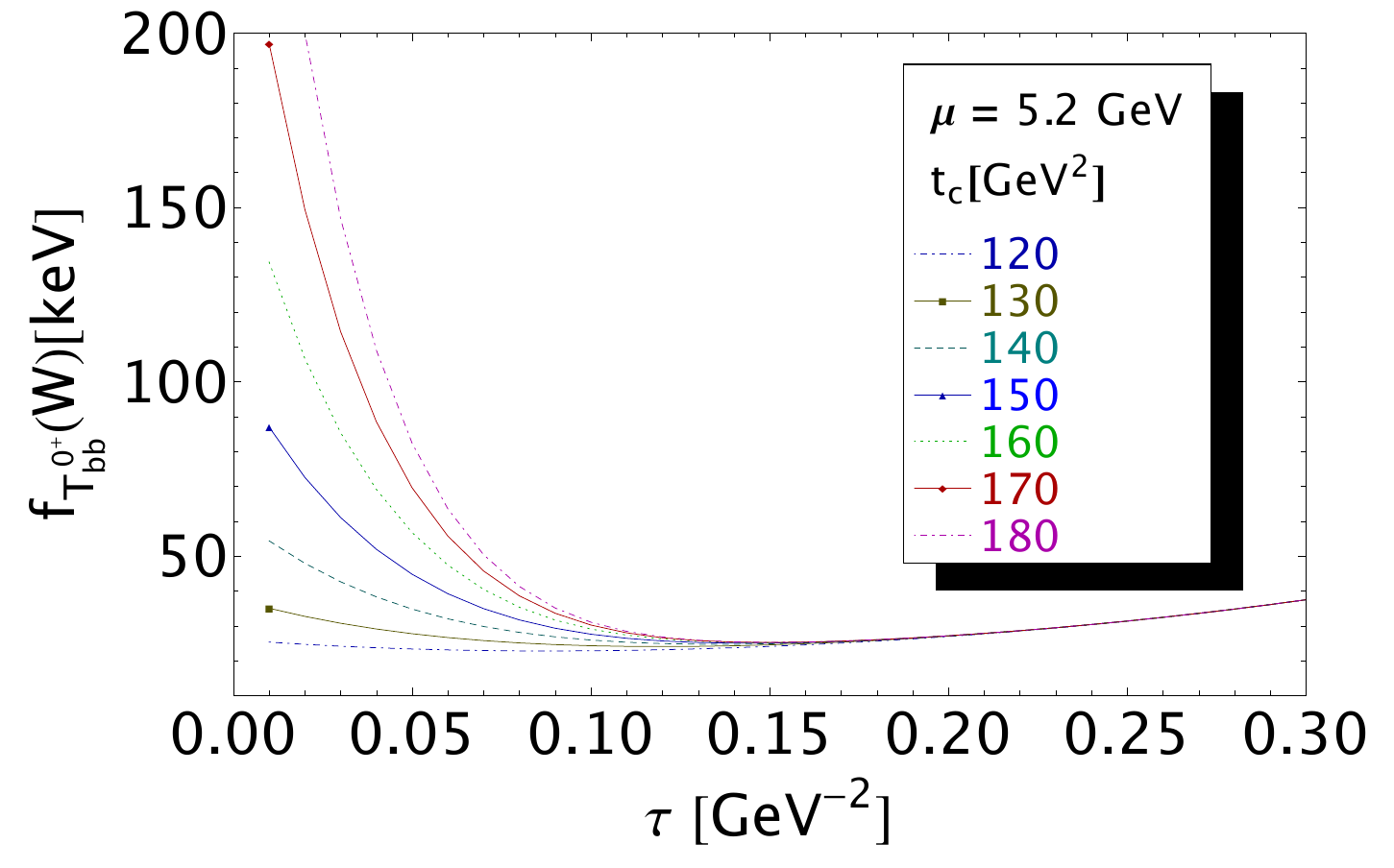}
\includegraphics[width=8cm]{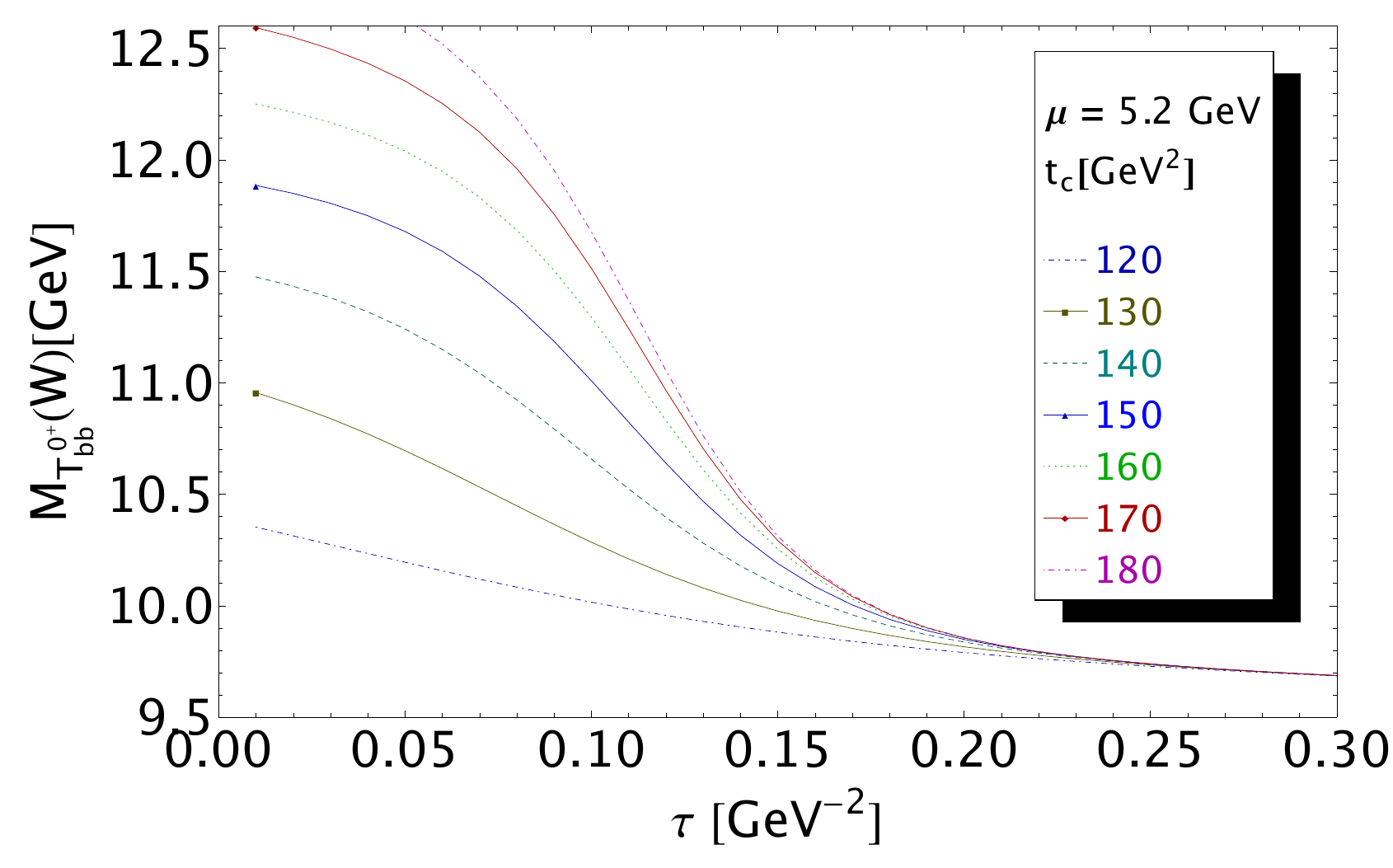}
\vspace*{-0.5cm}
\caption{\footnotesize  $f_{T^{0^+}_{bb}}$ and $M_{T^{0^+}_{bb}}$ as function of $\tau$  for \# values of $t_c$, for $\mu$=5.2 GeV and for the QCD inputs in Table\,\ref{tab:param}.} 
\label{fig:wang}
\end{center}
\vspace*{-0.5cm}
\end{figure}

-- One can also note  in Fig.\,\ref{fig:tbbf}, that the mass predictions for the $T^{0^+}_{bb\bar q\bar q}$ states ($ q\equiv d,s$) from \,\cite{WANG-Tb} are higher than ours by about 660 MeV. We look for the origin of this discrepancy by repeating the analysis using the (non corrected) expression of \,\cite{WANG-Tb}. The analysis is shown in Fig.\,\ref{fig:wang} where we have a nice $\tau$ stability for the coupling and an inflexion point for the mass. Both results also exhibit $t_c$-stability. We extract the optimal result for the set $(\tau,t_c)$ from (0.12,130) to (0.15,170) (GeV$^{-2}$,GeV$^2$) and deduce the central values:
\beq
f_{T^{0^+}_{bb}}\simeq 25~{\rm keV},~~~~~~~~~~M_{T^{0^+}_{bb}}\simeq 10.08~{\rm GeV},~~~~~~~~f_{T^{0^+}_{bb\bar s\bar s}}\simeq 24~{\rm keV},~~~~~~~~~~M_{T^{0^+}_{bb\bar s\bar s}}\simeq 10.28~{\rm GeV},
\eeq
lower than the ones quoted by\,\cite{WANG-Tb} :
\beq
M_{T^{0^+}_{bb}}\simeq (11.14\pm0.16)~{\rm GeV},~~~~~~~~~~~~~M_{T^{0^+}_{bb\bar s\bar s}}\simeq (11.32\pm 0.18)~{\rm GeV},
\eeq
but in lines with our results obtained from the (corrected) QCD expression summarized in Table\,\ref{tab:res}. We note that the range of $t_c$-values used by the authors are the same  as here while the value of $\tau$ is lower in\,\cite{WANG-Ta}  explaining their overestimate of the mass result.
 
\section{Checking the  $T^{1^+}_{QQ}$ results of  Agaev et al. in Ref.\,\cite{AGAEV-T}}
We see that the current used by\,\cite{AGAEV-T} is similar to the one used in Table\,\ref{tab:current} and $\eta_5^{1^+}$
(Eq.\,\ref{eq:eta1}) used by \cite{ZHU-T}. 
\subsection*{\b The $T^{1^+}_{cc}$ state}
The QCD expression is not given by the authors. However, inspecting the form of the propagator quoted in their review paper\,\cite{AGAEV-T}, 
we notice that it does not also induce the diagrams in Fig.\,\ref{fig:mixed} while for the numerical analysis, we notice that the optimal result is obtained for the set:
\beq
\tau\simeq (0.17-0.25)~{\rm GeV}^{-2},~~~~~~~~~~~~t_c\simeq  (19.5-21.5)~{\rm GeV}^2.
\eeq
where $t_c$ is below the beginning of $\tau$-stability of the coupling (Fig.\,\ref{fig:tpcc}), though the mass shows an apparent  $\tau\equiv 1/M^2$ stability (in reality, it increases with $M^2$)  in a narrow range of $\tau$ variation. As a result, the central value of the mass obtained by\,\cite{AGAEV-T} is slightly lower than ours and the LHCb data\,\cite{LHCb4}:
\beq
M_{T^{1^+}_{cc}}\simeq 3868(124)~{\rm MeV},
\eeq
though the errors are large.  
\subsection*{\b The $T^{1^+}_{bb}$ state}
The discrepancy is more pronounced in this case where the authors  extract their result using the sets:
\beq
\tau\simeq (0.077-0.111)~{\rm GeV}^{-2},~~~~~~~~~~~~t_c\simeq  (115-120)~{\rm GeV}^2.
\eeq
Looking at Fig.\,\ref{fig:tbb1p}, one can see like in the case of $T_{cc}$ that these sets of values are outside the (true) stability region.  As a result, the authors get :
\beq
M_{T^{1^+}_{bb}}\simeq 10035(260)~{\rm MeV},
\eeq
where the central value is much lower than ours in Table\,\ref{tab:res}. 
\section{Checking the  
results of  Du et al. in Ref.\,\cite{ZHU-T}\label{sec:zhu1}}
\subsection*{\b The $T_{QQ\bar q\bar q}$ $I(J^P)=1(1^+)$ state}
We complete the previous results from $\bar 3_c3_c$ currents in Table\,\ref{tab:current} by the ones from \,\cite{ZHU-T} where an 
exhaustive list is given.  In particular, we shall consider as a representative for the $1^+$ state the currents\,:
\beq
\eta_1^{1^+}\equiv\left(
 c_i^T\, C \gamma^\mu\, \gamma_5 \,c_j \right) \big{[} \left( \bar{q'}_i\, 
  C \,\bar{q}_j^T\right) +  \left( \bar{q'}_i\, 
  C \,\bar{q}_j^T\right)\big{]}~,~~~~\eta_5^{1^+}\equiv\left(
 c_i^T\, C \gamma^\mu\, c_j \right) \big{[} \left( \bar{q'}_i\, 
  C\gamma_5 \, \,\bar{q}_j^T\right) -  \left( \bar{q'}_i\, 
  C\gamma_5 \, \,\bar{q}_j^T\right)\big{]}~.
  \label{eq:eta1}
\eeq
(Eq. 5 of Ref.\,\cite{ZHU-T}) where  $\eta_1^{1^+}$ gives the highest mass prediction and $\eta_5^{1^+}$ is equivalent to ours in Table\,\ref{tab:current}.  

 By comparing our expression for the spectral function corresponding to $\eta_5^{1^+}$ given in\,\ref{app-a} with the one of\,\cite{ZHU-T},
we notice an agreement on the PT contribution. Our expressions for the $\la G^2\ra$ gluon and mixed $\la \bar qG q\ra$ condensates disagree. We notice an overall factor 3 (a misprint ?) in the four-quark contribution  while the $D=6$ $\la G^3\ra$ is missing.    

 For the $\eta_1^{1^+}$ current, our expression given in\,\ref{app-b}  agrees with the PT, $\la \bar qq\ra$ and four-quark condensates of\,\cite{ZHU-T} while there is a persisting disagreement for the $\la G^2\ra$ gluon and mixed $\la \bar qG q\ra$ condensates. The $\la G^3\ra$ contribution is also missing. 

 We interpret the origin of the discrepancy for $\la \bar qG q\ra$ as due to the expression of the propagator used in Ref.\,\cite{ZHU-T} which does not induce the contribution of the diagrams shown in Fig.\,\ref{fig:mixed}. 

 Doing the numerical analysis, we realize that\,:

-- The mixed quark-gluon condensate is parametrized with a wrong sign (a misprint ?).   

\begin{figure}[hbt]
\begin{center}
\centerline {\hspace*{-7.5cm} \bf a)\hspace{8cm} b)}
\includegraphics[width=8cm]{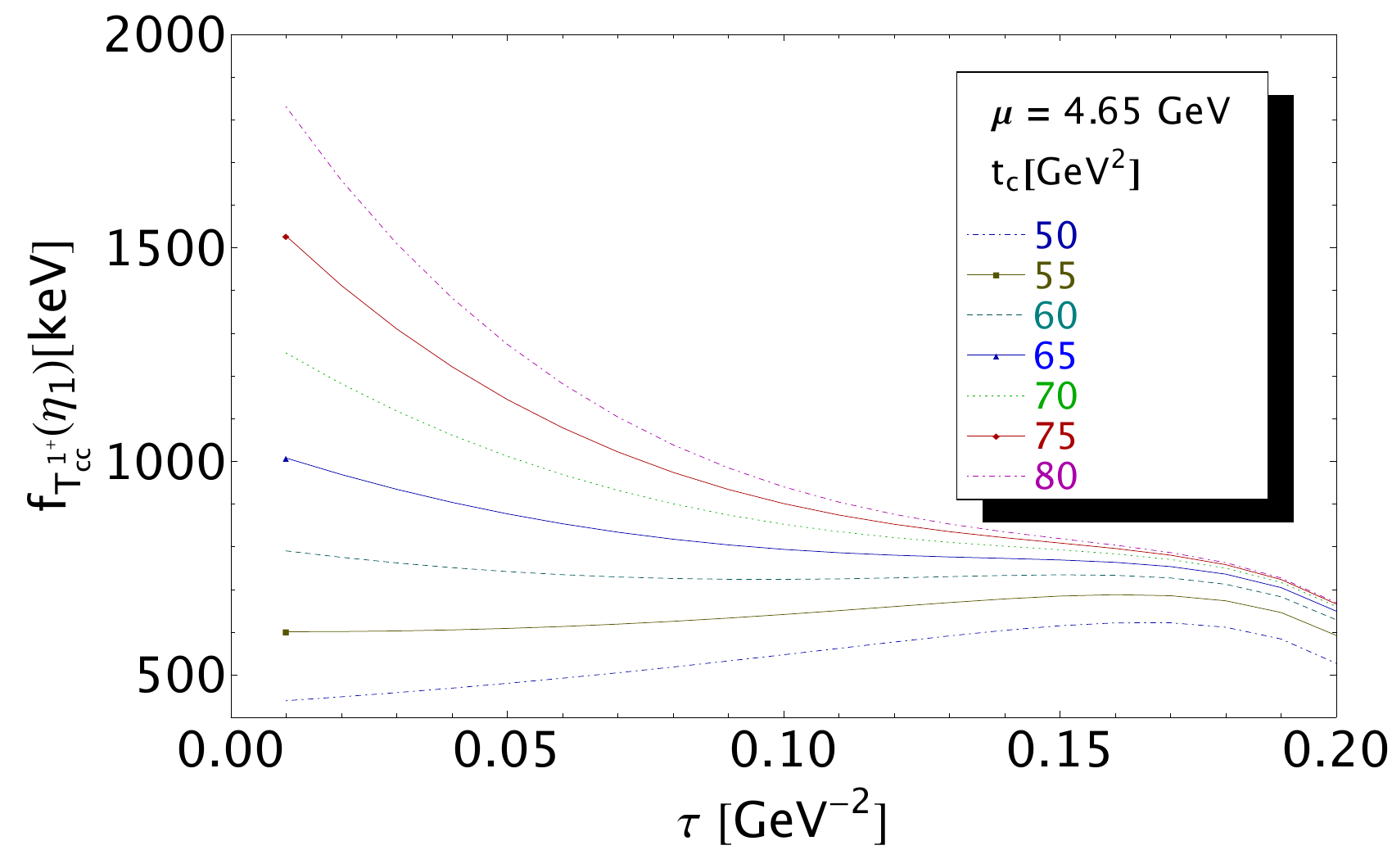}
\includegraphics[width=8cm]{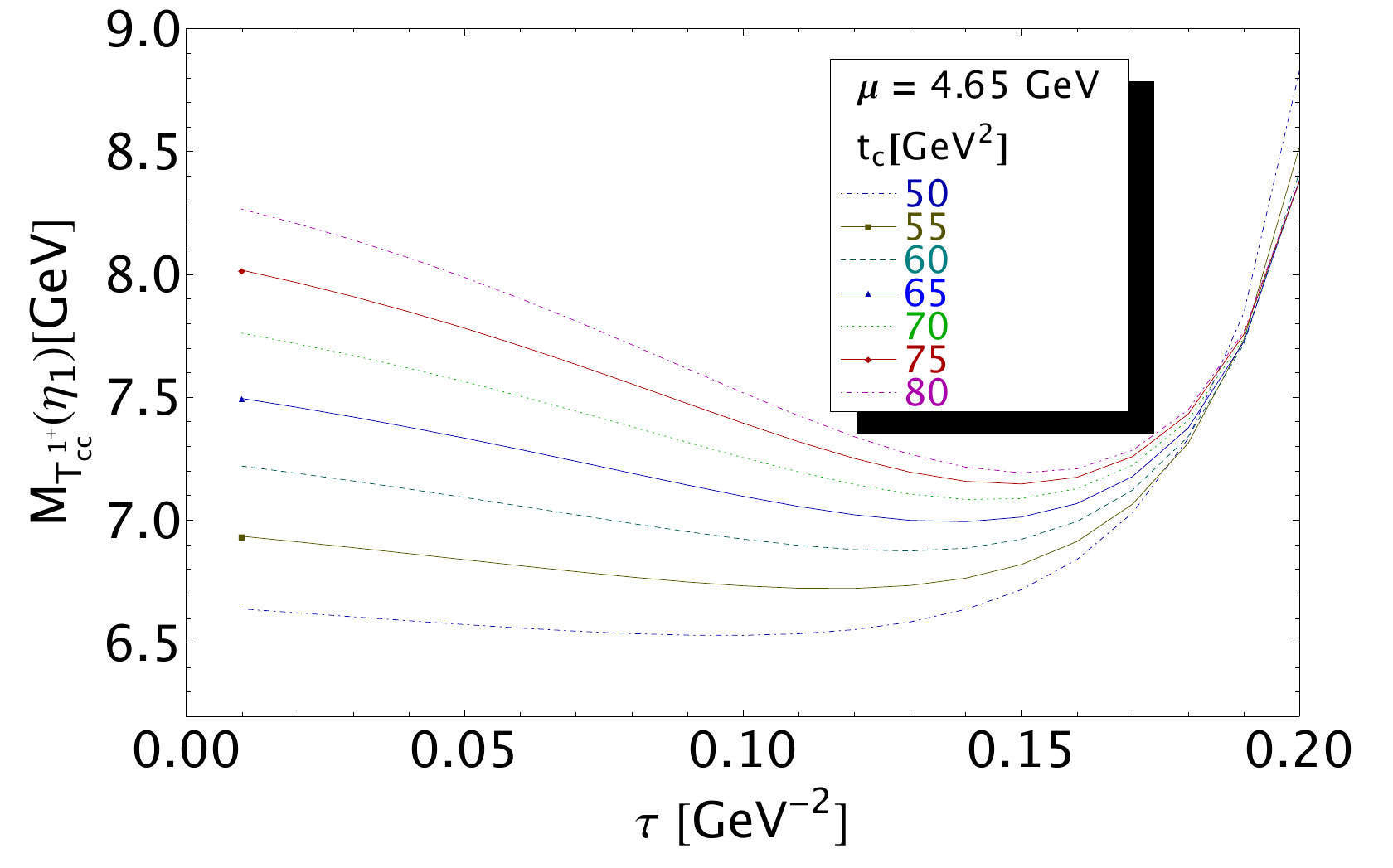}
\vspace*{-0.5cm}
\caption{\footnotesize  $f_{T^{1^+}_{cc}}$ and $M_{T^{1^+}_{cc}}$ as function of $\tau$  for \# values of $t_c$, for $\mu$=4.65 GeV and for the QCD inputs in Table\,\ref{tab:param}.} 
\label{fig:zhuc}
\end{center}
\vspace*{-0.5cm}
\end{figure}

\begin{figure}[hbt]
\begin{center}
\centerline {\hspace*{-7.5cm} \bf a)\hspace{8cm} b)}
\includegraphics[width=8cm]{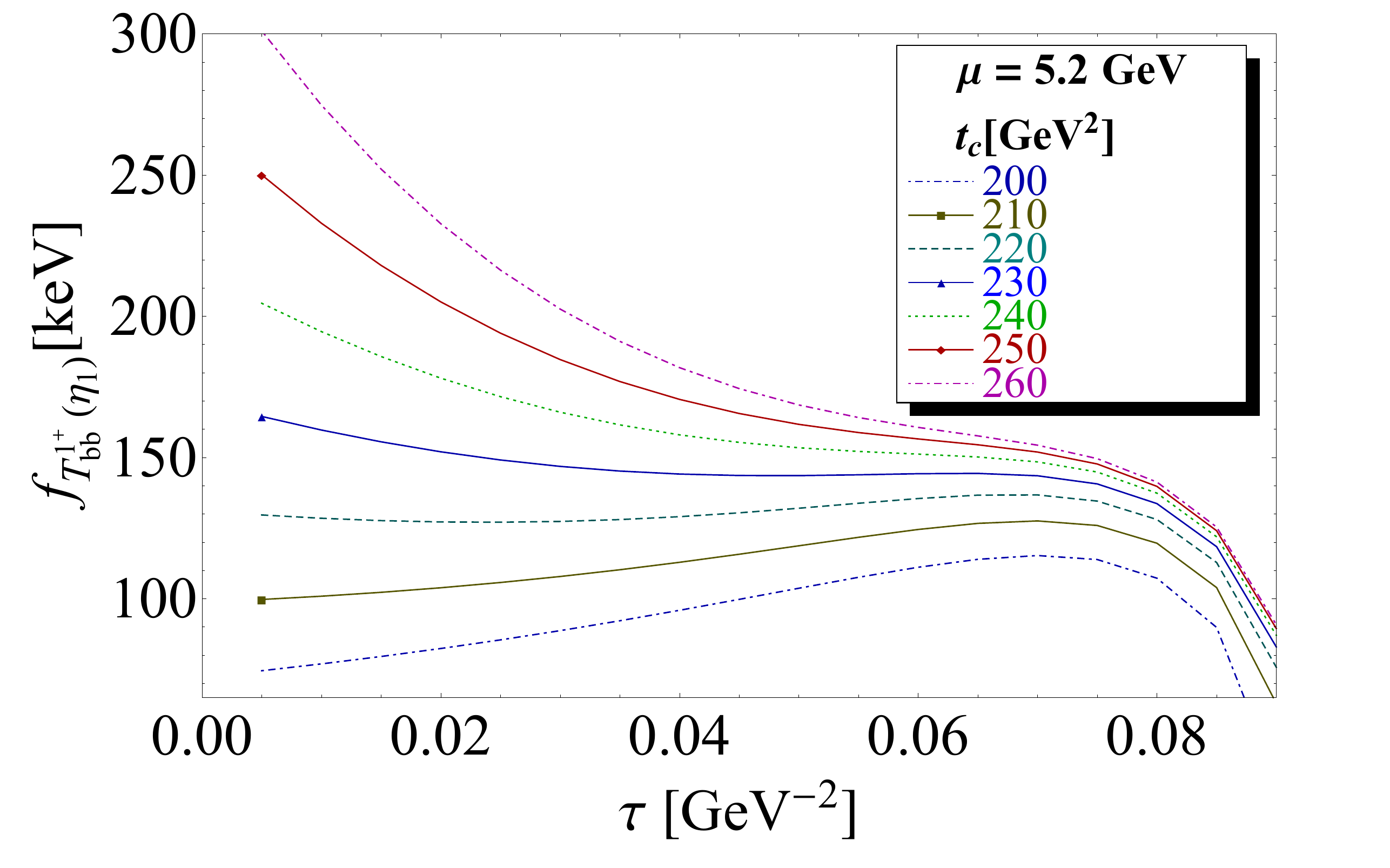}
\includegraphics[width=8cm]{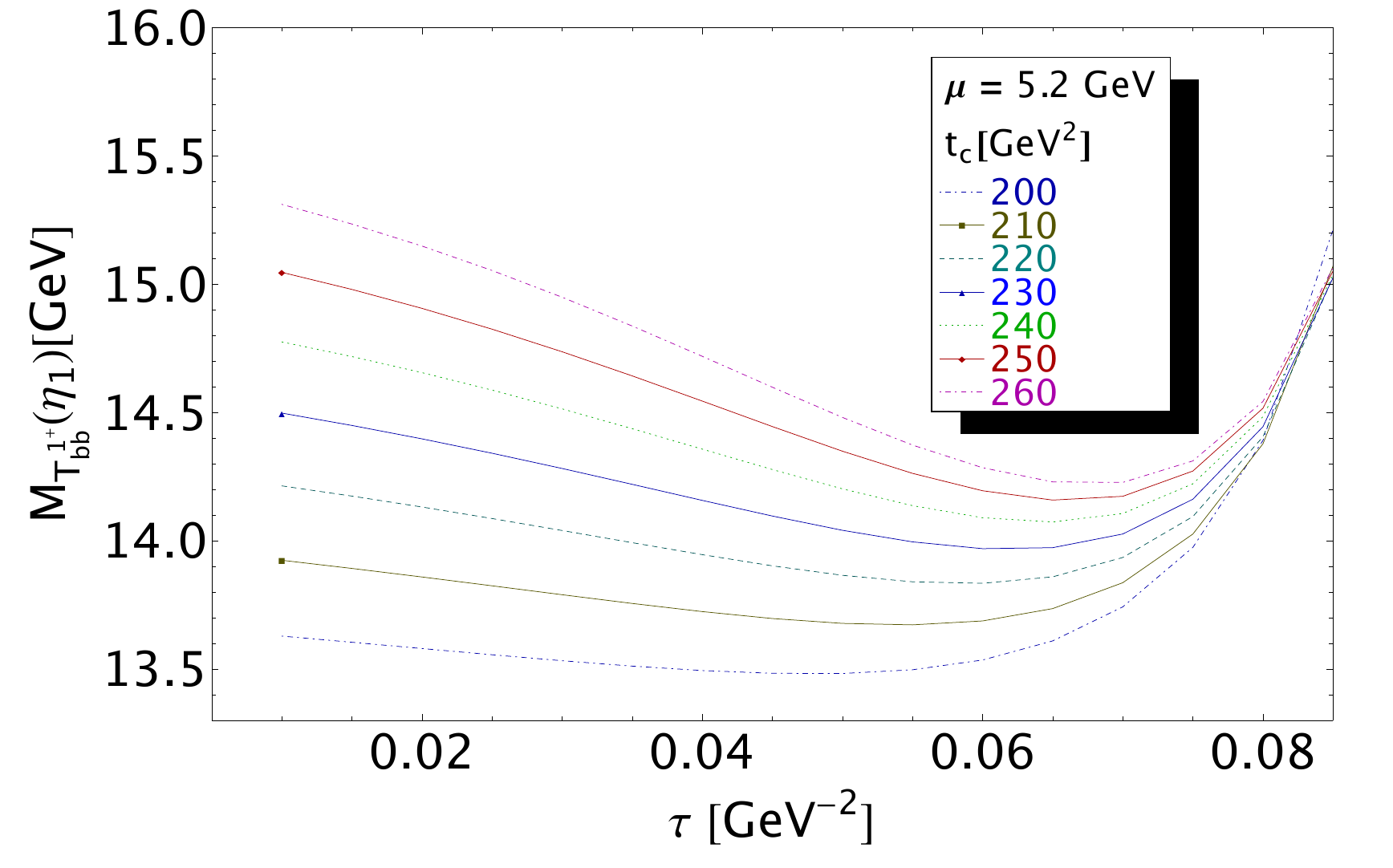}
\vspace*{-0.5cm}
\caption{\footnotesize  $f_{T^{1^+}_{bb}}$ and $M_{T^{1^+}_{bb}}$ as function of $\tau$  for \# values of $t_c$, for $\mu$=5.2 GeV and for the QCD inputs in Table\,\ref{tab:param}.} 
\label{fig:zhub}
\end{center}
\vspace*{-0.5cm}
\end{figure}

-- The choice of $t_c$ used by the authors are too low which is outside the beginning of the true $\tau$-stability region for the mass [$t_c\simeq$ 21-28 (resp. 115-125) GeV$^2$] for the charm (resp. beauty) channels (see Figs.\,\ref{fig:zhuc} and \ref{fig:zhub}).  Indeed, a (misleading) $\tau$-stability is obtained for the mass but at these low values of $t_c$ the coupling is not stable. 

 Using our QCD expression at NLO, we show the analysis  of the  $T_{cc}$ coupling and mass in Fig.\,\ref{fig:zhuc}.  We have not included the $D=8$ contribution due to $\la\bar qq\ra\la \bar s G s\ra$ obtained in\,\cite{ZHU-T}. Keeping (consistently) the term without $m_c^2$ and $m_c^4$  in this contribution which competes with the  $m_c^2\la\bar ss\ra^2$ dimension $D=6$ one, we find that it increases the mass prediction by about 40 MeV which is negligible compared to the errors of 224 MeV (see Table\,\ref{tab:zhuc}). 

 One can notice a $(\tau, t_c)$ stability for the sets (0.13, 60) to (0.15, 75) (GeV$^{-2}$, GeV$^2$) for the mass which allows to fix accurately the position of the inflexion point for the coupling.  
 
 The analysis of $T_{cc\bar s \bar s}$ gives a similar behaviour. An optimal result is obtained at the same  sets of $(\tau,t_c)$ values.  The result and the sources of the errors are given in Table\,\ref{tab:zhuc}.

Our results in this $J^P=1^+$ channel do not support the claims of\,\cite{ZHU-T}  on the non-existence of the $T_{cc}$ tetraquark state.

A similar analysis is done for the $T_{bb}$ coupling and mass which is shown in Fig.\,\ref{fig:zhub}.  The optimal results are obtained for the sets $(\tau , t_c)$=(0.055, 220) to (0.065, 250) (GeV$^{-2}$, GeV$^2$). The result and the sources of the errors are given in Table\,\ref{tab:zhub}.

  One can notice that\,:

-- The inclusion of the  NLO  PT corrections decreases the $T_{cc}$ and $T_{cc\bar s \bar s}$ masses by about 188 MeV and the $T_{bb}$ and $T_{bb\bar s \bar s}$ ones by 195 MeV.

--  The SU3 breaking decreases the central value of the $T_{cc\bar s \bar s}$ by 182 MeV relative to $T_{cc}$ and the one of $T_{bb\bar s \bar s}$ by 195 MeV relative to $T_{bb}$.

-- Our results for the masses are definitely higher than the ones obtained in \,\cite{ZHU-T} despite the large errors. 

-- The couplings from our analysis are also large. One can understand this increases by the exponential behaviour of the coupling in the LSR analysis\,:
$f_T\sim  (1/M_T^4)e^{\tau M_T^2/2}$
\begin{table}[hbt]
\setlength{\tabcolsep}{0.17pc}
{\scriptsize{
\begin{tabular}{ll ll  ll  ll ll ll ll ll l c}
\hline
\hline
                \bf Observables &\multicolumn{1}{c}{$\Delta t_c$}
					&\multicolumn{1}{c}{$\Delta \tau$}
					&\multicolumn{1}{c}{$\Delta \mu$}
					&\multicolumn{1}{c}{$\Delta \alpha_s$}
					&\multicolumn{1}{c}{$\Delta PT$}
					&\multicolumn{1}{c}{$\Delta m_s$}
					&\multicolumn{1}{c}{$\Delta m_c$}
					&\multicolumn{1}{c}{$\Delta \overline{\psi}\psi$}
					&\multicolumn{1}{c}{$\Delta \kappa$}					
					&\multicolumn{1}{c}{$\Delta G^2$}
					&\multicolumn{1}{c}{$\Delta M^{2}_{0}$}
					&\multicolumn{1}{c}{$\Delta \overline{\psi}\psi^2$}
					&\multicolumn{1}{c}{$\Delta G^3$}
					&\multicolumn{1}{c}{$\Delta OPE$}
					&\multicolumn{1}{c}{$\Delta M_{G}$}
					&\multicolumn{1}{r}{This work}
					&\multicolumn{1}{c}{Du et al.\,\cite{ZHU-T}}
\\
					
\hline
{\bf Coupling} [keV] &&&&&&&&&&&\\
$f_{T^{1^+}_{cc}}$ &74.4&11.0&1.86&8.45&3.21&$\cdots$&9.01&0.00&$\cdots$&0.67&0.00&12.3&0.07&5.18&61.1&734(99)&$\cdots$ \\
$f_{T^{1^+}_{ccsu}}$ &42.9&0.86&1.36&6.20&1.20&0.41&6.47&0.56&6.29&0.52&0.19&8.01&0.06&6.61&64.3&510(103)&132 \\
$f_{T^{1^+}_{ccss}}$ &45.2&7.49&2.00&9.15&1.64&1.09&9.46&3.14&13.5&0.85&0.57&11.3&0.11&7.20&50.3&710(72)&216 \\
\\
$f_{T^{0^+}_{cc}}$ &66.7&1.34&1.72&7.52&2.92&$\cdots$&11.9&0.00&$\cdots$&0.67&0.00&19.0&0.05&13.6&77.1&986(106)&$\cdots$ \\
$f_{T^{0^+}_{ccsu}}$ &35.3&0.69&1.28&4.24&3.25&0.49&11.9&0.67&6.66&0.53&0.29&14.2&0.04&7.78&54.9&677(69)&$\cdots$ \\
$f_{T^{0^+}_{ccss}}$ &34.9&0.72&1.89&8.33&2.75&1.23&12.3&1.73&19.8&0.84&0.84&17.2&0.07&14.4&82.8&928(96)&209 \\
{\bf Mass} [MeV] &&&&&&&&&&&&&&\\
$M_{T^{1^+}_{cc}}$&212&12.8&0.70&2.61&1.48&$\cdots$&9.57&0.00&$\cdots$&1.87&0.00&62.4&0.46&32.0&$\cdots$&6934(224)&$\cdots$\\
$M_{T^{1^+}_{ccsu}}$&183&11.7&0.88&3.46&1.50&1.91&9.89&2.62&33.2&2.11&1.34&62.3&0.55&34.8&$\cdots$&6846(200)&4960(110)\\
$M_{T^{1^+}_{ccss}}$&153&11.5&1.03&4.18&2.14&3.67&10.5&5.13&70.3&2.55&2.95&62.1&0.71&37.0&$\cdots$&6752(184)&5030(130)\\
\\
$M_{T^{0^+}_{cc}}$&161&12.2&0.02&0.93&5.80&$\cdots$&9.76&0.00&$\cdots$&1.42&0.00&72.5&0.23&41.8&$\cdots$&6758(182)&$\cdots$\\
$M_{T^{0^+}_{ccsu}}$&131&11.7&0.09&0.65&1.05&1.69&10.4&2.35&38.4&1.68&1.54&71.5&0.29&44.7&$\cdots$&6650(161)&$\cdots$\\
$M_{T^{0^+}_{ccss}}$&102&10.6&0.25&0.27&0.85&3.04&10.7&4.42&79.4&1.93&3.24&71.9&0.35&50.1&$\cdots$&6532(157)&5050(150)\\
\\
\hline
\hline
\end{tabular}
}}
 \caption{Sources of errors of the $T_{cc}$, $T_{ccsu}$, $T_{ccss}$ couplings and masses. The PT series is known to NLO and the OPE truncated at $D=6$ dimension  condensates. We take $\ve \Delta \mu\ve=0.05$ GeV and $\ve \Delta \tau\ve =0.01$ GeV$^{-2}$. Last column : results from\,\cite{ZHU-T}.}

\label{tab:zhuc}
\end{table}

\begin{table}[hbt]
\setlength{\tabcolsep}{0.17pc}
{\scriptsize{
\begin{tabular}{ll ll  ll  ll ll ll ll ll l c}
\hline
\hline
                \bf Observables &\multicolumn{1}{c}{$\Delta t_c$}
					&\multicolumn{1}{c}{$\Delta \tau$}
					&\multicolumn{1}{c}{$\Delta \mu$}
					&\multicolumn{1}{c}{$\Delta \alpha_s$}
					&\multicolumn{1}{c}{$\Delta PT$}
					&\multicolumn{1}{c}{$\Delta m_s$}
					&\multicolumn{1}{c}{$\Delta m_c$}
					&\multicolumn{1}{c}{$\Delta \overline{\psi}\psi$}
					&\multicolumn{1}{c}{$\Delta \kappa$}					
					&\multicolumn{1}{c}{$\Delta G^2$}
					&\multicolumn{1}{c}{$\Delta M^{2}_{0}$}
					&\multicolumn{1}{c}{$\Delta \overline{\psi}\psi^2$}
					&\multicolumn{1}{c}{$\Delta G^3$}
					&\multicolumn{1}{c}{$\Delta OPE$}
					&\multicolumn{1}{c}{$\Delta M_{G}$}
					&\multicolumn{1}{r}{This work}
					&\multicolumn{1}{c}{Du et al.\,\cite{ZHU-T}}\\
		
\hline
{\bf Coupling} [keV] &&&&&&&&&&&\\
$f_{T^{1^+}_{bb}}$ &15.4&1.09&0.33&2.41&0.26&$\cdots$&1.99&0.00&$\cdots$&0.04&0.00&2.37&0.00&1.78&17.2&144(24)&$\cdots$ \\
$f_{T^{1^+}_{bbsu}}$ &6.69&0.92&0.26&1.94&0.00&0.07&1.63&0.10&0.74&0.05&0.03&1.42&0.01&5.94&16.9&112(19)&12 \\
$f_{T^{1^+}_{bbss}}$ &9.46&0.75&0.33&2.40&0.07&0.89&1.97&0.20&1.77&0.06&0.07&2.12&0.01&1.24&21.0&132(24)&23 \\
\\
$f_{T^{0^+}_{bb}}$ &7.42&1.33&0.39&3.16&0.39&$\cdots$&3.23&0.00&$\cdots$&0.05&0.00&4.58&0.00&2.61&35.2&217(37)&$\cdots$ \\
$f_{T^{0^+}_{bbsu}}$ &3.54&3.63&0.27&2.17&0.10&0.07&2.20&0.10&1.28&0.04&0.05&2.85&0.00&1.88&23.9&141(25)&$\cdots$ \\
$f_{T^{0^+}_{bbss}}$ &3.28&0.83&0.37&2.96&0.10&0.16&2.95&0.24&3.87&0.06&0.13&3.45&0.00&2.57&31.8&181(33)&21 \\
{\bf Mass} [MeV] &&&&&&&&&&&&&&\\
$M_{T^{1^+}_{bb}}$&243&76.3&0.70&6.90&0.79&$\cdots$&9.23&0.00&$\cdots$&0.78&0.00&60.6&0.13&63.2&$\cdots$&14068(270)&$\cdots$\\
$M_{T^{1^+}_{bbsu}}$&199&135&0.73&7.83&0.84&2.15&12.0&3.00&41.8&1.28&1.58&73.4&0.23&88.8&$\cdots$&13960(270)&10700(300)\\
$M_{T^{1^+}_{bbss}}$&185&79.2&0.88&8.20&1.11&3.10&10.5&4.48&69.5&1.20&2.50&61.4&0.02&74.7&$\cdots$&13732(234)&11000(300)\\
\\
$M_{T^{0^+}_{bb}}$&125&90.3&0.08&4.90&0.17&$\cdots$&12.4&0.00&$\cdots$&0.73&0.00&85.9&0.05&115&$\cdots$&13647(211)&$\cdots$\\
$M_{T^{0^+}_{bbsu}}$&92.6&99.0&0.75&5.55&0.06&1.53&12.7&2.15&43.2&0.85&1.60&84.7&0.05&118&$\cdots$&13511(206)&$\cdots$\\
$M_{T^{0^+}_{bbss}}$&71.1&87.2&0.88&6.23&0.63&2.45&13.0&3.98&91.7&0.98&3.28&83.3&0.08&109&$\cdots$&13369(200)&11000(200)\\
\\
\hline
\hline
\end{tabular}
}}
 \caption{Same as in Table\,\ref{tab:zhuc} but for the $T_{bb}$, $T_{bbs}$, $T_{bbss}$ couplings and masses.}

\label{tab:zhub}
\end{table}

\subsection*{\b The $T_{QQ\bar q\bar q}$ $I(J^P)=1(0^+)$ state\label{sec:zhu0}}
For this state, we use the tetraquark current given in Eq.(3)  of Ref.\,\cite{ZHU-T}:
\begin{equation}
    \eta_1^{0^+} ~=~
    \: Q_i^T C Q_j
    \Big[ \bar{q}_i C \bar{q}_j^T
    + \bar{q}_j C \bar{q}_i^T \Big],
    ~~~~~~
        \eta_3^{0^+} ~=~
    \: Q_i^T C\,\gamma^\mu Q_j
    \Big[ \bar{q}_i C\,\gamma_\mu \bar{q}_j^T
    - \bar{q}_j C\,\gamma_\mu \bar{q}_i^T \Big],
    \label{eq:eta10}
\end{equation}
where we note that the current $\eta_3$ should give the same spectral function as the one from Table\,\ref{tab:current}. 

 By comparing the QCD expressions for $\eta_3$, we find an agreement for the PT, $\la\bar qq\ra$ and  $\la\bar qq\ra^2 $ contributions but not for the $\la G^2\ra$ gluon and $\la\bar qGq\ra$ mixed condensates. 

 A comparison of the spectral function of $\eta_1$ shows an agreement up to gluon condensate $\la G^2\ra$. The contribution of the mixed $\la\bar qGq\ra$ condensate is completely different and four-quark $\la\bar qq\ra^2 $ condensate differs from a minus sign. 

 One also notice a  parametrization of the mixed condensate with a (wrong) sign (Eq. 15) but it does not affect much the result while the (wrong) sign of the four-quark condensate changes completely the behaviour of the curves and the conclusion:

  We have checked that with the wrong sign, we (almost) reproduce the result of\,\cite{ZHU-T} where the coupling presents a $\tau$-stability in the region around 0.295 (resp 0.135) GeV$^{-2}$ and for low values  $t_c \simeq$ 28 (resp. 160) GeV$^2$ for $T^{0^+}_{cc\bar s\bar s}$ (resp.  $T^{0^+}_{bb\bar s\bar s}$). We obtain at LO\,:
\beq
M_{T^{0^+}_{cc\bar s\bar s}}\simeq 4.3~{\rm GeV},~~~f_{T^{0^+}_{cc\bar s\bar s}}\simeq 173~{\rm keV}~~~~~~{\rm and}~~~~~~
M_{T^{0^+}_{bb\bar s\bar s}}\simeq 11~{\rm GeV},~~~f_{T^{0^+}_{bb\bar s\bar s}}\simeq 12~{\rm keV},
\eeq

\begin{figure}[hbt]
\begin{center}
\centerline {\hspace*{-7.5cm} \bf a)\hspace{8cm} b)}
\includegraphics[width=8cm]{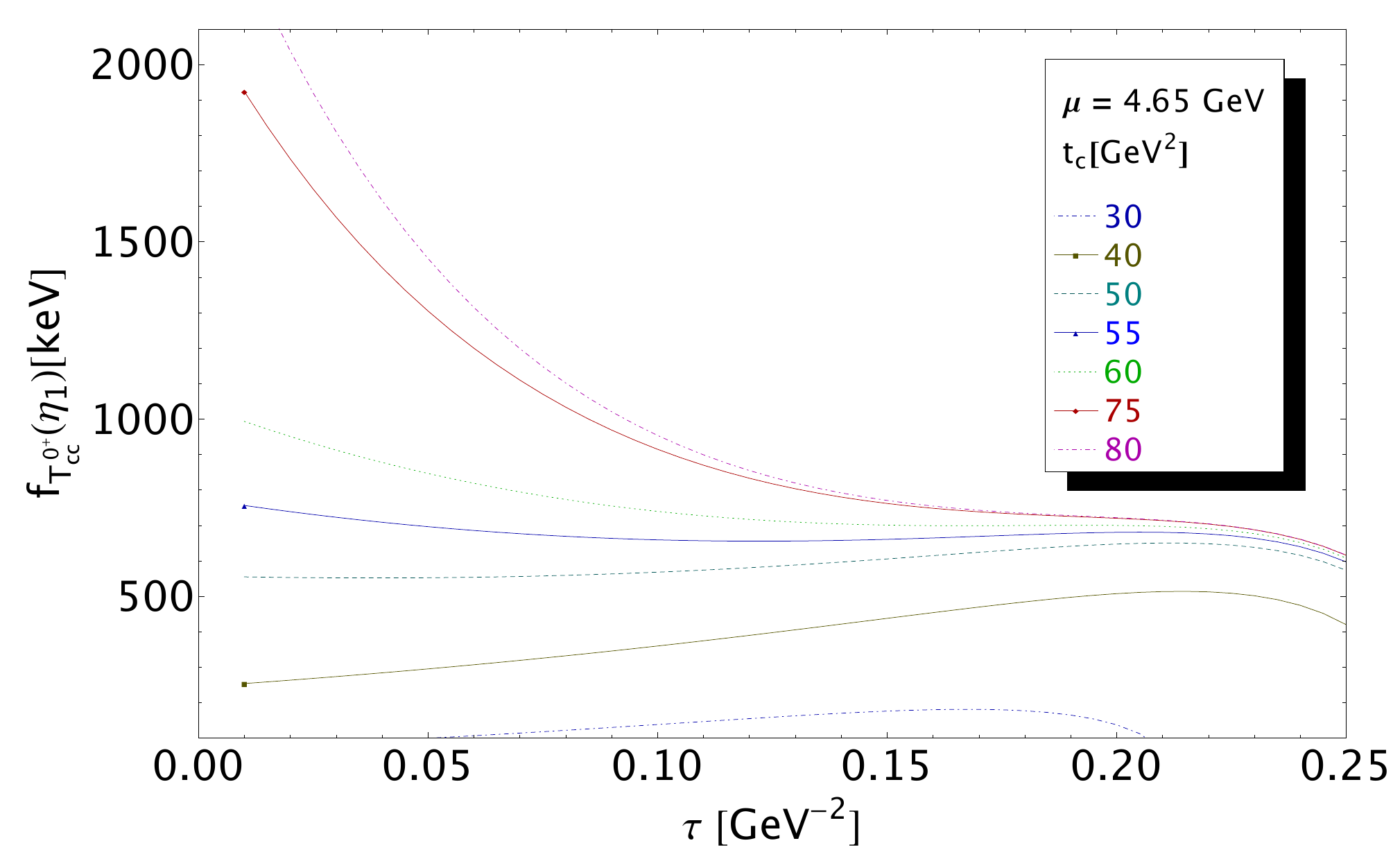}
\includegraphics[width=8cm]{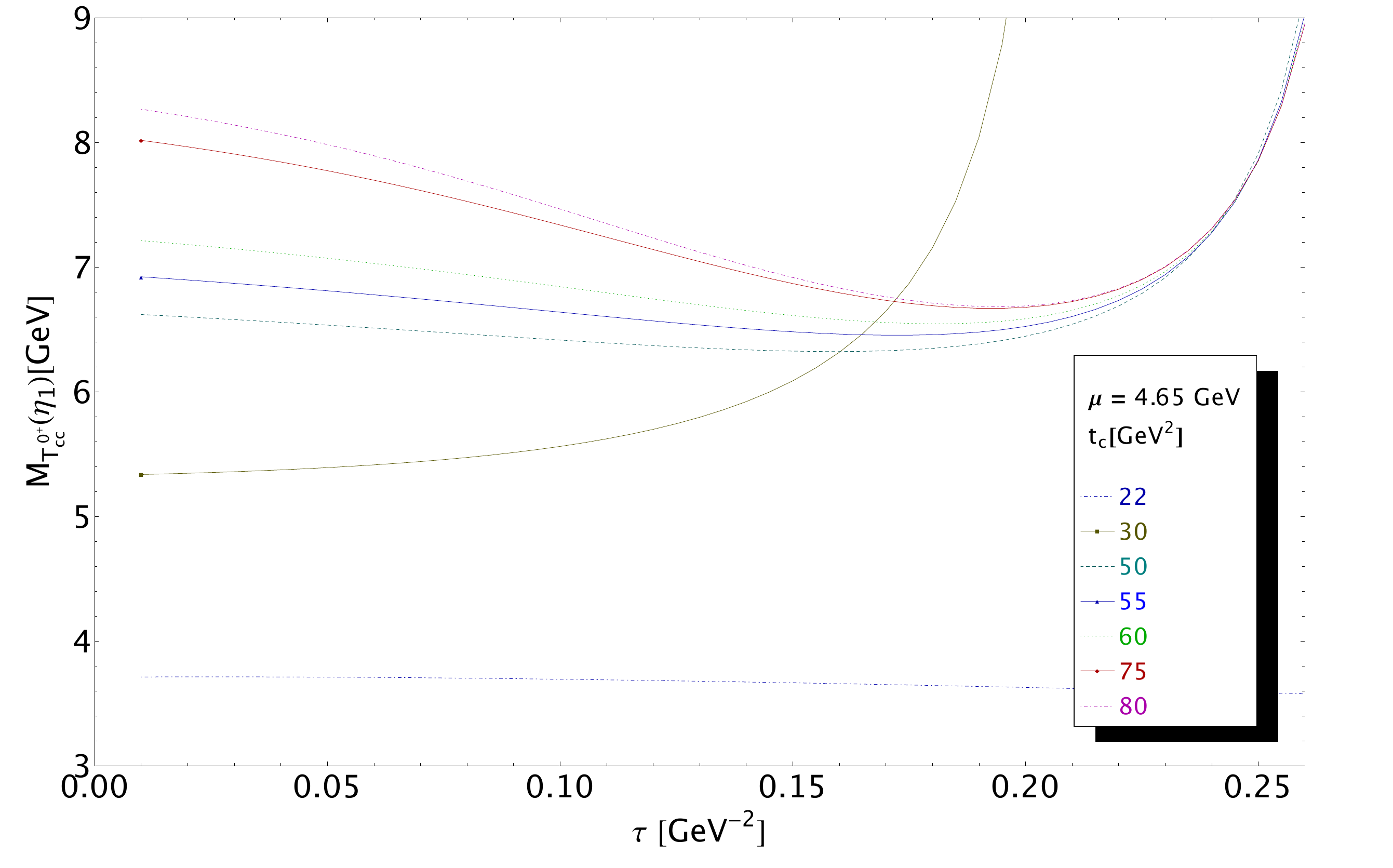}
\vspace*{-0.5cm}
\caption{\footnotesize  $f_{T^{0^+}_{cc}}$ and $M_{T^{0^+}_{cc}}$ as function of $\tau$  for \# values of $t_c$, for $\mu$=4.65 GeV and for the QCD inputs in Table\,\ref{tab:param}.} 
\label{fig:zhu-zeroc}
\end{center}
\vspace*{-0.5cm}
\end{figure}

\begin{figure}[hbt]
\begin{center}
\centerline {\hspace*{-7.5cm} \bf a)\hspace{8cm} b)}
\includegraphics[width=8cm]{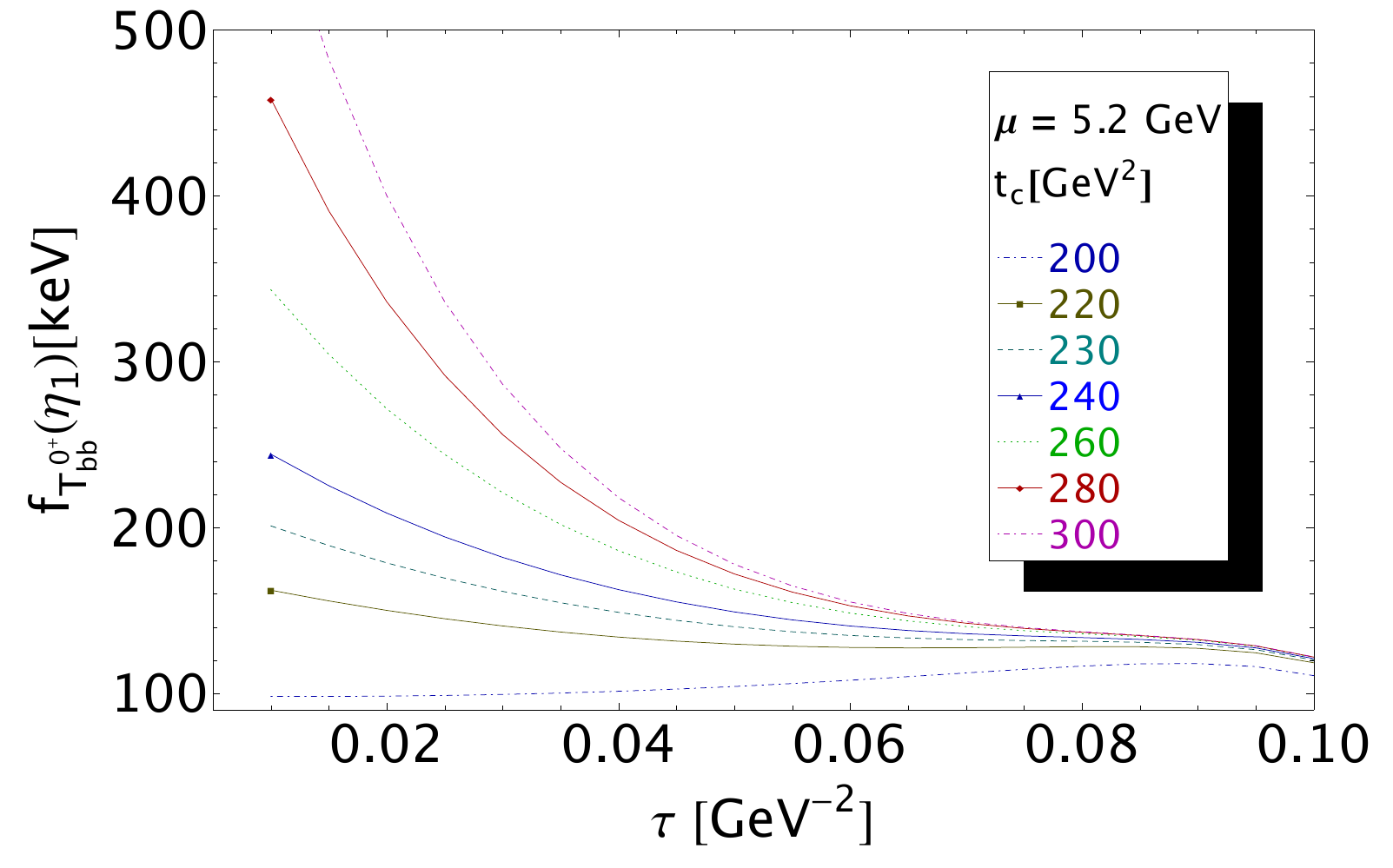}
\includegraphics[width=8cm]{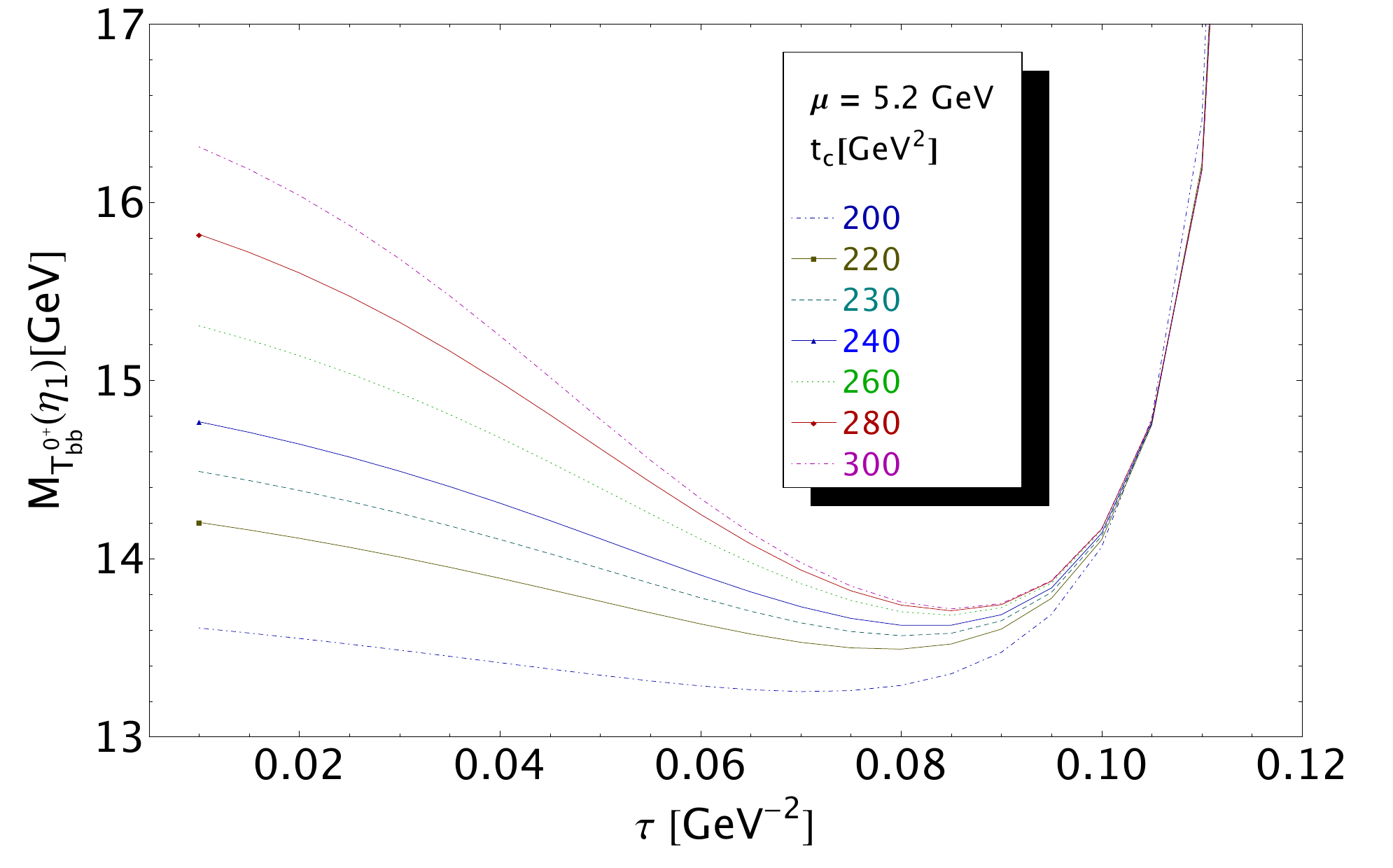}
\vspace*{-0.5cm}
\caption{\footnotesize  $f_{T^{0^+}_{bb}}$ and $M_{T^{0^+}_{bb}}$ as function of $\tau$  for \# values of $t_c$, for $\mu$=5.2 GeV and for the QCD inputs in Table\,\ref{tab:param}.} 
\label{fig:zhu-zerob}
\end{center}
\vspace*{-0.5cm}
\end{figure}

 For the corrected sign, we illustrate, in Figs.\,\ref{fig:zhu-zeroc} and \,\ref{fig:zhu-zerob}, the analysis in the chiral limit where one can notice that the optimal results are obtained at larger values of $t_c$ implying by duality larger values of the meson masses. Our results are summarized in Tables\,\ref{tab:zhuc} and \,\ref{tab:zhub} where we conclude that the masses of $T_{QQ\bar q'\bar q}$ states associated to the $\eta_1$ current are large and can be confused with the continuum. 

 Our results in this $J^P=0^+$ channel do not support the claims of\,\cite{ZHU-T}  on the non-existence of the $T_{cc}$ and $T_{cc\bar s\bar u}$  tetraquark states. 


\section{$T_{QQ\bar q\bar q'}$ state from the $\bar 8_c8_c$ current\,\cite{MALT-T}}\label{sec:malt}
\begin{figure}[hbt]
\begin{center}
\centerline {\hspace*{-7.5cm} \bf a)\hspace{8cm} b)}
\includegraphics[width=8cm]{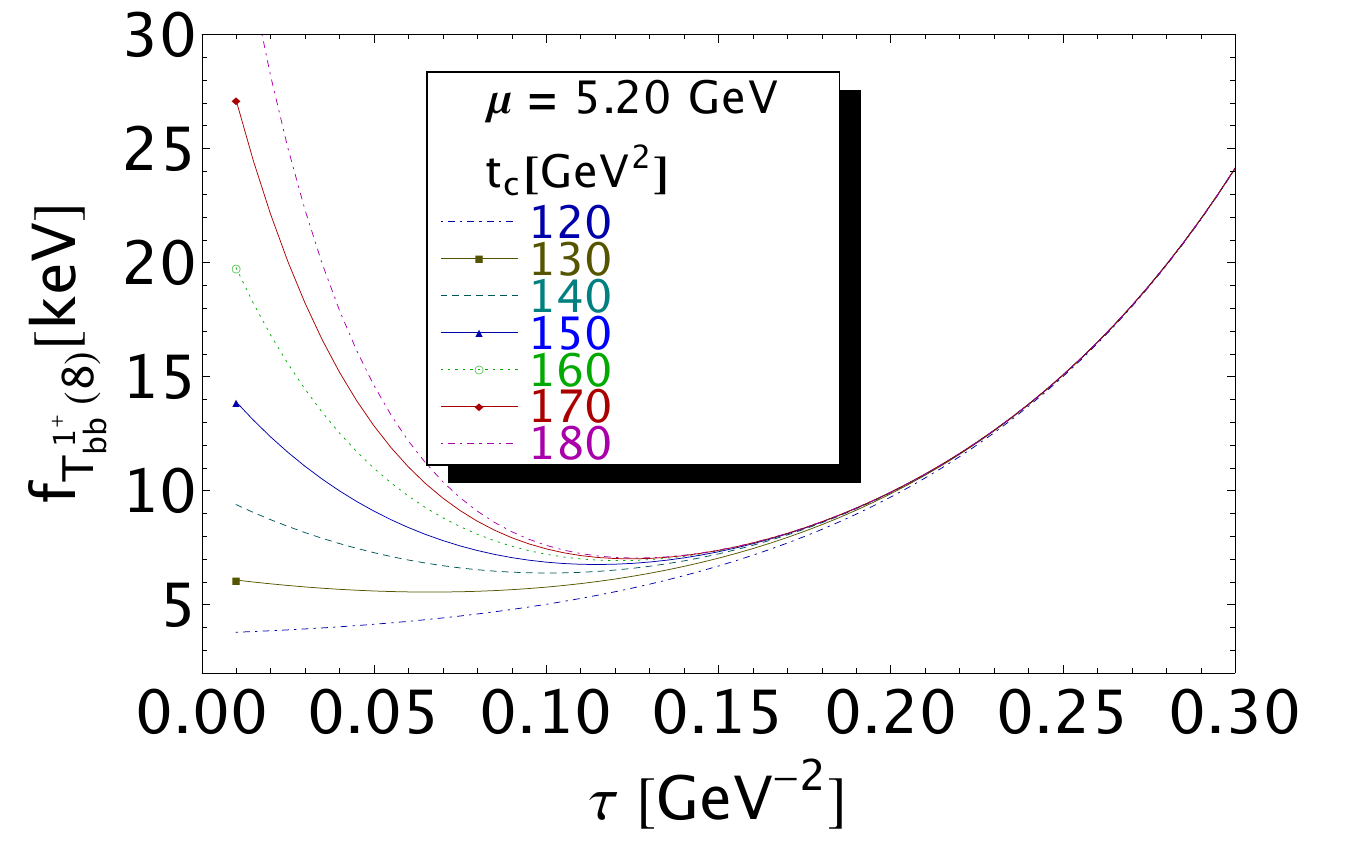}
\includegraphics[width=8cm]{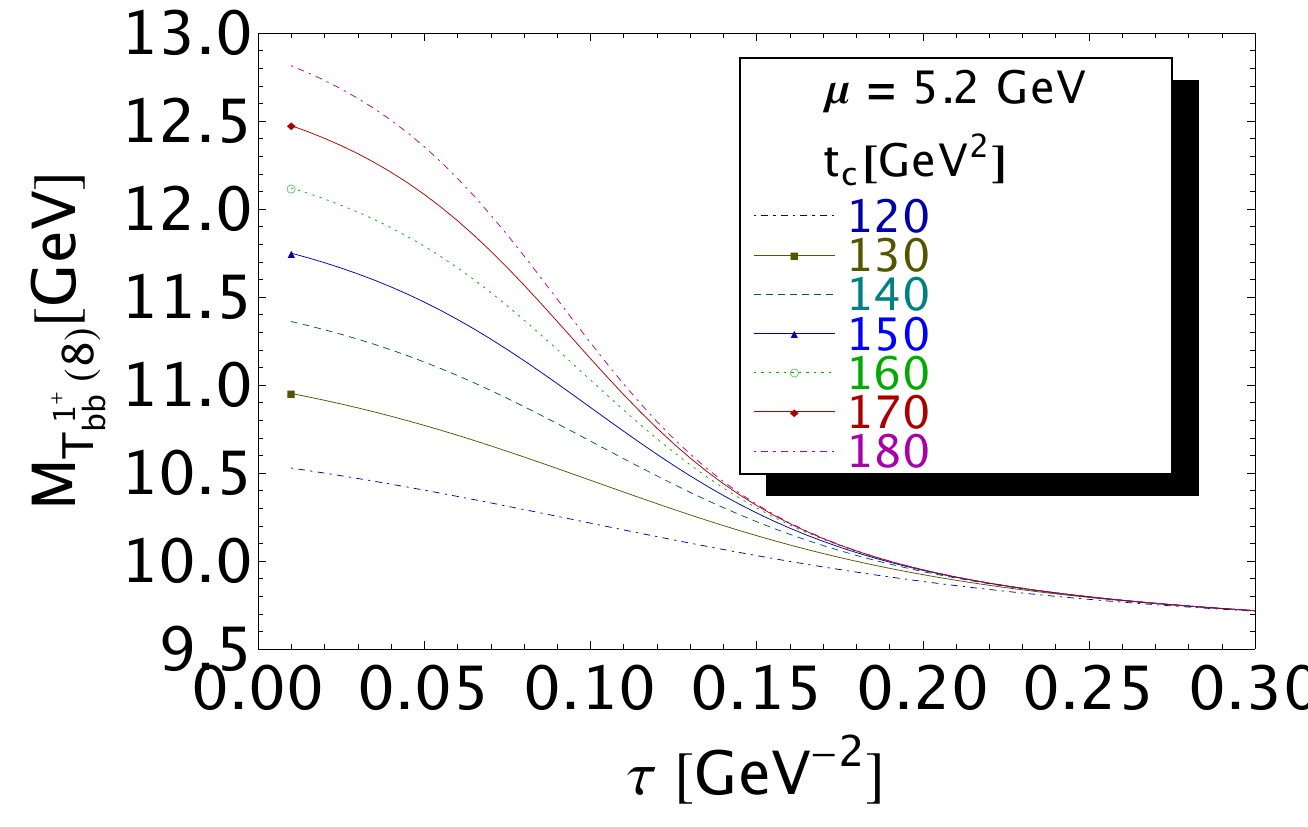}
\vspace*{-0.5cm}
\caption{\footnotesize  $f_{T^{0^+}_{bb}}$ and $M_{T^{0^+}_{bb}}$ as function of $\tau$  for \# values of $t_c$, for $\mu$=5.2 GeV and for the QCD inputs in Table\,\ref{tab:param}.} 
\label{fig:malt}
\end{center}
\vspace*{-0.5cm}
\end{figure} 
Here, we check the results of\,\cite{MALT-T} for the molecule state built from the octet $\bar Q\lambda_a q$ meson. We consider,
for instance, the interpolating current (Eq. 3 of\,\cite{MALT-T}):
\beq
{\cal O}_{T,1}^{8,\mu}=(\bar Q_j\gamma^\mu\frac{\lambda^{jk}_a}{2} s_k) (Q_m\,i\gamma_5\frac{\lambda^{mn}_a}{2} \bar q_n).
\label{eq:malt}
\eeq
Notice that a similar octet current has been used for testing the 4-quark nature of the $a_0$ light meson\,\cite{SNa0}
and of the $X$ state\,\cite{DRSR11a}.  Comparing our results with Ref.\,\cite{MALT-T}, we found that :

-- The input quark propagators used in their Eqs. 7 and 8 are correct except for the $\la G^3\ra$ condensate contribution 
which is incomplete.

-- The contribution coming from the trace for all 4 quark propagators in the spectral function is absent in Ref.\,\cite{MALT-T} as well as the contribution from the $\la\bar q'q'\ra$ condensate.

We show the analysis in Fig\,\ref{fig:malt} in the case of $T_{bb}$ and using the expression in\,\ref{app-c}\,:

-- The choice of the set $(\tau,t_c)=(0.08-0.11, 12)$ (GeV$^{-2}$, GeV$^2$) used by\,\cite{MALT-T} for $T^{1^+}_{bb}$ is just at the beginning of the stability region (see  Fig\,\ref{fig:malt}).  

-- Working with the spectral function of\,\cite{MALT-T} and the one in\,\ref{app-c}, the difference between the QCD expressions induces 
a small change of about 80 MeV when using the same QCD inputs. The contribution of the $\la G^3\ra$ condensate is negligible as well as of the $D=8,10$ condensates used in\,\cite{MALT-T}. However, the implicit use of factorization for the four-quark condensate increases the central value of the mass predictions by (815--829) MeV for $M_{T^{1^+}_{cc\bar q'\bar q}}$ and about (590-650) MeV for $M_{T^{1^+}_{bb\bar q'\bar q}}$.  

-- The difference between the values of the couplings can be understood by its  sum rule behaviour :  $f_T\sim  (1/M_T^4)e^{\tau M_T^2/2}$ which introduces a suppression factor of about 0.5 on our results. 

 From the analysis in Fig.\,\ref{fig:malt}, we deduce the results quoted in Table\,\ref{tab:malt}. 
\begin{table}[hbt]
\setlength{\tabcolsep}{0.17pc}
{\scriptsize{
\begin{tabular}{ll ll  ll  ll ll ll ll ll l ll}
\hline
\hline
                \bf Observables &\multicolumn{1}{c}{$\Delta t_c$}
					&\multicolumn{1}{c}{$\Delta \tau$}
					&\multicolumn{1}{c}{$\Delta \mu$}
					&\multicolumn{1}{c}{$\Delta \alpha_s$}
					&\multicolumn{1}{c}{$\Delta PT$}
					&\multicolumn{1}{c}{$\Delta m_s$}
					&\multicolumn{1}{c}{$\Delta m_Q$}
					&\multicolumn{1}{c}{$\Delta \overline{\psi}\psi$}
					&\multicolumn{1}{c}{$\Delta \kappa$}					
					&\multicolumn{1}{c}{$\Delta G^2$}
					&\multicolumn{1}{c}{$\Delta M^{2}_{0}$}
					&\multicolumn{1}{c}{$\Delta \overline{\psi}\psi^2$}
					&\multicolumn{1}{c}{$\Delta G^3$}
					&\multicolumn{1}{c}{$\Delta OPE$}
					&\multicolumn{1}{c}{$\Delta M_{G}$}
					&\multicolumn{1}{l}{This work}
					&\multicolumn{1}{l}{Ref.\,\cite{MALT-T}}
\\
					
\hline
{\bf Coupling} [keV] &&&&&&&&&&&\\
$f_{T^{1^+}_{cc}}$ &6.00&0.06&0.36&1.70&0.14&$\cdots$&0.98&0.79&$\cdots$&0.02&1.02&3.08&$\cdots$&3.63&0.74&76(8)&$\cdots$\\
$f_{T^{1^+}_{ccsu}}$ &5.33&0.07&0.29&1.55&0.03&0.03&0.84&0.84&1.72&0.02&1.08&2.54&$\cdots$&4.60&1.18&64(8)&$\cdots$ \\
$f_{T^{1^+}_{bb}}$ &0.73&0.04&0.03&0.19&0.03&$\cdots$&0.11&0.08&$\cdots$&0.00&0.07&0.25&$\cdots$&0.24&0.58&7(1)&18(1) \\
$f_{T^{1^+}_{bbsu}}$ &0.66&0.04&0.03&0.51&0.01&0.00&0.10&0.08&0.14&0.00&0.08&0.21&$\cdots$&0.78&0.53&6(1)&20(6)\\

{\bf Mass} [MeV] &&&&&&&&&&&&&&\\
$M_{T^{1^+}_{cc}}$&6.79&40.9&2.09&5.93&0.07&$\cdots$&4.25&5.95&$\cdots$&0.11&4.40&3.62&$\cdots$&42.3&$\cdots$&3905(60)&4720(130)\\
$M_{T^{1^+}_{ccsu}}$&7.68&39.8&5.18&13.9&0.03&2.66&4.16&13.8&18.8&0.14&6.41&13.6&$\cdots$&95.1&$\cdots$&3931(108)&4760(140)\\
$M_{T^{1^+}_{bb}}$&8.60&115&1.80&11.8&0.02&$\cdots$&6.65&7.58&$\cdots$&0.05&3.78&17.3&$\cdots$&67.7&$\cdots$&10690(136)&11280(150)\\
$M_{T^{1^+}_{bbsu}}$&2.50&109&1.93&12.3&0.00&1.88&6.75&7.13&15.7&0.05&6.05&16.8&$\cdots$&53.0&$\cdots$&10706(125)&11360(160)\\

\\
\hline
\hline
\end{tabular}
}}
 \caption{Sources of errors and estimates of the masses and couplings of the  $T^{1^+}_{QQ}$ and $T^{1^+}_{QQsu}$ $(Q\equiv c,b)$ states for the $\bar 8_c8_c$ currents. We take $\ve \Delta \mu\ve=0.05$ GeV and $\ve \Delta \tau\ve =0.01$ GeV$^{-2}$.}

\label{tab:malt}
\end{table}
 
\section{Comparison  of different LSR results\label{sec:comparison}} 
\begin{figure}[hbt]
\begin{center}
\includegraphics[width=16.5cm]{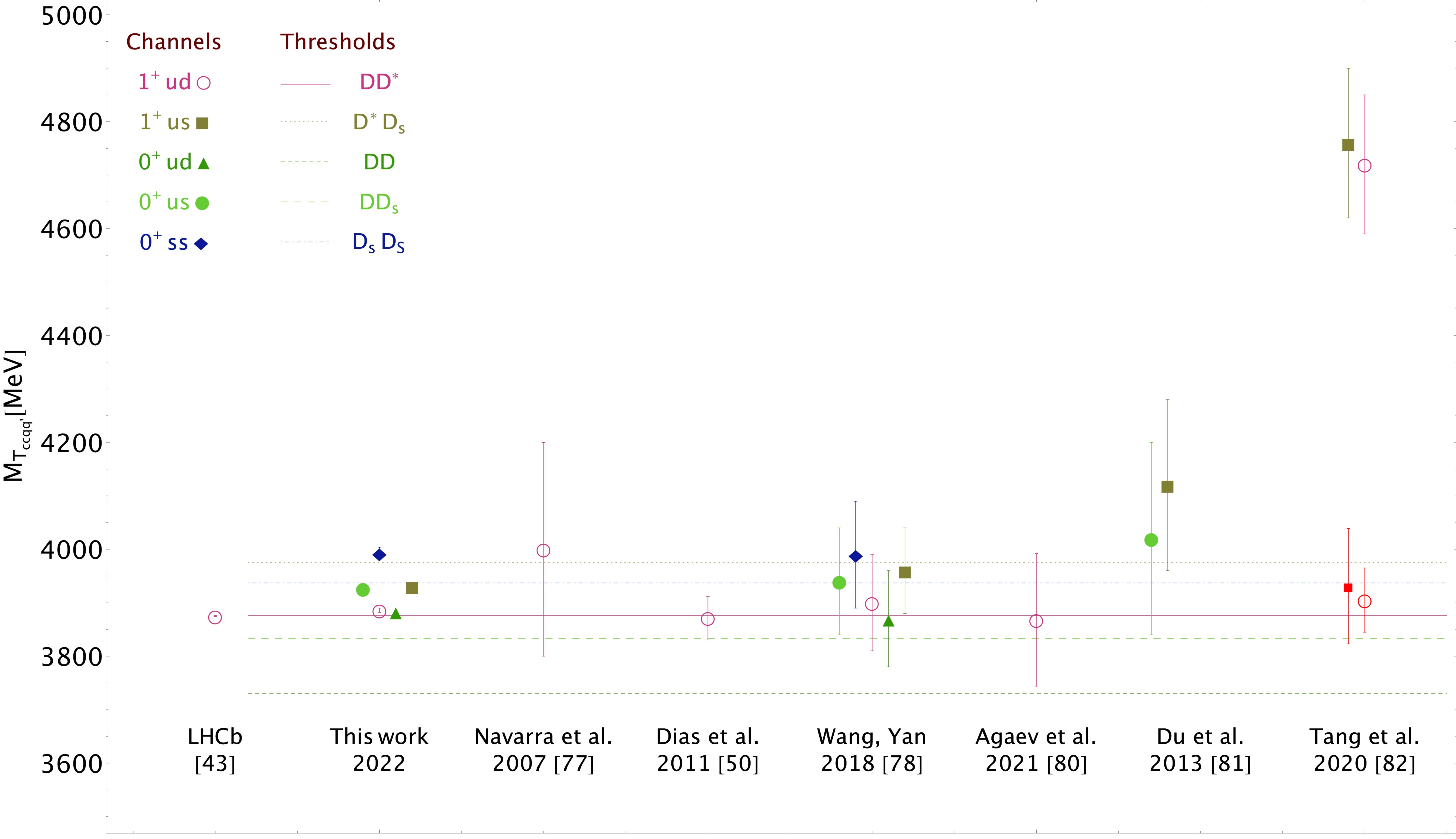}
\vspace*{-0.5cm}
\caption{\footnotesize   Different determinations of $T^{1^+,0^+}_{ccqq'}$ from LSR. The horizontal lines are physical thresholds. Comments and corrections of some results are given in the text. The predictions of Du et al. and ours for the $\eta_1$ current quoted in Table\,\ref{tab:zhuc} are too high and are not shown here. 
The red rectangle and open circle below the ones of Tang et al. are our predictions for the same $\bar 8_c8_c$ current. } 
\label{fig:tccf}
\end{center}
\vspace*{-0.5cm}
\end{figure} 
\subsection*{\b Results}
We compare the different published LSR results  in Figs\,\ref{fig:tccf} and \,\ref{fig:tbbf} for the $T_{QQ}^{1^+}$ and   $T_{QQ}^{0^+}$ states. The quoted results come from the original works. However, from our previous checks, we have 
realized that results from\,\cite{WANG-Ta,WANG-Tb,AGAEV-T,ZHU-T} are not exactly correct which explain the divergence of some results.
\subsection*{\b QCD expressions of the spectral function}
 For the QCD expressions of the spectral function, we notice that the propagator used in\,\cite{WANG-Ta,WANG-Tb,AGAEV-T,ZHU-T}
 does not generate the diagrams in Fig.\,\ref{fig:mixed} which induces a different result for the mixed $\la \bar qG q\ra$ condensate. A discrepancy  is also noticed for the contribution  of the gluon condensate $\la G^2\ra$ from the diagram in Fig.\,\ref{fig:gluon} while the contribution of the $\la G^3\ra$ condensate is often missing. 
 Inconsistently some authors include the contribution of some classes of high-dimension condensates. 
 
  We could not (unfortunately) check the QCD expression of the spectral function used in \,\cite{AGAEV-T} which is not given. 

 For Ref.\,\cite{ZHU-T}, where many configurations are considered,  we only quote the (uncorrected) lowest masses given in their Tables from the currents $\eta_5$ for the $1^+$ and $\eta_3$ for $0^+$ which are similar to our currents in Table\,\ref{tab:current}. The (corrected ) predictions from the $\eta_1$ current  are given in Tables\,\ref{tab:zhuc} and\,\ref{tab:zhub}  but not  in Figs\,\ref{fig:tccf} and \,\ref{fig:tbbf} because the mesons masses are too high.  
 
\subsection*{\b Concluding remarks}
 Despite the previous caveats on the QCD expressions and on the choice of sum rule windows, one may conclude that  (within the errors) \,:

There are (almost) good agreements among different determinations and with the data for the $1^+~ T_{cc\bar q\bar q'}$ states from the $\bar 3_c3_c$ interpolating currents. Most of the approaches predict the $T_{bb}$ states to be below the hadronic thresholds. 




  The $0^+~ T_{bb}$ and $0^+~ T_{bb\bar ss}$ states\cite{WANG-Tb} predicted at relatively high masses using $\bar 3_c3_c$ currents  were quite surprising compared to the charm analogue. We have corrected these predictions in Section\,\ref{sec:wang}. 

  The predictions of \cite{AGAEV-T} are usually lower than the other ones due to the fact that the authors extract the mass at lower values of $t_c$ outside the $\tau$-stability of the coupling. The corrected values are given in our predictions in Table\,\ref{tab:res}.  

 The high-value of the masses from the $\eta_3,~\eta_5$ currents quoted by\,\cite{ZHU-T} shown in Fig.\ref{fig:tccf} are due to the wrong sign of the four-quark condensate contribution. The corrected results are given in Tables\,\ref{tab:zhuc} and \,\ref{tab:zhub}. Ours do not also support the argument of\,\cite{ZHU-T} for the non-existence of the $1^+,0^+~T_{cc}$ and $0^+~T_{cc\bar s\bar u} $ states. 

\begin{figure}[hbt]
\begin{center}
\includegraphics[width=16.2cm]{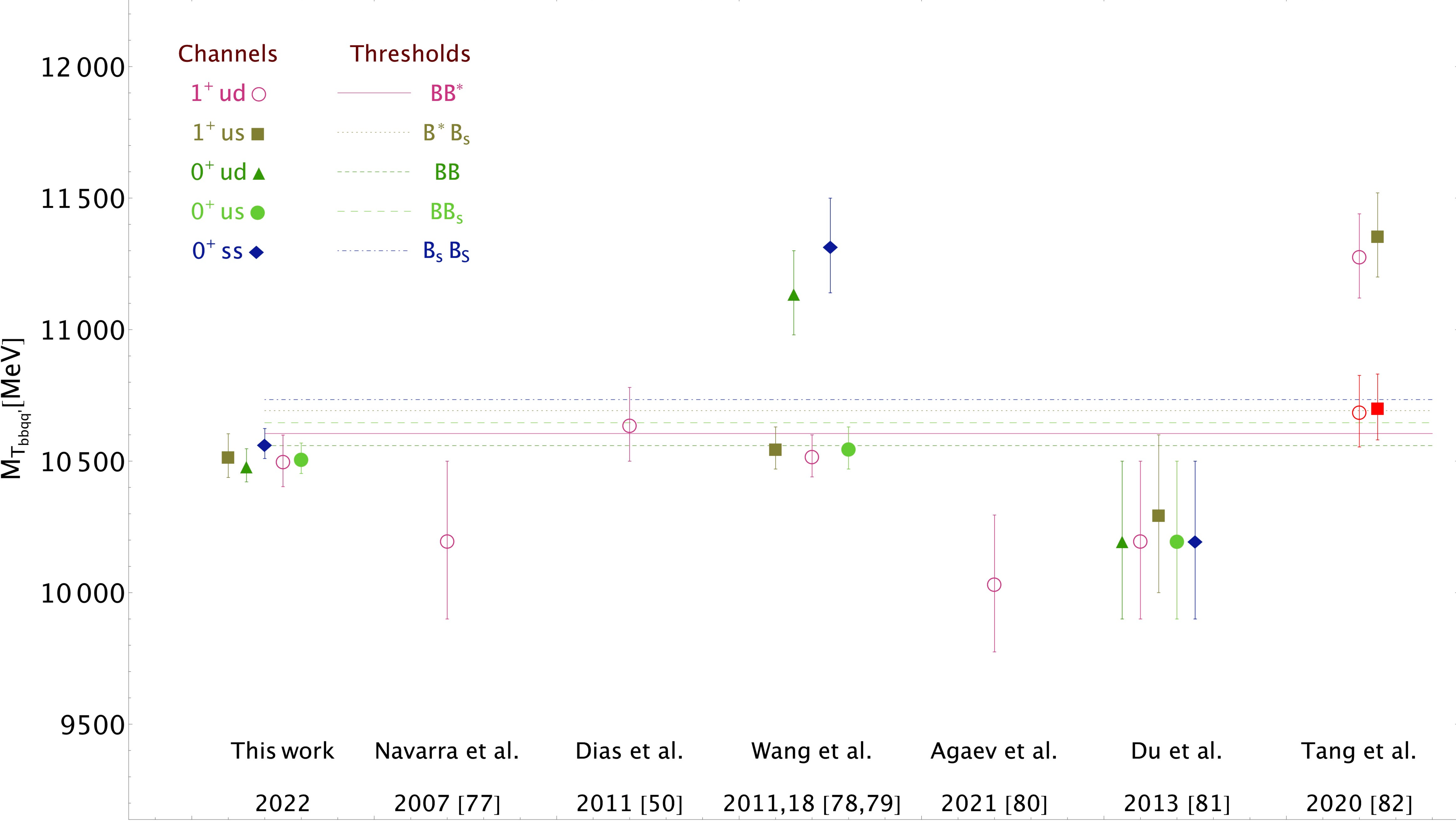}
\vspace*{-0.25cm}
\caption{\footnotesize   Same as Fig.\,\ref{fig:tccf} but for $T^{1^+,0^+}_{bbqq'}$. } 
\label{fig:tbbf}
\end{center}
\vspace*{-0.5cm}
\end{figure}

 The high central values of the masses obtained by\,\cite{MALT-T} shown in Figs.\ref{fig:tccf} and \ref{fig:tbbf} from the $\bar 8_c8_c$ current are essentially due to the implicit use of four-quark condensate factorization. Results not using this assumption are given in Table\,\ref{tab:malt} which go in lines with the other ones from $\bar 3_c3_c$ current. A such conclusion is somewhat expected if one looks at the result for $X$ in Ref.\,\cite{DRSR11a} where $\bar 8_c8_c$ and $\bar 3_c3_c$ current has been also used. 

\begin{figure}[hbt]
\begin{center}
\includegraphics[width=16cm]{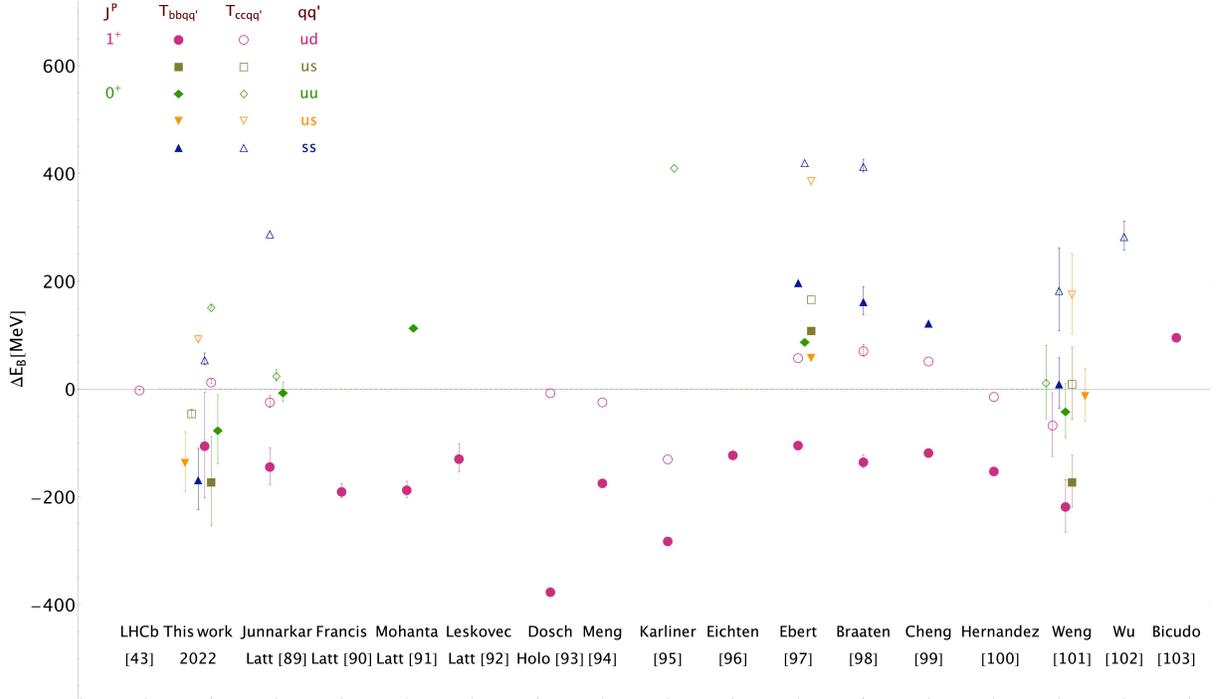}
\vspace*{-0.5cm}
\caption{\footnotesize   Confronting  the LSR $\oplus$ DSR results of $T^{1^+,0^+}_{QQqq'}$ masses  with some  estimates from lattice and quark models.} 
\label{fig:tcc-rev}
\end{center}
\vspace*{-0.5cm}
\end{figure} 
\section{LSR $\oplus$ DRSR  confronted to some other approaches}
\subsection*{\b Comparison of different results}

In Fig.\,\ref{fig:tcc-rev}, we confront our results from LSR $\oplus$ DRSR with the ones from different approaches in the literature (lattice calculations\,\cite{JUN,MALT,MOHAN,LESK},  light front holographic\,\cite{DOSCH2}, quark and potential models $\oplus$heavy quark symmetry\,\cite{MENG,ROSNER,QUIGG,BRAATEN,CHENG,RICHARD2,ZHUMODEL,WU,BICUDO}).
We refrain to comment on the technical details of the estimates from different approaches being non-experts in these fields. 
 However, for a more meaningful comparison, we regret that most of the predictions from quark and potential models $\oplus$heavy quark symmetry are quoted without any estimated errors. 

One can notice from\, Fig.\ref{fig:tcc-rev}, that there is (almost) a consensus for the predictions of the axial-vector $1^+$masses from different approaches: the $T_{cc}$ state is expected to be around  the physical threshold while the $T_{bb}$ one is below the threshold and then stable against strong interactions.  However, the recent LHCb data for the $1^+~T_{cc}$ candidate does not favour the models of\,\cite{LUCHA,BRAATEN,CHENG} which  predict a too high $1^+~T_{cc}$ mass. 

For the $T_{bb}$ $0^+$ scalar state, the situation is quite similar. This state is expected to be below the hadronic threshold by different approaches  except the lattice result of \cite{MOHAN} and the quark model of\,\cite{LUCHA}. 


\subsection*{\b Some comments on our results}
 Our predictions for different $1^+$ and $0^+$ states including SU3 breakings states are clustered  in the range $-250$ to $+150$ MeV of the hadronic thresholds.

 From our approach,  the mass shifts due to SU3 breakings are postive but tiny. Therefore, our results for the masses of different states are grouped around the physical thresholds. This is not often the case of some other approaches. In particular, a lattice calculation\,\cite{JUN} and some quark models\,\cite{LUCHA,BRAATEN,CHENG,ZHUMODEL,WU,BICUDO} expect a mass of the $T_{cc\bar s \bar s}$ and $T_{bb\bar s \bar s}$ $0^+$ states well above the physical threshold while in our case the  $T_{bb\bar s \bar s}$ state lies below the physical threshold and the one of the $T_{cc\bar s \bar s}$ $0^+$ state is slightly above (see Table 5). This peculiar feature of SU3 breakings for exotic states 
needs to be checked experimentally.
\vfill\eject
\appendix
\section{ :\, Spectral functions corresponding to the currents in Table\,\ref{tab:current}\,\label{app-a}}
 In this appendix\,\footnote{For the $X$ and $Z$ states, we have used the expressions of the spectral functions given respectively in\,\cite{Zc} and \cite{DRSR07,DRSR11,DRSR11a,MOLE16} which will not be reported here.}, we shall give the compact integrated QCD expressions of the spectral functions of the $T_{QQ\bar q\bar q'}$ states associated to the interpolating currents given in Table\,\ref{tab:current}. Compared to the existing non-integrated ones given in the current literature, our expressions are more compact, less horrible and easier to handle in the numerical analysis. Checks of some existing expressions in the literature have been discussed in Section\,\ref{sec:zhu1} and given in\,\ref{app-b}. 

  We shall define
 \beq
\mathcal{L}_v=\text{Log}\left[\frac{1+v}{1-v}\right], ~~
{\rm and}~~
\rho(t)\equiv \frac{1}{\pi} \mbox{Im}~\Pi(t)~~({\rm see~Eq.}\,\ref{eq:lsr})~,
 \eeq
 where $x$ and $v=\sqrt{1-x}$ have been defined in Eq.\,\ref{eq:def}.
 We use the short-handed notations:
 \beq
 \la G^2\ra\equiv  \la g^2 G_a^{\mu\nu}G^a_{\mu\nu}\ra,~~~~~~\la \bar qG q\ra\equiv  g\la \bar qG^{\mu\nu}(\lambda_a/2) q\ra,~~~~~~ \la G^3\ra \equiv  \la g^3 f_{abc}G^{a,\mu\nu}G^b_{\nu\rho}G^{c,\rho}_\mu\ra, 
 \eeq
\section*{\b $T_{QQ\bar{u}\bar{q}}$ or  $T_{QQ\bar{u}\bar{s}}$: $J^{P} =1^+$ } 
Its spectral function is associated to the current $ {\cal O}_{T^{1^+}_{QQuq}}$: : $q\equiv d,s$, $Q\equiv c,b$ (see Table\,\ref{tab:current}) We keep the $m_q$-linear mass term corrections. 
\vspace*{-0.25cm}
\begin{eqnarray*}
    \rho^{pert}(t) &=& \frac{m_Q^8}{
    5\cdot 3^2\cdot 2^{12} \,\pi^6} \bigg[ 
    v \Big(840x + 7340 + 52528/x + 5796/x^2 
    - 62/x^3 + 5/x^4 \Big) + \\ 
    &&
    + 120{\cal L}_v \Big( 14x^2 + 120x + 207 - 
    18(9 + 4/x) \log(x) - 320/x - 15/x^2 \Big) + 
    - 4320 {\cal L_+} \Big(9 + 4/x \Big)
    \bigg],
    \\ 
    %
    \rho^{\langle \bar{q}q \rangle}(t) &=& 
    \frac{m_q m_Q^4}{3 \cdot 2^8 \,\pi^4} 
    \Big( \langle \bar{q}q \rangle 
    - 2 \langle \bar{u}u \rangle \Big) 
    \bigg[ v \Big( 12x + 50 - 2/x + 3/x^2 \Big) 
    + 24 {\cal L}_v \Big( x^2 + 4x - 3 \Big) \bigg], \\
    %
    %
    \rho^{\langle G^2 \rangle}(t) &=& 
    \frac{m_Q^4 \langle G^2 \rangle}{3^3 \cdot 2^{11} \pi^6}
    \bigg[ v \Big( 102x + 557 + 538/x + 18/x^2 \Big) 
    \\ &&
    + 6 {\cal L}_v \Big( 34x^2 + 180x - 123 - 63 \log(x) 
    - 44/x \Big) - 756 {\cal L}_+ \bigg],
    \\ 
    %
    \rho^{\langle \bar{q}Gq \rangle}(t) &=&
    \frac{m_q m_Q^2}{3^2 \cdot 2^6 \,\pi^4}
    \Big( \langle \bar{q}Gq \rangle 
    + 6 \langle \bar{u}Gu \rangle \Big) \:
    v \Big( 2 + 1/x \Big),
    \\ 
    %
    \rho^{\langle \bar{q}q \rangle^2}(t) &=& 
    \frac{m_Q^2 \langle \bar{u}u \rangle
    \langle \bar{s}s \rangle}{18 \pi^2} 
    \:v \Big( 2 + 1/x \Big),
    \\ 
    %
    \rho^{\langle G^3 \rangle}(t) &=& 
    \frac{m_Q^2 \langle G^3 \rangle}
    {5\cdot 3^5 \cdot 2^{12} \pi^6} \bigg[ 
    v \Big(4020x - 19490 - 21070/x - 495/x^2 + 
    36m_c^2\tau (5/x^3 + 58/x^2) \Big) + \\ 
    &&
    + 24{\cal L}_v \Big( 335x^2 - 1680x + 945 + 
    60(6 + 1/x) \log(x) + 440/x - 18m_c^2\tau 
    (1/x + 3/x^2) \Big) + \\
    &&
    + 2880 {\cal L_+} \Big(6 + 1/x \Big) \bigg].
\end{eqnarray*}
\section*{\b $T_{QQ\bar{s}\bar{s}}$ : $J^{PC}=1^+$}
We note that its spectral function is identically zero. 
\section*{\b $T_{QQ\bar{q}\bar{q}}$  : $J^{P}=0^+$}
Its spectral function is associated to the current $ {\cal O}_{T^{0^+}_{QQ\bar u\bar d}}$ or  $ {\cal O}_{T^{0^+}_{QQ\bar s\bar s}}$: $q\equiv u,d,s$, $Q\equiv c,b$ (see Table\,\ref{tab:current}). 
\vspace*{-0.25cm}
\begin{eqnarray*}
    \rho^{pert}(t) &=& \frac{m_Q^8}{
    5\cdot 3\cdot 2^8 \,\pi^6} \bigg[ 
    v \Big(1080 + 5400/x + 306/x^2 
    - 28/x^3 + 1/x^4 \Big) + \\ 
    &&
    + 120{\cal L}_v \Big( 18x + 15 - 6(3 + 1/x) \log(x) 
    - 32/x \Big) -1440 {\cal L_+} \Big(3 + 1/x \Big)
    \bigg],
    \\ 
    %
    \rho^{\langle \bar{q}q \rangle}(t) &=&
    - \frac{ 3 m_q m_Q^4 \langle \bar{q}q \rangle}{4\pi^4} 
    \bigg[ v \Big( 2 + 1/x \Big) + 4{\cal L}_v \Big( x-1 \Big)
    \bigg],
    \\ 
    %
    \rho^{\langle G^2 \rangle}(t) &=& 
    \frac{m_Q^4 \langle G^2 \rangle}{3 \cdot 2^{8} \pi^6}
    \bigg[ v \Big( 6 + 17/x + 1/x^2 \Big) + 
    12 {\cal L}_v \Big( x - \log(x) - 1/x \Big) 
    - 24 {\cal L}_+ \bigg], \\
      \rho^{\langle \bar{q}Gq \rangle}(t) &=& 
    \frac{m_q m_Q^2 \langle \bar{q}G q \rangle}
    {3\cdot 2^{3}\,\pi^4}
    \bigg[ v \Big( 8 + 1/x \Big) + 
    3 {\cal L}_v \Big( 2x - 1 \Big) \bigg],
    \\ 
  \end{eqnarray*}
    %
    \begin{eqnarray*}
    %
    \rho^{\langle \bar{q}q \rangle^2}(t) &=& 
    \frac{m_Q^2 \langle \bar{q}q \rangle^2}{3 \pi^2} 
    \:v \Big( 2 + 1/x \Big),
    \\ 
    %
    \rho^{\langle G^3 \rangle}(t) &=& 
    \frac{m_Q^2 \langle G^3 \rangle}
    {5\cdot 3^2\cdot 2^{8} \pi^6} \bigg[ 
    v \Big(310 + 415/x - 6/x^2 + 2m_c^2\tau
    (1/x^3 + 8/x^2) \Big) + \\ 
    &&
    + 4{\cal L}_v \Big( 155x - 53 - 10(6 + 1/x) \log(x) 
    - 55/x - 3m_c^2\tau/x^2 \Big) - 80 {\cal L_+} 
    \Big(6 + 1/x \Big)\bigg].
\end{eqnarray*}
\section*{\b $T_{QQ\bar{u}\bar{s}}$ : $J^{P}=0^+$}
Its spectral function is associated to the current $ {\cal O}_{T^{0^+}_{QQ\bar u\bar s}}$ (see Table\,\ref{tab:current}). 
\vspace*{-0.25cm}
\begin{eqnarray*}
    \rho^{pert}(t)&=& \frac{m_Q^8}{
    5\cdot 3\cdot 2^9 \,\pi^6} \bigg[ 
    v \Big(1080 + 5400/x + 306/x^2 
    - 28/x^3 + 1/x^4 \Big) + \\ 
    &&
    + 120{\cal L}_v \Big( 18x + 15 - 6(3 + 1/x) \log(x) 
    - 32/x \Big) -1440 {\cal L_+} \Big(3 + 1/x \Big)
    \bigg],
    \\ 
    %
    \rho^{\langle \bar{q}q \rangle}(t) &=& - 
    \frac{m_s m_Q^4}{3 \cdot 2^4\pi^4} 
    \bigg[ v \Big(  \langle \bar{q}q \rangle 
    (24 + 2/x + 1/x^2) -  \langle \bar{s}s \rangle 
    (6 - 7/x + 1/x^2) \Big) + 
    \\ &&
    - 12{\cal L}_v \Big( \langle \bar{q}q \rangle 
    (3 - 4x) + \langle \bar{s}s \rangle \,x \Big) \bigg], \\
    %
    %
    \rho^{\langle G^2 \rangle}(t) &=& 
    \frac{m_Q^4 \langle G^2 \rangle}{3 \cdot 2^{9} \pi^6}
    \bigg[ v \Big( 6 + 17/x + 1/x^2 \Big) + 
    12 {\cal L}_v \Big( x - \log(x) - 1/x \Big) 
    - 24 {\cal L}_+ \bigg],
    \\ 
    %
    \rho^{\langle \bar{q}Gq \rangle}(t) &=& 
    \frac{m_s m_Q^2}{3\cdot 2^{6}\,\pi^4}
    \bigg[ v \Big( 3\langle \bar{q}G q \rangle 
    - \langle \bar{s}G s \rangle \Big) (8 + 1/x) +
    6 {\cal L}_v \Big( \langle \bar{q}G q \rangle 
    (4x - 1) - 2\langle \bar{s}G s \rangle x \Big) 
    \bigg], 
    \\ 
    %
    \rho^{\langle \bar{q}q \rangle^2}(t) &=& 
    \frac{m_Q^2 \langle \bar{q}q \rangle^2}{6 \pi^2} 
    \:v \Big( 2 + 1/x \Big),
    \\ 
    %
    \rho^{\langle G^3 \rangle}(t) &=& 
    \frac{m_Q^2 \langle G^3 \rangle}
    {5\cdot 3^2\cdot 2^{9} \pi^6} \bigg[ 
    v \Big(310 + 415/x - 6/x^2 + 2m_c^2\tau
    (1/x^3 + 8/x^2) \Big) + \\ 
    &&
    + 4{\cal L}_v \Big( 155x - 53 - 10(6 + 1/x) \log(x) 
    - 55/x - 3m_c^2\tau/x^2 \Big) - 80 {\cal L_+} 
    \Big(6 + 1/x \Big)\bigg],
\end{eqnarray*}
\section{Spectral function corresponding to the $\eta_1$ currents used in\,\cite{ZHU-T}\label{app-b}}

\section*{ \b $T_{QQ\bar{q}\bar{q}}$ : $J^P=1^+$ : $q\equiv u,d,s$}
\noindent
For this state, we use the tetraquark current given in Eq.\,\ref{eq:eta1}. 
\begin{eqnarray*}
    \rho^{pert} &=& \frac{m_Q^8}{
    5\cdot 3^2 \cdot 2^{10} \,\pi^6} \bigg[ 
    v \Big(840x - 7060 - 42302/x - 8124/x^2 
    - 302/x^3 + 5/x^4 \Big) + \\ 
    &&
    + 120{\cal L}_v \Big( 14x^2 - 120x - 177 
    + 18(7 + 4/x) \log(x) + 256/x + 33/x^2 \Big) 
    + 4320 {\cal L_+} \Big( 7 + 4/x \Big)
    \bigg] \, ,
    \\
    \rho^{\langle \bar{q}q \rangle}(t) &=&
    \frac{m_q m_Q^4 \langle \bar{q}q \rangle}
    {2^5 \pi^4} \bigg[ v \Big( 12x - 46 - 50/x + 3/x^2 \Big)
    + 24{\cal L}_v \Big( x^2 - 4x + 5 \Big) \bigg]\, ,
    \\
    \rho^{\langle G^2 \rangle}(t) &=& 
    \frac{m_Q^4 \langle G^2 \rangle}{3^3 \cdot 2^{10} \pi^6}
    \bigg[ v \Big( 78x - 815 - 1192/x - 69/x^2 \Big) + \\
    &&
    + 6 {\cal L}_v \Big( 26x^2 - 276x + 111 + 135\log(x) + 
    128/x \Big) + 1620{\cal L_+} \bigg]\, ,
    \\
    %
    \rho^{\langle \bar{q}Gq \rangle}(t) &=&
    \frac{19 m_q m_Q^2 \langle \bar{q}Gq \rangle}
    {3^2 \cdot 2^4 \pi^4} \:v \Big( 4 - 1/x \Big)\, ,
    \\
    \rho^{\langle \bar{q}q \rangle^2}(t) &=& 
    \frac{2m_Q^2 \langle \bar{q}q \rangle^2}{9 \pi^2} 
    \:v \Big( 4 - 1/x \Big)
    \\
    \rho^{\langle G^3 \rangle}(t) &=& 
    -\frac{m_Q^2 \langle G^3 \rangle}
    {5\cdot 3^5 \cdot 2^{10} \pi^6} \bigg[ 
    v \Big( 2820x - 12850 - 15350/x - 225/x^2 + 
    108m_Q^2\tau\,(1/x^3 + 26/x^2) \Big) + \\ 
    &&
    + 24{\cal L}_v \Big( 235x^2 - 1110x + 810 + 
    30(6 + 1/x) \log(x) + 265/x - 54m_c^2\tau
    \,(1/x^2 + 1/x) \Big) + \\
    &&
    1440{\cal L_+} \Big(6 + 1/x \Big)\bigg]\, .
\end{eqnarray*}
\section*{\b   $T_{QQ\bar{u}\bar{s}}$ : $J^P=1^+$}
\vspace*{-0.5cm}
\begin{eqnarray*}
    \rho^{pert}(t) &=& \frac{m_Q^8}{
    5\cdot 3^2 \cdot 2^{11} \,\pi^6} \bigg[ 
    v \Big(840x - 7060 - 42302/x - 8124/x^2 
    - 302/x^3 + 5/x^4 \Big) + \\ 
    &&
    + 120{\cal L}_v \Big( 14x^2 - 120x - 177 
    + 18(7 + 4/x) \log(x) + 256/x + 33/x^2 \Big) 
    + 4320 {\cal L_+} \Big( 7 + 4/x \Big)
    \bigg]\, ,
    \\
    \rho^{\langle \bar{q}q \rangle}(t) &=&
    \frac{m_s m_Q^4}{3 \cdot 2^7 \pi^4} \Big(
    2\langle \bar{q}q \rangle + \langle \bar{s}s \rangle \Big)
    \bigg[ v \Big( 12x - 46 - 50/x + 3/x^2 \Big)
    + 24{\cal L}_v \Big( x^2 - 4x + 5 \Big) \bigg] \, ,
    \\
    \rho^{\langle G^2 \rangle}(t) &=& 
    \frac{m_Q^4 \langle G^2 \rangle}{3^3 \cdot 2^{11} \pi^6}
    \bigg[ v \Big( 78x - 815 - 1192/x - 69/x^2 \Big) + \\
    &&
    + 6 {\cal L}_v \Big( 26x^2 - 276x + 111 + 135\log(x) + 
    128/x \Big) + 1620{\cal L_+} \bigg] \, ,
    \\
    \rho^{\langle \bar{q}Gq \rangle}(t) &=&
    \frac{m_s m_Q^2}{3^2 \cdot 2^6 \pi^4} \Big( 
    12 \langle \bar{q}Gq \rangle + 
    7\langle \bar{s}Gs \rangle \Big) 
    \:v \Big( 4 - 1/x \Big) \, ,
    \\
    \rho^{\langle \bar{q}q \rangle^2}(t) &=& 
    \frac{m_Q^2 \langle \bar{q}q \rangle 
    \langle \bar{s}s \rangle}{9 \pi^2} 
    \:v \Big( 4 - 1/x \Big) \, ,
    \\
    \rho^{\langle G^3 \rangle}(t) &=& 
    -\frac{m_Q^2 \langle G^3 \rangle}
    {5\cdot 3^5 \cdot 2^{11} \pi^6} \bigg[ 
    v \Big( 2820x - 12850 - 15350/x - 225/x^2 + 
    108m_Q^2\tau\,(1/x^3 + 26/x^2) \Big) + \\ 
    &&
    + 24{\cal L}_v \Big( 235x^2 - 1110x + 810 + 
    30(6 + 1/x) \log(x) + 265/x - 54m_c^2\tau
    \,(1/x^2 + 1/x) \Big) + \\
    &&
    1440{\cal L_+} \Big(6 + 1/x \Big)\bigg]\, .
\end{eqnarray*}
\section*{\b $T_{QQ\bar{q}\bar{q}}$ or $T_{QQ\bar{s}\bar{s}}$ : $J^P=0^+$ }
\noindent
For this state, we use the tetraquark current given in Eq.\,\ref{eq:eta10}. We obtain: 
\begin{eqnarray*}
    \rho^{pert}(t) &=& -\frac{m_Q^8}{
    5\cdot 3\cdot 2^9 \,\pi^6} \bigg[ 
    v \Big(720 + 6420/x + 1434/x^2 
    + 58/x^3 - 1/x^4 \Big) + \\ 
    &&
    + 120{\cal L}_v \Big( 12x + 33 - 6(3 + 2/x) \log(x) 
    - 40/x -6/x^2\Big) -1440 {\cal L_+} \Big(3 + 2/x \Big)
    \bigg]\, ,
    \\
    \rho^{\langle \bar{q}q \rangle}(t) &=&
    - \frac{m_q m_Q^4 \langle \bar{q}q \rangle}
    {8 \pi^4} \bigg[ v \Big(12 + 16/x - 1/x^2 \Big)
    + 12{\cal L}_v \Big( 2x - 3 \Big) \bigg] \, ,
    \\
    \rho^{\langle G^2 \rangle}(t) &=& 
    - \frac{m_Q^4 \langle G^2 \rangle}{3 \cdot 2^{9} \pi^6}
    \bigg[ v \Big( 30 + 31/x + 5/x^2 \Big) + 6 {\cal L}_v
    \Big( 10x - 6 - (5 - 1/x)\log(x) - 4/x \Big) 
    - 12 {\cal L_+}\Big( 5 - 1/x \Big) \bigg]\, ,
    \\
    \rho^{\langle \bar{q}Gq \rangle}(t) &=&
    \frac{19 m_q m_Q^2 \langle \bar{q}Gq \rangle}
    {3 \cdot 2^5 \pi^4} v \Big( 4 - 1/x \Big)\, ,
    %
    \end{eqnarray*}
    \begin{eqnarray*}
    \rho^{\langle \bar{q}q \rangle^2}(t) &=& 
    \frac{m_Q^2 \langle \bar{q}q \rangle^2}{3 \pi^2} 
    \:v \Big( 4 - 1/x \Big)\, ,
    \\
    \rho^{\langle G^3 \rangle}(t) &=& 
    -\frac{m_Q^2 \langle G^3 \rangle}
    {5\cdot 3^3\cdot 2^{8} \pi^6} \bigg[ 
    v \Big(430 + 495/x - 6/x^2 + 3m_Q^2\tau\,
    (1/x^3 + 26/x^2) \Big) + \\ 
    &&
    + {\cal L}_v \Big( 860x - 232 - 5(87 + 10/x) \log(x) 
    - 285/x - 36m_c^2\tau\, (1/x^2 + 1/x) \Big) - 
    10 {\cal L_+} \Big(87 + 10/x \Big)\bigg]\, .
\end{eqnarray*}
\section*{\b $T_{QQ\bar{q}\bar{s}}$ : $J^P=0^+$ }
For this state, we use the tetraquark current given in Eq.\,\ref{eq:eta10}. We obtain: 
\begin{eqnarray*}
     \rho^{pert}(t) &=& -\frac{m_Q^8}{
    5\cdot 3\cdot 2^{10} \,\pi^6} \bigg[ 
    v \Big(720 + 6420/x + 1434/x^2 
    + 58/x^3 - 1/x^4 \Big) + \\ 
    &&
    + 120{\cal L}_v \Big( 12x + 33 - 6(3 + 2/x) \log(x) 
    - 40/x -6/x^2\Big) -1440 {\cal L_+} \Big(3 + 2/x \Big)
    \bigg]\, ,
    \\
    \rho^{\langle \bar{q}q \rangle}(t) &=&
    - \frac{m_s m_Q^4}{3 \cdot 2^5 \pi^4} \Big( 
    2\langle \bar{q}q \rangle + \langle \bar{s}s \rangle \Big)
    \bigg[ v \Big(12 + 16/x - 1/x^2 \Big)
    + 12{\cal L}_v \Big( 2x - 3 \Big) \bigg] \, ,
    \\
    \rho^{\langle G^2 \rangle}(t) &=& 
    - \frac{m_Q^4 \langle G^2 \rangle}{3 \cdot 2^{10} \pi^6}
    \bigg[ v \Big( 30 + 31/x + 5/x^2 \Big) + 6 {\cal L}_v
    \Big( 10x - 6 - (5 - 1/x)\log(x) - 4/x \Big) 
    - 12{\cal L_+} \Big( 5 - 1/x \Big) \bigg]\, ,
    \\
    \rho^{\langle \bar{q}Gq \rangle}(t) &=&
    \frac{m_s m_Q^2}{3 \cdot 2^7 \pi^4} \Big( 
    12 \langle \bar{q}Gq \rangle + 7 \langle \bar{s}Gs \rangle
    \Big) \:v \Big( 4 - 1/x \Big)\, ,
    \\
    \rho^{\langle \bar{q}q \rangle^2}(t) &=& 
    \frac{m_Q^2 \langle \bar{q}q \rangle 
    \langle \bar{s}s \rangle}{6 \pi^2} 
    \:v \Big( 4 - 1/x \Big)\, ,
    \\
    \rho^{\langle G^3 \rangle}(t) &=& 
    -\frac{m_Q^2 \langle G^3 \rangle}
    {5\cdot 3^3\cdot 2^{9} \pi^6} \bigg[ 
    v \Big(430 + 495/x - 6/x^2 + 3m_Q^2\tau\,
    (1/x^3 + 26/x^2) \Big) + \\ 
    &&
    + {\cal L}_v \Big( 860x - 232 - 5(87 + 10/x) \log(x) 
    - 285/x - 36m_c^2\tau\, (1/x^2 + 1/x) \Big) - 
    10 {\cal L_+} \Big(87 + 10/x \Big)\bigg]\, .
\end{eqnarray*}

\section{Spectral function of $T^{1^+}_{QQ\bar{q}\bar{s}}$ corresponding to the  current in Eq.\,\ref{eq:malt} used in\,\cite{MALT-T}\label{app-c}}

\begin{eqnarray*}
    \rho^{pert}(t) &=& \frac{m_Q^8}{3^4\cdot 2^{16} \,\pi^6} 
    \bigg[ v \Big( 2520x + 420 + 19128/x - 2016/x^2 
    - 454/x^3 + 13/x^4 \Big) + \\ 
    &&
    + 24{\cal L}_v \Big( 210x^2 + 363 - 6(51 + 8/x) \log(x) 
    - 592/x + 99/x^2\Big) - 288 {\cal L_+} \Big(51 + 8/x 
    \Big) \bigg] + \\
    &&
    -\frac{m_s m_Q^7}{3^3 \cdot 2^{12} \,\pi^6} 
    \bigg[ v \Big( 60 + 130/x - 18/x^2 - 1/x^3 \Big) + 
    12{\cal L}_v \Big( 10x - 4 - 6\log(x) - 6/x + 1/x^2 \Big) 
    - 144 {\cal L_+} \bigg] \, , \\
    \rho^{\langle \bar{q}q \rangle}(t) &=&
    \frac{7m_Q^5}{3^4 \cdot 2^{10} \pi^4} 
    \bigg[ \langle \bar{q}q \rangle \Big( 
    v (60x + 10 - 34/x - 9/x^2) + 24{\cal L}_v 
    ( 5x^2 - 3 + 2/x) \Big) + \\
    &&
    +12 \langle \bar{s}s \rangle \Big( 
    v (6 - 5/x - 1/x^2) + 6{\cal L}_v 
    ( 2x - 2 + 1/x) \Big) \bigg] - 
    \frac{m_s m_Q^4}{3^3 \cdot 2^{12} \pi^4} \bigg[ 
    8\langle \bar{q}q \rangle \Big( v (156 + 68/x + 1/x^2) + \\
    &&
    + 12{\cal L}_v ( 26x - 25 ) \Big) - 
    \langle \bar{s}s \rangle \Big( v (204x + 34 - 322/x + 
    39/x^2) + 24{\cal L}_v ( 17x^2 + 11 ) \Big)\bigg] \, ,
    \\
    \rho^{\langle G^2 \rangle}(t) &=& 
    \frac{m_Q^4 \langle G^2 \rangle}{3^5 \cdot 2^{16} \pi^6}
    \bigg[ v \Big( 2130x + 355 - 712/x -234/x^2 \Big) + 
    6 {\cal L}_v \Big( 710x^2 - 369 - 27\log(x) + 200/x \Big) 
    - 324 {\cal L_+} \bigg]\, ,
    %
    \end{eqnarray*}
    \begin{eqnarray*}
    \rho^{\langle \bar{q}Gq \rangle}(t) &=&
    -\frac{m_Q^3}{3^4 \cdot 2^{12} \pi^4} \bigg[ 
    2\langle \bar{q}Gq \rangle \Big( v(174x + 29 - 53/x) 
    + 6 {\cal L}_v ( 58x^2 + 36 - 7/x) \Big)  + \\
    &&
    - \langle \bar{s}Gs \rangle \Big( v(48x + 14 + 493/x) 
    + 12 {\cal L}_v ( 8x^2 + x - 84) \Big) \bigg]
    +\frac{m_s m_Q^2}{3^4 \cdot 2^{12} \pi^4} \bigg[ 
    \langle \bar{q}Gq \rangle \Big( 9v( 68 + 5/x) + \\
    &&
    - 6 {\cal L}_v ( 84x - 39) \Big)
    - \langle \bar{s}Gs \rangle \Big( 8v(3x - 10 + 7/x) 
    + 6 {\cal L}_v ( 8x^2 + 4x + 9) \Big) \bigg] \, ,
    \\
    \rho^{\langle \bar{q}q \rangle^2}(t) &=& 
    \frac{m_Q^2 \langle \bar{q}q \rangle 
    \langle \bar{s}s \rangle}{3^3 \cdot 2^6 \pi^2} 
    \:v \Big( 24 + 1/x \Big)
    - \frac{7m_s m_Q \langle \bar{q}q \rangle 
    \langle \bar{s}s \rangle}{3^3 \cdot 2^6 \pi^2}
    \:v \,s \,\tau\,, \\
     \rho^{\langle G^3 \rangle}(t) &=& 
    -\frac{m_Q^2 \langle G^3 \rangle}{ 5\,\cdot 3^8 \cdot\,2^{16} \pi^6}\bigg{[}
    v \Big{(} 5 (211680 x^3 - 277200 x^2 + 87288 x - 
           77680 - 89312/x - 1215/x^2) - \nnb\\
&&        12 m_Q^2 \tau (14700 x^2 - 10150 x + 2170 - 105/x + 792/
           x^2 - 144/x^3)\Big{)} +   \nnb\\  
&&  12 {\cal L}_v \Big{(} 5  \ga 35280 x^4 - 52080 x^3 + 22052 x^2 - 15138 x + 4593 + 
           6 (729 + 116/x) \log(x)  + 4274/x\dr - 
           \nnb\\
   &&     24 m_Q^2 \tau (1225 x^3 - 1050 x^2 + 315 x - 35 - 54/x + 9/
           x^2)\Big{)}  + 720 (729 + 116/x) {\cal L}_+\bigg{]}.
\end{eqnarray*}

Note that the spectral function related to the current:
\beq
{\cal O}_{T,2}^{8,\mu}=(\bar Q_j\gamma_5\frac{\lambda^{jk}_a}{2} u_k) (Q_m\,i\gamma^\mu\frac{\lambda^{mn}_a}{2} \bar d_n)\nnb
\eeq
used in Eq. 3 of Ref.\,\cite{MALT-T} can be deduced from the previous expression
by changing $s$ to $d$, $q$ to $u$ and by taking $m_s\equiv m_d=0$.


\vfill\eject

\end{document}